\documentclass[iop,revtex4]{emulateapj}
\usepackage{amsmath}
\usepackage{pdflscape}
\usepackage{natbib}
\usepackage{graphicx}
\usepackage[usenames]{xcolor}

\newcommand{\madcows}{MaDCoWS} % Project Acronym
\newcommand{\afm}{\altaffilmark} % Affiliation Mark for Author List
\newcommand{\spitzer}{\emph{Spitzer}}
\newcommand{\chandra}{\emph{Chandra}}
\newcommand{\hst}{\emph{HST}}
\newcommand{\irac}{IRAC}
\newcommand{\wise}{\emph{WISE}}
\newcommand{\neowise}{NEOWISE}
\newcommand{\planck}{\emph{Planck}}
\newcommand{\allwise}{AllWISE}

\newcommand{\rosat}{\emph{ROSAT}}

\newcommand{\msun}{~$\rm{M}_{\odot}$} % solar mass
\newcommand{\chone}{[$3.6${}]}
\newcommand{\chtwo}{[$4.5${}]}
\newcommand{\wone}{W1}
\newcommand{\wtwo}{W2}
\newcommand{\nspecz}{38}
\newcommand{\nspecztot}{39}
\newcommand{\sigspecz}{0.036}

\newcommand{\micr}{$\mu$m}
\newcommand\rfive{$r_{500}$}
\newcommand\mfive{$M_{500}$}
\newcommand\mstar{$M^\ast$}
\newcommand\lstar{$L^\ast$}
\newcommand\rtwo{$r_{200}$}

\newcommand\mtwo{$M_{200}$}

\newcommand\nspitzerpanstarrs{$1723$}
\newcommand\nspitzersupercosmos{$86$}
\newcommand\nsupercosmosdes{$64$}
\newcommand\nsdss{$156$}
\newcommand\npanstarrs{$2433$}
\newcommand\nsupercosmos{$250$}
\newcommand\panstarrs{Pan-STARRS}
\newcommand\ncarma{14}

\begin{document}
\title{The Massive and Distant Clusters of \wise\ Survey. I: Survey Overview and a Catalog of $>2000$ Galaxy Clusters at $z\simeq1$ }

%% Authorship
\author{%
  Anthony H. Gonzalez\afm{1}, 
  Daniel P. Gettings\afm{1}, 
  Mark Brodwin\afm{2}, %
  Peter R. M. Eisenhardt\afm{3}, 
  S. A. Stanford\afm{4}$^,$\afm{5}, 
  Dominika Wylezalek\afm{6}, %
  Bandon Decker\afm{2},
  Daniel P. Marrone\afm{7},
  Emily Moravec\afm{1},
  Christine O'Donnell\afm{7},
  Brian Stalder\afm{8},
  Daniel Stern\afm{3}, %
  Zubair Abdulla\afm{9,10},
  Gillen Brown\afm{2,11},
  John Carlstrom\afm{9,10},
  Kenneth C. Chambers\afm{12},
  Brian Hayden\afm{13},
  Yen-ting Lin\afm{14}, %
  Eugene Magnier\afm{12},
  Frank J. Masci\afm{15}, %
  Adam B. Mantz\afm{16,17}, %
  Michael McDonald\afm{18},
  Wenli Mo\afm{1},
  Saul Perlmutter\afm{13,19},
  Edward L. Wright\afm{20}, %
  Gregory R. Zeimann\afm{21} %
}

\altaffiltext{1}{Department of Astronomy, University of Florida, 211 Bryant Space Center, Gainesville, FL 32611, USA}
\altaffiltext{2}{Department of Physics and Astronomy, University of Missouri, 5110 Rockhill Road, Kansas City, MO 64110, USA}
\altaffiltext{3}{Jet Propulsion Laboratory, California Institute of Technology, Pasadena, CA 91109, USA}
\altaffiltext{4}{Institute of Geophysics and Planetary Physics, Lawrence Livermore National Laboratory, Livermore, CA 94550, USA}
\altaffiltext{5}{Department of Physics, University of California, One Shields Avenue, Davis, CA 95616, USA}
\altaffiltext{6}{European Southern Observatory, Karl Schwarzschild Stra{\ss}e 2, D-85748, Garching bei M{\"u}nchen, Germany}
\altaffiltext{7}{Steward Observatory, University of Arizona, 933 North Cherry Avenue, Tucson, AZ 85721, USA}
\altaffiltext{8}{LSST, 950 N. Cherry Ave, Tucson, AZ 85719, USA}
\altaffiltext{9}{Kavli Institute for Cosmological Physics, University of Chicago, 5640 South Ellis Avenue, Chicago, IL 60637, USA}
\altaffiltext{10}{Department of Astronomy and Astrophysics, University of Chicago, 5640 South Ellis Avenue, Chicago, IL 60637, USA}
\altaffiltext{11}{Department of Astronomy, University of Michigan, Ann Arbor, MI 48109, USA}
\altaffiltext{12}{Institute for Astronomy, University of Hawaii at Manoa, 2680 Woodlawn Drive, Honolulu, HI 96822, USA}
\altaffiltext{13}{Lawrence Berkeley National Laboratory, 1 Cyclotron Road, MS 50B-4206, Berkeley, CA 94720, USA}
\altaffiltext{14}{Institute of Astronomy and Astrophysics, Academia Sinica, Taipei 10617, Taiwan}
\altaffiltext{15}{Infrared Processing and Analysis Center, Caltech 100-22, Pasadena, CA 91125, USA}
\altaffiltext{16}{Kavli Institude for Particle Astrophysics and Cosmology, Stanford University, 382 Via Pueblo Mall, Stanford, CA 94305-4060, USA}
\altaffiltext{17}{Department of Physics, Stanford University, 382 Via Pueblo Mall, Stanford, CA 94305-4060, USA}
\altaffiltext{18}{Kavli Institute for Astrophysics and Space Research, Massachusetts Institute of Technology, 77 Massachusetts Avenue, Cambridge, MA 02139, USA}
\altaffiltext{19}{Department of Physics, University of California Berkeley, Berkeley, CA 94720, USA}
\altaffiltext{20}{UCLA Astronomy, P.O. Box 951547, Los Angeles, CA 90095-1547, USA}
\altaffiltext{21}{Hobby Eberly Telescope, University of Texas, Austin, TX 78712, USA}

\begin{abstract}
  We present the Massive and Distant Clusters of \wise\ Survey (\madcows), a search for galaxy clusters at $0.7\la z\la 1.5$ based upon data from
the {\emph Wide-field Infrared Survey Explorer} (\wise) mission. \madcows\ is the first cluster survey capable of discovering massive clusters at these
redshifts over the full extragalactic sky. The search is divided into two regions -- the region of the  extragalactic sky covered by \panstarrs\ ($\delta>-30^\circ$) and
the remainder of the southern extragalactic sky at $\delta<-30^\circ$ for which shallower optical data from SuperCOSMOS Sky Survey are available.
In this paper we describe the search algorithm, characterize the sample, and present the first \madcows\ data release  -- catalogs of the \npanstarrs\ highest amplitude detections in the \wise---\panstarrs\ region and the \nsupercosmos\ highest amplitude detections in the \wise---SuperCOSMOS region. 
A total of \nspitzerpanstarrs\ of the detections from the \wise---\panstarrs\ sample have also been observed with the \spitzer\ {\emph Space Telescope}, providing photometric redshifts and richnesses, and
an additional 64 detections within the \wise---SuperCOSMOS region also have photometric redshifts and richnesses.
Spectroscopic redshifts for \nspecz\ \madcows\ clusters with IRAC photometry demonstrate that the photometric redshifts have an uncertainty of $\sigma_z/(1+z)\simeq0.036$. Combining the richness measurements
with Sunyaev-Zel'dovich observations of \madcows\ clusters, we also present a preliminary mass-richness relation that can be used to infer the approximate mass distribution of the full sample. The estimated median mass for the \wise---\panstarrs\ catalog is \mfive\ $=1.6^{+0.7}_{-0.8}\times10^{14}$ M$_\odot$, with  the Sunyaev-Zel'dovich data confirming that we detect clusters with masses up to  \mfive$\sim5\times10^{14}$ M$_\odot$ (\mtwo\ $\sim10^{15}$M$_\odot$).
\end{abstract}


%% =========================================================================================================
%% Section 1  -- Introduction
%% =========================================================================================================
\section{Introduction}\label{intro}
Clusters of galaxies have historically been used as powerful probes of cosmology and galaxy evolution, providing
such landmark results as evidence for the existence of dark matter \citep[e.g.][]{zwicky1937, clowe2004, clowe2006}, and demonstration of the importance of environment in galaxy evolution \citep{dressler1980}.
Other notable results include early evidence
for a low-density universe \citep{white1993,luppino1995,carlberg1997}, constraints on the dark matter self-interaction cross-section \citep{arabadjis2002,markevitch2004,randall2008,harvey2015},
and competitive constraints on cosmological parameters \citep[e.g.,][]{vikhlinin2009,allen2011,mantz2014,bocquet2015,dehaan2016}. 
Most of the results listed above are based upon observations of the highest mass galaxy clusters %(\mtwo\ $>10^{15}$ \msun;a(
(\mfive\ $>5\times10^{14}$ \msun) -- and are primarily at low 
redshifts where well-characterized samples exist.
There are multiple reasons to expect that investigations of the massive cluster population at higher redshift have the potential to further  our understanding of both fundamental physics and galaxy formation.

The first detailed cluster investigations to extend
to $z\ga 1.5$ have yielded intriguing results on the formation and evolution of cluster galaxies.  While observations indicate that the bulk of the stellar population in these systems form at $z>2$ \citep[e.g.,][]{eisenhardt2008,mancone2010,snyder2012,andreon2013,foltz2015,cooke2015,muldrew2018}, some studies also suggest that significant galaxy assembly and star formation can continue to later times. 
For example, \citet{webb2015b} find that at $z=1-1.8$ star formation is an important and possibly dominant contributor to the growth of brightest cluster galaxies, with the Phoenix cluster \citep{mcdonald2015} providing one example of ongoing substantial BCG growth via star formation at lower redshift ($z=0.596$).
Several programs also find an inversion of the star formation $-$ density relation at $z>1.3$ \citep{tran2010,hilton2010,fassbender2011,brodwin2013,alberts2014,santos2015,ma2015,alberts2016}, with cluster cores having a significant population of strongly star-forming, luminous infrared galaxies and star formation densities exceeding the field level.

More generally, there are multiple lines of evidence (galaxy colors, infrared star formation rates, evolution of the luminosity function) consistent with $z\sim1.3-1.5$ being a transition epoch in the evolution of cluster galaxies for the clusters that have thus far been studied at this epoch \citep{brodwin2013, fassbender2014}.
These systems, however, are typically drawn from relatively small-area surveys  \citep[e.g. the IRAC Shallow Cluster Survey, the {\it XMM-Newton} Distant Cluster Project, and the {\it XMM} Cluster Survey;][]{eisenhardt2008,fassbender2011,ld2011} that lack the comoving volume necessary to discover signifanct numbers of massive clusters (\mfive\ $\ga 3\times10^{14}$ M$_\odot$) at this epoch.
As a consequence, they provide little leverage on the dependence of this transition epoch upon cluster mass. 
Samples of more massive clusters from wider area surveys at the same $z \sim 1.3$ epoch can be used to directly test the mass-dependence of this transition epoch.

For cosmology, the unique leverage provided by galaxy clusters comes primarily from their extreme mass 
 and late time growth that continues through the present epoch. Because of this late time growth, evolution of the cluster mass function is a very sensitive growth of structure test, which has been exploited by a number of groups to constrain cosmological parameters and place upper limits on neutrino masses \citep[e.g.,][]{benson2013,planckszcosmo2014}. The abundance of the most extreme mass clusters is also sensitive to details of the initial density fluctuations from inflation. Multiple groups have investigated whether the existing known massive clusters at high redshift are consistent with Gaussian density fluctuations at the end of inflation, or instead require primordial non-Gaussianity on cluster scales  \citep{cayon2011, enqvist2011, hoyle2011,williamson2011,harrison2012,hoyle2012}. 
Evidence favors the null hypothesis, but a definitive answer remains elusive due to small number statistics. 
For standard $\Lambda$CDM with Gaussian fluctuations,  there should only be $\sim15$ clusters over the entire sky at $z>1$ with \mtwo\ $>10^{15}$ \msun\ 
-- consistent with the single $>10^{15}$ \msun\ cluster known at this epoch prior to \madcows, which is from the South Pole Telescope \citep[SPT,][]{foley2011}.\footnote{This value is calculated for WMAP9 \citep{hinshaw2013} and Planck \citep{planck2016_13} cosmologies using {\tt hmf} \citep{murray2013} with a \citet{tinker2010} mass function.} 

Complementary to abundance-based constraints, measurement of the X-ray emitting gas mass fraction, $f_{gas}$, in the largest, dynamically-relaxed galaxy clusters has been used to provide an independent constraint on dark energy \citep{ettori2009,mantz2014}. Constraints on the dark energy equation of state from this method are competitive with other techniques \citep{mantz2014}, but are presently limited by the small number of massive, relaxed clusters known at high redshift. \citet{allen2013} demonstrate that doubling the size of the best current sample, which  includes $\sim 10$ relaxed clusters at $z>1$, can improve the figure of merit for the dark energy equation of state by more than an order of magnitude.

Wide area surveys provide the opportunity to identify well-defined samples of the most massive, rarest galaxy clusters.
The \rosat\ All-Sky Survey produced several catalogs of massive X-ray selected galaxy clusters to moderate redshifts \citep[e.g. BCS at $z<0.3$ and MACS at $z<0.7$;][]{ebeling1998,ebeling2001}, and the SDSS yielded large catalogs of nearby clusters spanning a wider cluster mass range \citep[e.g.][ $z\la0.5$]{koester2007,rozo2015}.  {\footnote{As we were submitting this paper, we became aware of a new paper by \citet{wh2018}, which presents a catalog of cluster candidates at a median redshift of $z=0.75$ -- a higher redshift than previous SDSS searches. They identify candidates by searching near spectroscopically-confirmed Luminous Red Galaxies at $z>0.65$ from SDSS for overdensities of \wise\ sources. While this is a fundamentally different approach than the one employed in this paper, it highlights the value of \wise\ for extending the redshift baseline of wide-area cluster searches}}.
  The \planck\ mission also provides an all-sky catalog of massive galaxy clusters extending to somewhat higher redshift \citep[50\% completeness limit of \mfive$\simeq 6\times10^{14} M_\odot$ at $z=1$,][]{planck2016_27}, while the SPT, Atacama Cosmology Telescope (ACT), and Dark Energy Survey (DES) can provide complementary samples reaching to $z\ga1$ drawn from 2000-5000 deg$^2$ \citep{hasselfield2013,bleem2015,des2016,hilton2017}.

The NASA {\emph Wide-field Infrared Survey Explorer} \citep[\wise;][]{wright2010} provides the means to conduct the first search for massive galaxy clusters at $z\sim1$ covering the full extragalactic sky. \wise\ is an infrared survey mission covering the entire sky in four bands,
3.4, 4.6, 12 and 22 \micr\ (designated W1-W4). The sensitivity in W1 is sufficient to detect $L^\ast$ galaxies to $z\ga1$ and the brightest galaxies in clusters out to $z\sim2$. Using the \wise\ W1 and W2 data, we have undertaken the Massive and Distant Clusters of \wise\ Survey (MaDCoWS) to identify the most massive high-redshift clusters at $0.7\la z\la 1.5$. The only other planned comparably wide-area survey at this epoch is eROSITA \citep[planned launch in 2019;][]{predehl2006}.  

The first cluster discovered by the \madcows\ survey, at $z=0.99$, was presented in \citet{gettings2012}. Subsequently, we have published spectroscopic redshift determinations for twenty clusters in \citet{stanford2014}, Sunyaev-Zel'dovich masses for five clusters in \citet{brodwin2015}, and confirmation of the second most massive cluster known at $z>1$ (\mtwo$\simeq10^{15}$ M$_\odot$, $z=1.19$) in \citet{gonzalez2015}.  In \citet{mo2018} and \citet{moravec2018} we also investigate the AGN populations associated with these clusters.
In this paper we describe the details of our cluster search, and release catalogs of both the top \npanstarrs\ cluster candidates identified using the combination of \wise\ and \panstarrs\ data at $\delta>-30^\circ$ and the top \nsupercosmos\ cluster candidates identified using the combination of \wise\ and SuperCOSMOS data at $\delta<-30^\circ$. 
We begin in section \ref{sec:data} by describing the catalogs used as inputs for the \madcows\ search and ancillary data acquired to characterize the sample. In section \ref{sec:algorithm} we discuss the algorithm employed in the search. We next discuss the detailed implementation of this algorithm in section \ref{sec:implementation}. In section \ref{sec:catalog} we present the catalog of the \npanstarrs\ highest significance detections within the \panstarrs\ region, and discuss candidate properties derived directly from the survey data. In this section we also provide catalogs of cluster candidates from our \wise---SDSS and \wise---SuperCOSMOS searches for which we have obtained assorted follow-up data. 
Section \ref{sec:characterization} then explores the properties of the \wise-\panstarrs\ catalog as characterized from follow-up observations.
Finally, in section \ref{sec:summary} we summarize the main results from this work.
Throughout this paper we use Vega magnitudes for \wise\ bands and AB magnitudes for optical bands unless otherwise stated. We use the \citet{planck2016_13} cosmological parameters  assuming a flat cosmology ($H_0=67.7$ km s$^{-1}$, $\Omega_0=0.307$). In this paper \rtwo\ (\rfive) refers to radius within which the enclosed density is 200 (500) times critical density, and \mtwo\  (\mfive) is the corresponding enclosed mass.

%% =========================================================================================================
%% Section 2 -- Survey Data  (WISE and SDSS)
%% =========================================================================================================
\section{Data Sets}\label{sec:data}

Conducting the \madcows\ search requires catalogs based upon \wise\ imaging coupled with catalogs derived from optical surveys. In this section we describe the input data sets used for \madcows. In addition, we present \spitzer/IRAC data used to better characterize the resultant cluster sample.
  
  %% ------------------
  %% WISE Data
  %% ------------------
  \subsection{\wise\ Data}\label{sec:data_wise}

  \wise\ \wone\ and \wtwo\ data are the foundation for the \madcows\ cluster search.
  For a description of the \wise\ satellite and survey strategy we refer the reader to \citet{wise_paper}.
  Our cluster search uses the \wise\ project data products created and distributed by NASA/IPAC, available at the Infrared Science Archive. Initial work to develop
  the \madcows\ algorithm was based upon the \wise\ All-Sky Data Release of 14 March 2012 \citep{wise_allsky_exsup}. The first \madcows\ clusters were discovered using the All-Sky Data Release \citep[e.g.]{gettings2012,stanford2014}. 
For
  this data release the survey scanning strategy yielded approximately 12 exposures at positions along the ecliptic plane, and a published 5$\sigma$ photometric sensitivity in the ecliptic plane
of 68~$\mu$Jy and 111~$\mu$Jy (16.63 and 15.47 mag Vega) in the W1 and W2 bands.  
Sensitivity improves toward the ecliptic poles due to the denser coverage and lower zodiacal background \citep{wright2010}.

The current
  search is based upon the updated \allwise\ Data Release from 13 November 2013,
with approximately twice the coverage depth in W1 and W2 \citep{wise_allwise_exsup}.
  Full descriptions of the data processing and catalog constructions for each are contained in the Explanatory Supplements \citep{wise_allsky_exsup,wise_allwise_exsup}.
  The \allwise\ release yields both improved sensitivity and uniformity of coverage relative to the earlier All-Sky release, and also significantly reduces
the flux underestimation bias that impacted the All-Sky release.  The 5$\sigma$ 
depths for the \allwise\ release are 54~$\mu$Jy and 71~$\mu$Jy (16.96 and 15.95
 mag Vega) in the W1 and W2 bands for low coverage sky regions (23 exposures) along the ecliptic plane \citep{wise_allwise_exsup}. For regions away from the Galactic plane that are not confusion-limited, the \allwise\ release  enables uniform selection down to these magnitudes.

  The primary data used in the cluster search comes from the \allwise\ source catalog, which provides positions and profile-fitting-derived fluxes for over 747 million sources over the full sky.
  In the public catalogs provided by IPAC, the position and flux information are derived from a combination of the deep coadds in the AllWISE Image Atlas and the single-exposure (L1b) frames.
  The initial source positions for the catalog are derived from the deep coadds using a multi-wavelength $\chi^2$ technique that combines information from all four bands simultaneously \citep{marsh2012}.
  Based on this initial list, procedures for profile-fitting photometry and source deblending are performed on the L1b frames at each source position.
  We note that the resolution of \wise\ ($6.1^{\prime\prime}$ in W1 and $6.4^{\prime\prime}$ in W2) effectively suppresses detection of sources within 10$^{\prime\prime}$ of one another due to blending. As shown   in section VI.2.c.iv (Figure 27) of the All-Sky Explanatory Supplement \citep{wise_allsky_exsup},
  few sources are detected within 10$^{\prime\prime}$ of another source. We discuss the impact of blending on the search in \S \ref{sec:physicalmotivation}.

  %% ------------------
  %% SDSS Data
  %% ------------------
  \subsection{Optical Data}
  In addition to the \wise\ photometry, we also use data from ground-based optical surveys to reject foreground galaxies (as described below). For the initial phase of this program, including clusters published in \citet{gettings2012} and \citet{stanford2014}, we used the Sloan Digital Sky Survey \citep[SDSS;][]{sdss_paper}, which restricted our search to the SDSS footprint. The SDSS data set has now been superseded by the \panstarrs\ $3\pi$ survey  \citep{chambers2016}, which extends to $\delta=-30^\circ$. At more southern latitudes we have also investigated use of the SuperCOSMOS Sky Survey \citep{hambly2001a,hambly2001b}. While we present details of all three surveys here, \panstarrs\ provides the primary optical data for the current \madcows\ search.

%% ---------------
%% Models vs. z
%% ---------------
\begin{figure*}
 \includegraphics[width=0.95\textwidth,keepaspectratio=true]{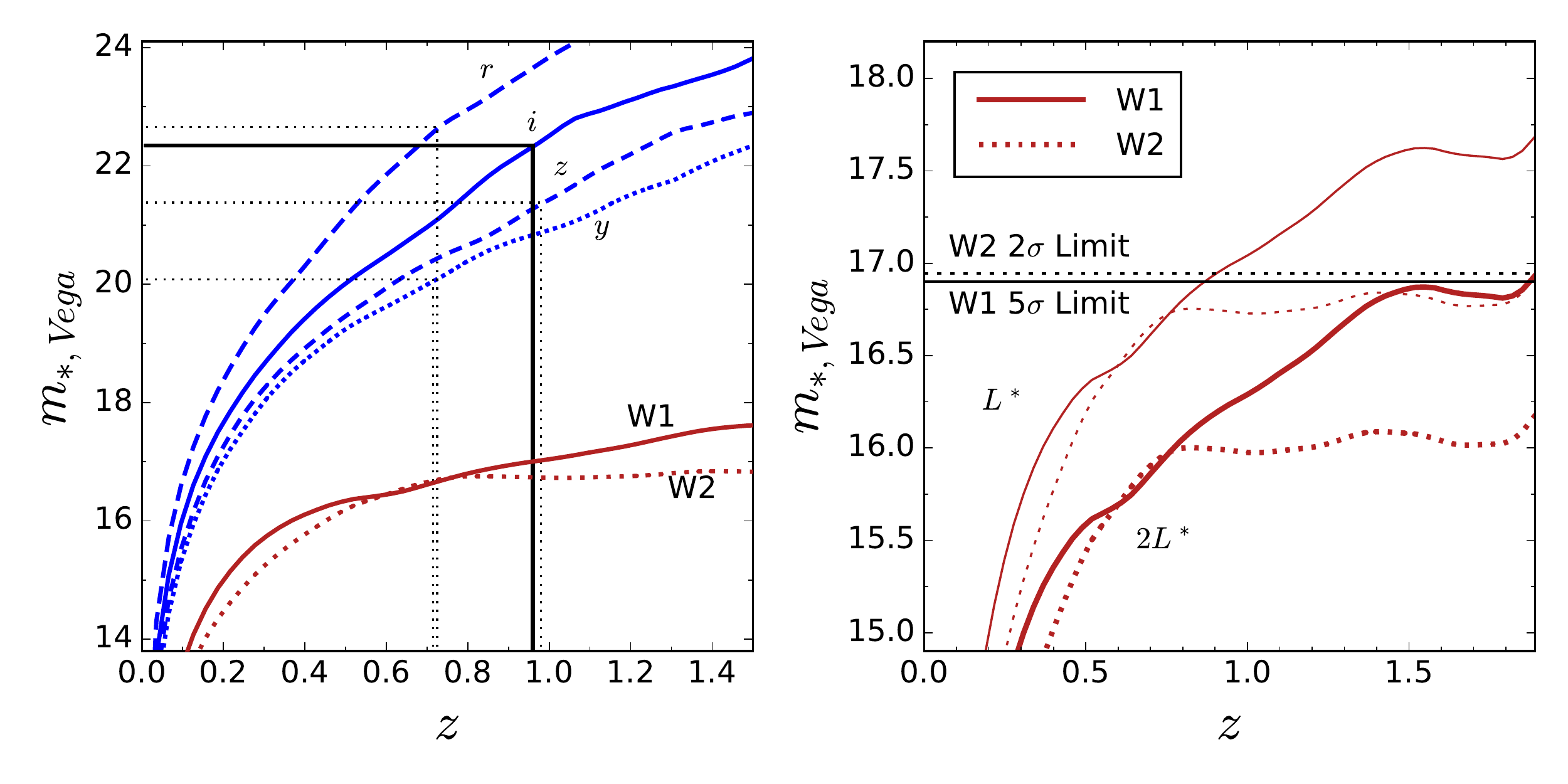}
 \caption{The apparent magnitude of a passively evolving \lstar\ galaxy as a function of redshift in various passbands. The plotted curves are for Vega magnitudes and are based upon a \citet{conroy2009} model with
 a single $\tau=0.1$~Gyr exponential burst of star formation at $z=3$ and a Chabrier IMF normalized using the $z=1.1$  \spitzer\ \chone\ luminosity function from \citet{mancone2010}. Similar results are obtained for a \citet{bc03} model.
   {\it Left panel:} $m^*$ versus $z$ in \wise\ W1 and W2, and in the four reddest bands provided by \panstarrs\ (Vega magnitudes in all bands).  We also plot horizontal lines denoting the \panstarrs\ 50\% completeness limits in each band. These are the galaxy completeness limits, which are taken to be 0.4 mag brighter than the point source completeness limits \citep{metcalfe2013}. The vertical lines indicate the corresponding redshift reach in each filter for the model galaxy.
 Among the \panstarrs\ passbands, the $i-$band and $z-$band have the greatest redshift reach. 
For \madcows\ we use the $i-$band both to provide a greater wavelength lever arm relative to the \wise\ bands and for consistency with our preliminary SDSS search.  The W1 and W2 curves are much flatter with redshift due to 
negative e+k corrections offsetting the impact of increasing luminosity distance at these wavelengths.
   {\it Right panel:} The horizontal lines show the relevant sensitivity thresholds in W1 and W2 at the median depth of the AllWISE Survey. Galaxies detected at $5\sigma$ in W1 are included in the source catalog; these sources are considered non-detections in W2 if their fluxes fall below a 2$\sigma$ threshold. 
For comparison we plot the apparent magnitudes of \lstar\ and 2 \lstar\ galaxies in the \wise~bands. Individual \lstar\ galaxies at $z \ga 1.1$ are not detected at 5$\sigma$ in W1 for the AllWISE survey depth, but blends of two \lstar\ galaxies are detectable over the full redshift range shown. 
}
 \label{fig:sdss_wise_phot_depth}
\end{figure*}

  \subsubsection{The SDSS Eighth Data Release}
  
  The Eighth Data Release of the SDSS \citep[DR8;][]{aihara2011} covers $14,555$ deg$^2$, mostly in the northern hemisphere, in five optical bands \citep[$ugriz$;][]{fukugita1996}. The 95\% completeness limits in these bands are $u, g, r, i, z = 22.0, 22.2, 22.2, 21.3, 20.5$ \citep[AB;][]{abazajian2009}.
As discussed below, the most important filter for \madcows\ is $i$-band, for which the median seeing is $\sim1.4^{\prime\prime}$.\footnote{\url{http://www.sdss3.org/dr8/imaging/other\_info.php}}

For the \madcows\ program, we use data from a more restricted area (hereafter referred to as the \wise---SDSS region). Specifically, we avoid regions at low Galactic latitude ($b<25^\circ$), and restrict our use of SDSS data to Galactic cap areas with large, contiguous coverage, avoiding areas with only thin strips of imaging \citep[e.g. see Figure 1 in ][]{aihara2011}. With these restrictions, the remaining SDSS area corresponds to 10,959 deg$^2$. After also considering area lost to masking, due to issues such as bright stars and low coverage by \wise, the net effective area in the \wise---SDSS region  is 10,290 deg$^2$ (see \S \ref{subsec:maps}). 

   \subsubsection{\panstarrs}
The \panstarrs\ PS1 3$\pi$ Steradian Sky Survey \citep{chambers2016} is designed to provide complete coverage for $\delta>-30^\circ$ in $grizy$ with better than 1\% photometry in the $grizy$ bands.
This data set supercedes SDSS in both area and depth for the \madcows\ search.
There have been three internal releases (processing versions; PV) of stacked $3\pi$ catalogs, plus the Public Data Release DR1 \citep{flewelling2016}, which corresponds to PV3.  For \madcows\ we are using $i-$band data from the PV2 catalog. This catalog uses the same input image set as PV3/DR1, but differs slightly in how the PSF photometry and star/galaxy flags are implemented.
We refer the reader to \citet{laevens2015} for details on the differences between the different preliminary versions, noting that PV2 is sufficient for the \madcows\ search since we are only concerned with galaxy photometry in the $i-$band. Most relevant for \madcows, the $i-$band data are $\sim0.9$ mag deeper than SDSS \citep[$5\sigma$,][]{metcalfe2013}, yielding lower photometric uncertainties and hence cleaner selection of input galaxies for the cluster search.  As with our initial SDSS analysis, with \panstarrs\ we avoid regions at low Galactic latitude. Specifically, we require Galactic latitude $|b|>25^\circ$, increasing this limit to $|b|>30^\circ$ for Galactic longitude within 60$^\circ$of the Galactic center. 
These limits correspond to an extragalactic sky area of
23,290 deg$^2$, with \panstarrs\ covering 18,120 deg$^2$ (78\% of the extragalactic sky).  After removal of masked regions, areas with low coverage from \wise, and the region near the Galactic plane, the net effective area for the \panstarrs\ search is 17,668 deg$^2$, or 76\% of the extragalactic sky.

\subsubsection{The SuperCOSMOS Sky Survey}
\label{sec:supercosmos}

While there are multiple ongoing large optical surveys designed to map large areas of the extragalactic sky extending beyond the \panstarrs\ footprint \citep[e.g.][]{keller2007, shanks2013, des2016}, 
no surveys with depth comparable to \panstarrs\  (or SDSS) yet provide uniformly calibrated catalogs over a large fraction of the sky. At $\delta<-30^\circ$ we have therefore undertaken a shallow search using optical data from the SuperCOSMOS Sky Survey \citep{hambly2001a,hambly2001b}. The SuperCOSMOS project digitized photographic plates from multiple Schmidt telescopes, initially in the southern hemisphere, with coverage subsequently extended to the entire sky \citep{hambly2009}. 
We use SuperCOSMOS data from the UK Schmidt Telescope Red Southern and Equatorial Surveys and the Palomar-II Oschin Schmidt Telescope Red (IIIaF) plates, which are quoted as having nominal depths of $R_F\simeq21.5$ (Vega).\footnote{\url{http://www-wfau.roe.ac.uk/sss/surveys.html}} From our own testing of the data, we find that it is necessary to restrict our attention to $R_F<20.5$ to avoid non-uniformity due to depth variations between plates. Subsequent to our search, \citet{peacock2016} have constructed an updated all-sky SuperCOSMOS catalog. From the calibration in their analysis, $R_F=20.35$ corresponds to a 4$\sigma$ detection; we therefore are using a threshold slightly below 4$\sigma$.
The astrometry for this data is accurate to $\pm0.3^{\prime\prime}$ at this depth \citep{hambly2001c}, which is sufficient for our program. The total and net effective areas at $\delta<-30^\circ$ for the SuperCOSMOS  search are 4,260 deg$^2$ and  3,828 deg$^2$, respectively. Between \panstarrs\ and SuperCOSMOS,  we are able to detect clusters across the entire extragalactic sky, with a total combined area of 21,814 deg$^2$ after accounting for masking. As discussed in subsequent sections, use of SuperCOSMOS   data does yield a significant degradation of the search due to less effective rejection of lower redshift galaxies. 

\subsection{The Dark Energy Survey}

The Dark Energy Survey is in the process of mapping $\sim5000$ deg$^2$ in the region of the southern Galactic cap in the $grizY$ passbands \citep{desdr1}. Over half of the DES footprint lies south of $\delta=-30^\circ$ and hence outside the \panstarrs\ area, thus providing complementary optical imaging. The photometric depths for the first data release (DES DR1) are $g=24.33$, $r=24.08$, $i=23.44$, $z=22.69$, and $Y=21.44$ \citep[10$\sigma$;][]{desdr1}. While the DES DR1 was not availble in time to incorporate into the current \madcows\ cluster search, we use the $i-$band photometry in \S \ref{sec:redshifts} to derive photometric redshifts for the subset of cluster candidates that lie within the DES footprint but outside the \panstarrs\ survey area.

\subsection{The Spitzer Space Telescope}
 Our team was awarded \spitzer\ time during Cycles 9, 11, and 12 to obtain \chone\ and \chtwo\ imaging for $1959$ cluster candidates (PI: Gonzalez, PIDs 90177 and 11080), enabling photometric redshift and richness estimates. 
For the Cycle 9 program, we targeted 200 candidates 
from a preliminary \wise---SDSS search. In the Cycle 11 -- 12 snapshot program we targeted an additional  $1759$ clusters, selected by peak amplitude in the \wise---\panstarrs\ and \wise---SuperCOSMOS\ searches  (see \S \ref{sec:extract}), that were not previously observed in Cycle 9. We obtain total exposure times of 180 s in each band using a $6\times30$ s cycling dither pattern.  These two programs, both conducted during the \spitzer\  ``warm'' mission,  and existing archival data together yield IRAC \chone\ and \chtwo\ imaging for $1967$ \madcows\ clusters. Of these, \nspitzerpanstarrs\ are in the \wise---\panstarrs\ catalog presented in this paper,  and \nspitzersupercosmos\ are in the \wise---SuperCOSMOS catalog. The remainder are within the \wise---\panstarrs\ footprint, but are detected at lower significance.

Data were reduced using the MOPEX \citep{makovoz2005} package and source extraction was performed using SExtractor \citep{bertin1996} in dual image mode with the \chtwo\ image serving as the detection image. 
During the \spitzer\ warm mission the FWHM values for the point spread function are 1.95$^{\prime\prime}$ and 2.02$^{\prime\prime}$ for \chone\ and \chtwo, respectively, providing a factor of three improvement in spatial resolution relative to \wise.
Following the methodology of \citet{wylezalek2013}, we determine that at 10$\mu$Jy the recovered source density in our fields is 95\% that of SpUDS at the same threshold.
For subsequent analysis we include only sources with $f_{4.5} > 10~\mu$Jy. % with this limit determined following the methodology of \citet{wylezalek2013}. 
We measure our completeness by comparing the \madcows\ \spitzer\ number counts with number counts from the  \spitzer\ UKIDSS Ultra Deep Survey (SpUDS, PI: J. Dunlop) survey. The SpUDS survey is a \spitzer\ Cycle 4 legacy program that observed $\sim$1 deg$^2$ in the UKIDSS UDS field with \irac\ and the Multiband Imaging Spectrometer \citep{rieke2004}, reaching $5\sigma$ depths of $\sim 1~\mu$Jy at 3.6$\mu$m. 

\subsection{CARMA}
We were awarded time with the Combined Array for Research in Millimeter-wave Astronomy (CARMA)\footnote[1]{mmarray.org} between 2012 and 2014 (PIDs c0884, c1128, c1197, c1272, c1303) to observe a selection of the richest cluster candidates at 31 GHz. We were also awarded time in 2014 September (PID c1272) to target $\sim 150$ cluster candidates with very short exposures to identify very massive clusters. Most of the observations were made with the array in the 23-element CARMA-23 mode, with the exception of the 2012 observations from c0884 and c1128, which used only the eight element SZA. Detections from the pilot run in 2012 and 2013 are presented in \citet{brodwin2015}, and MOO J1142+1527, observed in 2014, is presented in \citet{gonzalez2015}. In the present work, the data were re-reduced using a new version of the SZA MATLAB pipeline \citep{muchovej2007} updated to handle 23-element data and produce uv-fits files and the CLIMAX software was used to fit pressure profiles from \citet{arnaud2010} to the data. The spherically-integrated Comptonization was measured from the Arnaud model, and \mfive, \rfive, and $Y_{500}$ were calculated by forcing consistency with the \citet{andersson2011} scaling relation. A more detailed description of the observations and analysis is given in \citet{brodwin2015} and \citet{decker2018}.

%% =========================================================================================================
%% Section 3 -- Cluster Finding with WISE: The Algorithm
%% =========================================================================================================
\section{Cluster Finding with \wise : The Algorithm}
\label{sec:algorithm}

  %% ----------------------------------------
  %% Physical Motivation
  %% ----------------------------------------
  \subsection{Physical Motivation For Search Algorithm}
\label{sec:physicalmotivation}

  A number of authors have demonstrated that the stellar mass content of cluster galaxies is tightly correlated with the total cluster mass \citep[e.g.][and reference therein]{Lin2004,Lin2012,mulroy2014}. \citet{mulroy2014} for example find an intrinsic scatter of only $\sim 10$\% between the $K-$band luminosity and weak lensing mass for nearby clusters. The \wone\ and \wtwo\ \wise\ bands, which probe approximately rest-frame $H$ and $K_s$ at $z\simeq 1$, trace the total stellar mass content, while being relatively insensitive to the age of the stellar population.
In the \wise\ bands the apparent magnitude of an \lstar\ galaxy is only weakly dependent upon redshift at $z\gtrsim 0.7$ due to e+k corrections that offset the impact of increasing luminosity distance (Figure \ref{fig:sdss_wise_phot_depth}). Consequently, a magnitude-limited galaxy sample selected with \wise\ has a roughly uniform luminosity limit within this redshift range.
Photometry from \wise\ therefore provides a proxy for stellar mass that is relatively unbiased by star-formation history, and the uniform luminosity limit translates to a uniform selection in stellar mass.

The primary observable for galaxy-based cluster searches is the projected overdensity of galaxies. The luminosity function of cluster galaxies is
well-parameterized by the Schechter function \citep{schechter1976},
and recent papers have demonstrated that at \wise\ wavelengths high-redshift galaxy clusters have relatively flat faint end slopes \citep[e.g.][]{mancone2012}. Combined with the rising number counts of
the field population at faint magnitudes, a cluster will have the greatest density contrast relative to the background population when the limiting magnitude of the input galaxy catalog is slightly below \lstar.
Thus, while \wtwo\ offers the more uniform stellar mass limit with redshift, for \madcows\ we use a \wone$-$selected galaxy sample due to the greater depth relative to \lstar\ in this band. 
For $z\simeq1$ the \wone\ \allwise\ imaging reaches 1.1\lstar\ at 5$\sigma$, while the [4.6] imaging only
reaches approximately 2.1\lstar\ at 5$\sigma$ (\mstar$-0.8$), or 0.85\lstar\ at 2$\sigma$ (\mstar$+0.2$; Figure \ref{fig:sdss_wise_phot_depth}). 
With \madcows\, we are therefore effectively searching for $z\sim1$ galaxy clusters via overdensities of galaxies with luminosities of approximately \lstar\ or greater.

\begin{figure*}
\includegraphics[width=0.99\textwidth,keepaspectratio=true]{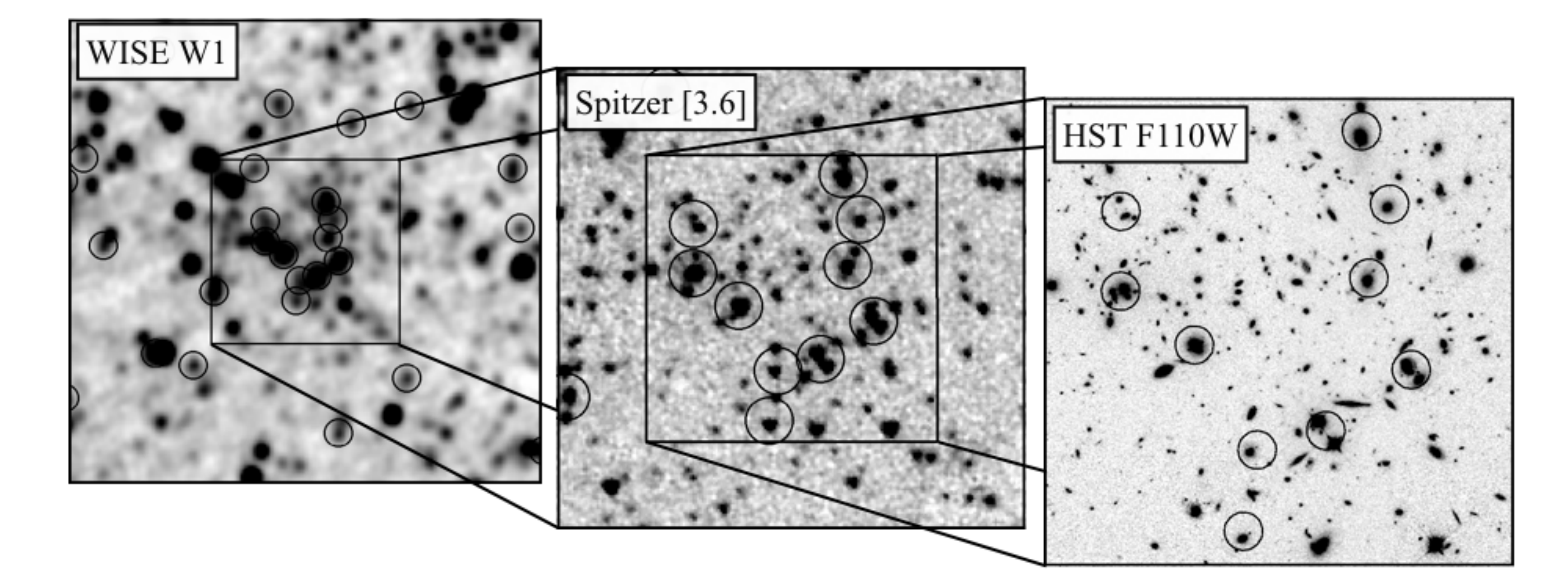}
\caption{Progressively zoomed images for the cluster MOO J1514+1346 \citep[$z=1.059$,][]{brodwin2015}. The left panel is a $5^\prime\times5^\prime$ W1 image from the \allwise\ survey, while the middle panel is a deeper \spitzer\ \chone\ image from our Cycle 9 program of the central $2^\prime$ region. In the right panel we show an \hst\ F110W image of the central $75^{\prime\prime}$ (555 s exposure; Program 14456, PI: Brodwin), with black circles indicating the locations of the objects from the \wise\ catalog that contributed to detection of the cluster. The diameter of each circle is  $6.1^{\prime\prime}$ in the \spitzer\ and \hst\ images, equivalent to the FWHM of the \wise\ W1 PSF. For the \wise\ image we use a larger 9$^{\prime\prime}$ circle for clarity. This sequence of images illustrates the impact of blending upon detection. While bright cluster galaxies are the main source of signal for cluster detection, unresolved blends of cluster galaxies also contribute to cluster detection.}
\label{fig:blending_comparison}
\end{figure*}

In practice, one additional consideration that impacts the effective depth is source blending in \wise\ due to the large PSF. Blending affects the number of observed galaxies in two competing ways. First, blending 
decreases the number of individual detections for galaxies brighter than the apparent magnitude limit. Second,
blending leads to detections arising from blends of galaxies that are individually fainter than the detection
limit. For the general field population the net impact of these two factors will be a uniform shift in the number counts as a function of magnitude, which does not impact our cluster search. For clusters, both factors
will have the greatest effect in the core region where the projected density is highest. For \madcows, because the magnitude limit is close to \lstar, the second effect will generally dominate due to the higher surface density of galaxies with $L<$\lstar\ compared to super-\lstar\ galaxies. 
The \madcows\ search therefore ends up benefitting from inclusion of blended galaxies that are individually somewhat fainter than the nominal \wise\ detection limit.   For illustration, we show \wise\ and \spitzer\ imaging for one of the spectroscopically confirmed \madcows\ clusters in Figure \ref{fig:blending_comparison}.

  %% ----------------------------------------
  %% Algorithm Details
  %% ----------------------------------------
  \subsection{Algorithm Details} %%
  The concept for the \madcows\ algorithm, though different in detail, is in the spirit of %a similar vein to
previous cluster searches using \spitzer\ data. The basic idea is to first isolate the distant galaxy population, using color and magnitude cuts to minimize foreground contamination, and then use wavelet filtering to
identify the most significant overdensities on cluster scales.  The color and magnitude selections, as described below, are similar to those employed 
by \citet{papovich2008} and \citet{muzzin2013},
while the wavelet technique draws upon the legacy of the ISCS and IDCS \citep{eisenhardt2008,stanford2012}. 

%% ---------------
%% W1-W2 vs. z and Source Selection
%% ---------------
\begin{figure*}
  \includegraphics[width=0.99\textwidth,keepaspectratio=true]{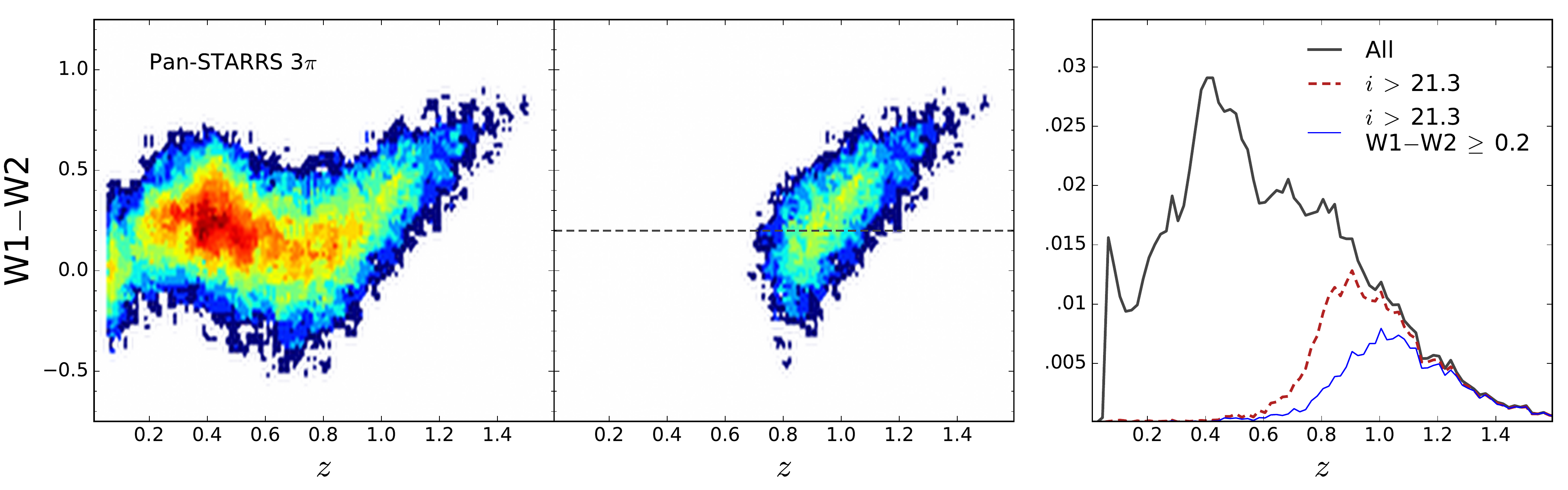}
  \caption{Illustration of the impact of the color and magnitude cuts using a sample of \wise\ sources matched to the %\irac-selected
    photometric redshift catalog for \irac-selected sources in the NDWFS region \citep{brodwin2006}. 
    {\it Left:} Density map of the \wise\ color distribution as a function of photometric redshift for all galaxies with \wone$<16.9$.
    {\it Center:} Density map showing the redshift distribution of galaxies from the left panel with $i>21.3$. The optical magnitude cut, based upon \panstarrs\ photometry, effectively removes foreground galaxies at $z<0.8$. The dashed line shows the \wise\ color cut. Use of SDSS rather than \panstarrs\ photometry in the initial search yields a nearly identical selection.
    {\it Right:} The redshift distribution of the full galaxy sample from the left panel (black), galaxies with $i>21.3$ (red), and those passing both the optical magnitude and \wise\ color cuts (blue). The addition of the \wone$-$\wtwo\ color cut increases the mean redshift of the remaining galaxy population, but has minimal impact on foreground removal. 
  }
  \label{fig:wise_color_z}
\end{figure*}

\begin{figure*}
  \includegraphics[width=0.99\textwidth,keepaspectratio=true]{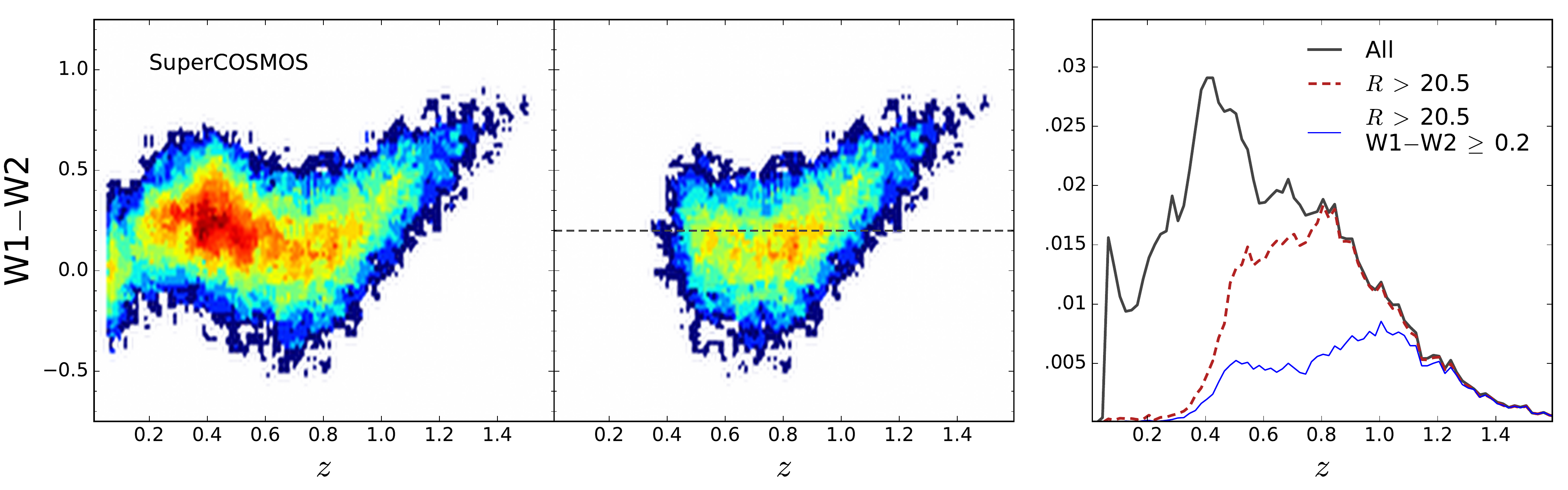}
\caption{The panels in this figure are the same as in Fig. \ref{fig:wise_color_z}, but now using %SDSS (top) and SuperCOSMOS (bottom) 
SuperCOSMOS photometry for optical rejection. 
this figure illustrates that the optical magnitude and \wise\ color cuts are less effective for the shallower SuperCOSMOS catalog, resulting in a significantly higher fraction of low-redshift interloper galaxies.}
\label{fig:wise_supercosmos_color_z}
\end{figure*}

\subsubsection{Galaxy Selection}
\label{sec:galcuts}

  For the \madcows\ cluster search, we start with the full \wise\ catalog of all sources detected at 5$\sigma$
in \wone. We then impose a magnitude cut \wone$<$16.9 to enforce uniformity of depth for the galaxy catalog.\footnote{From the AllWISE Explanatory Supplement, this magnitude corresponds to 5.3$\sigma$ depth in typical low sky coverage regions.} 
  
  The optical magnitude criterion is applied next. Within the \panstarrs\ region
we reject sources with $i<21.3$ ($i<20.93$ Vega). In Figure \ref{fig:wise_color_z} we cross-match \wise\ sources within the NOAO Deep Wide-Field (NDWFS) region with a photometric redshift catalog for \irac-selected sources from \citet{brodwin2006} to illustrate the impact of our cuts. As can been seen in this Figure, the optical rejection
effectively removes galaxies at $z\la 0.8$. In Figure \ref{fig:wise_color_z} we also show the redshift distribution in the \wise\ bands of all sources surviving this cut.  The $i-$band magnitude of this cut is predominantly empirical based upon the data shown in Figure \ref{fig:wise_color_z}, but set at a physical level where no cluster galaxies, except potentially BCGs at $z\simeq0.8$ are removed. For the same evolutionary model as in Figure \ref{fig:sdss_wise_phot_depth}, this magnitude limit corresponds to a 1.8 \lstar\ galaxy at $z=0.8$. Use of a brighter magnitude cut increases foreground contamination, while using a significantly fainter cut would diminish the cluster signal.
Outside the \panstarrs\ region we reject sources with $R_F<20.5$ from SuperCOSMOS, a shallower cut that is less effective at removing low-redshift interloper galaxies. In Figure \ref{fig:wise_supercosmos_color_z}, we illustrate the approximate impact of this cut by applying an $R>20.5$ cut within NDWFS. These interlopers decrease the density contrast between clusters and the field -- and hence larger scatter between detection amplitude and richness, and also result in higher contamination of the sample by low-redshift clusters (see \S \ref{sec:zdist}).

Subsequent to the optical cut, we impose a \wise\ color cut, rejecting objects with \wone$-$\wtwo$<0.2$. 
As a precaution at this stage we also reject galaxies not detected at 2$\sigma$ in \wtwo. 
The \wise\ color cut preferentially removes galaxies at $z<0.8$ from the galaxy population remaining after the optical rejection.
For the \wise---\panstarrs\ region, the median redshift increases from 0.93  to 1.01 with the addition of the \wise\ color cut  %  galaxy redshift distribution (top right panel of Figure \ref{fig:highzsdssgals});
 (see the redshift distributions in the rightmost panel of Figure \ref{fig:wise_color_z}). % [verify exact numbers]. 
As a result, clusters at $z\la1$ are downweighted in the \wise\---\panstarrs\ search. 
Outside the \panstarrs\ region,
because of the shallower SuperCOSMOS optical cut, this color cut is vital for reducing contamination from galaxies at $0.5\la z\la 0.8$. This can be seen in the center and right panels of Figure \ref{fig:wise_supercosmos_color_z}. It is worth emphasizing that even with the \wise\ color cut, the lack of SDSS- or \panstarrs-quality optical data has a 
detrimental impact on the search at $\delta<-30^\circ$. % as the surface density of galaxies after these cuts is a factor of XX higher than within the \wise---SDSS region. 
We discuss in \S \ref{sec:future} prospects for an improved southern search.

\subsubsection{Identifying Galaxy Overdensities}

\begin{figure}
\includegraphics[width=0.45\textwidth,keepaspectratio=true]{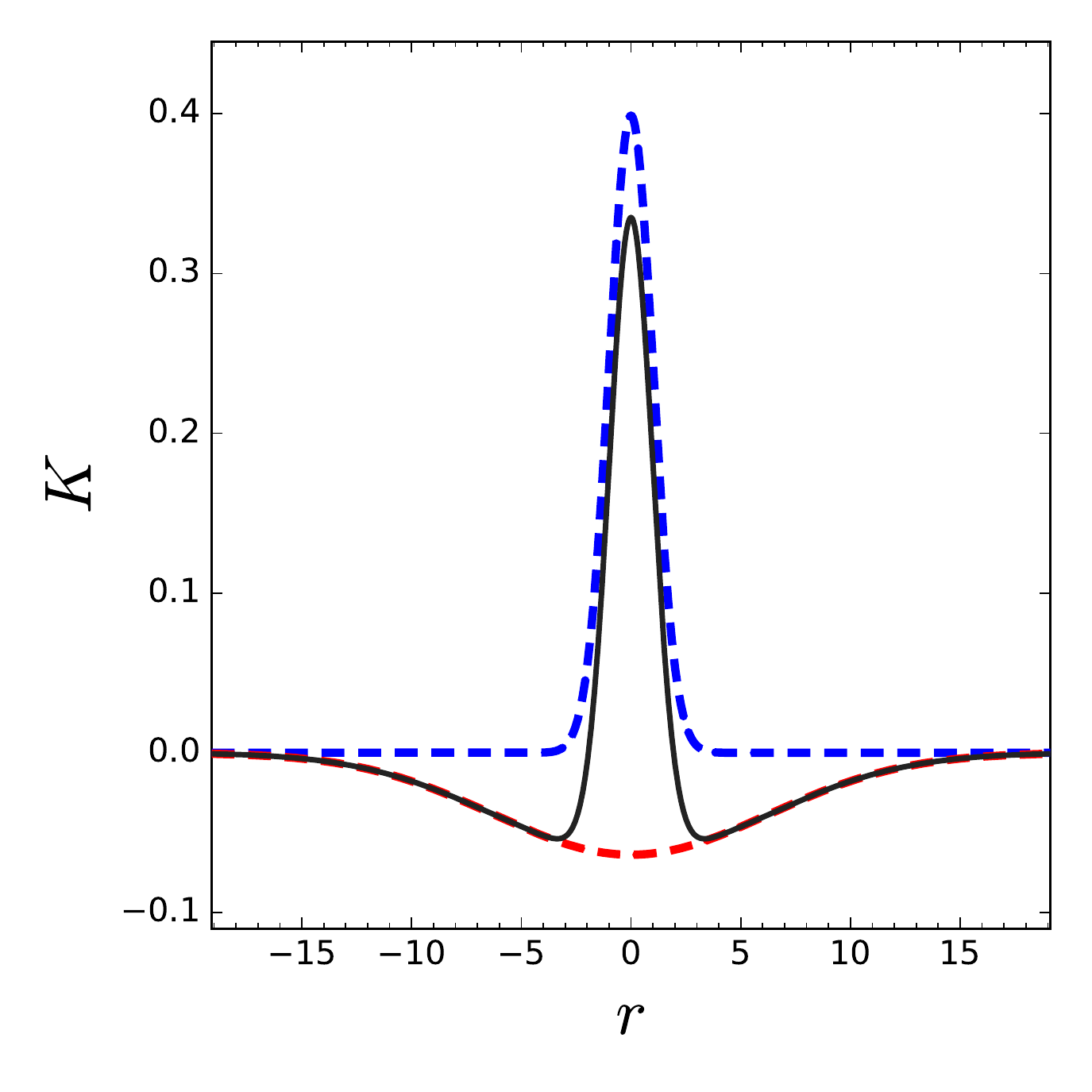}
\caption{A cross-sectional representation of a normalized, two-dimensional Difference-of-Gaussians kernel (black, solid curve). For illustration, we plot a kernel with $\sigma_1=1$ and a 2.5:1 scale ratio. The two dashed curves show the inner (blue) and outer (red) Gaussians used to construct the kernel. This kernel acts as a bandpass filter. Structures on scales smaller than the inner kernel are smoothed out, while those on scales larger than the outer component are effectively removed as a background component. }
\label{fig:kernel}
\end{figure}

From the filtered galaxy catalogs, we construct density maps with a resolution of 15$^{\prime\prime}$. These density maps are filtered with a Difference-of-Gaussians kernel (similar to a Mexican hat kernel) to identify cluster-scale overdensities. The functional form for this kernel is
\begin{equation}
K=\frac{1}{2\pi\sigma_1^2\sigma_2^2}\left(\sigma_2^2 \exp(-\frac{r^2}{2\sigma_1^2})  - \sigma_1^2\exp(-\frac{r^2}{2\sigma_2^2})\right),
\label{eq:dog}
\end{equation}
where $\sigma_1$ and $\sigma_2$ are the scales of the inner and outer Gaussians, respectively.
 This kernel functions as a bandpass filter (much like the filters in the SZ surveys), removing contributions to the density map from large scale structure and other sources of gradients in the projected galaxy density on large scales. The form of the kernel is shown in Figure \ref{fig:kernel}. Details regarding the specific scales set for the kernel are presented in \S \ref{subsec:maps}.

%% =========================================================================================================
%% Section 4 -- Cluster Finding with WISE: Implementation
%% =========================================================================================================
\section{Cluster Finding with \wise: Implementation}
\label{sec:implementation}

   \subsection{Catalog Cleaning}
   Both the \wise\ and optical catalogs contain quality flags for each source. For \wise, the catalog contains information
on sources that are flagged as contaminants in \texttt{cc\_flags}, which can arise from optical ghosts, diffraction spikes, persistence effects, or scattered light.  We reject sources with
\texttt{cc\_flags}$\neq 0$ in W1 or W2, as non-zero flags are indicative that the source detection may be unreliable or measurements for that source may be contaminated.  We also reject sources that are flagged as optical ghosts in either W3 or W4 as a precaution. While we are not using W3 and W4 photometry, the detection of an optical ghost in these bands is indicative of potential contamination from ghosts at shorter wavelengths -- which may not always be flagged.  

The above criteria are designed to maximize the purity of the \wise\ catalog, and hence minimize spurious cluster detections. For the optical catalogs, the more important factor is completeness because the optical photometry is used to reject low-redshift interlopers. Put simply, it is better to be able to use the existence of an optically bright source with some quality issues to identify a \wise\ source as low-redshift than to allow that interloper to contribute to the density map. We therefore minimize the rejection due to flagging in the optical catalogs to the extent possible.
For the SDSS catalog, we require that all sources are \texttt{primary} for the initial SQL query when downloading the data from CASJobs, but apply no additional filters. For \panstarrs\, we apply no filters to the source catalog. For SuperCOSMOS we reject sources for which the $R_F$-band quality flag indicates a severe defect. 

   \subsection{Matching \wise\ and Optical Catalogs}
\begin{figure}
\includegraphics[width=0.45\textwidth,keepaspectratio=true]{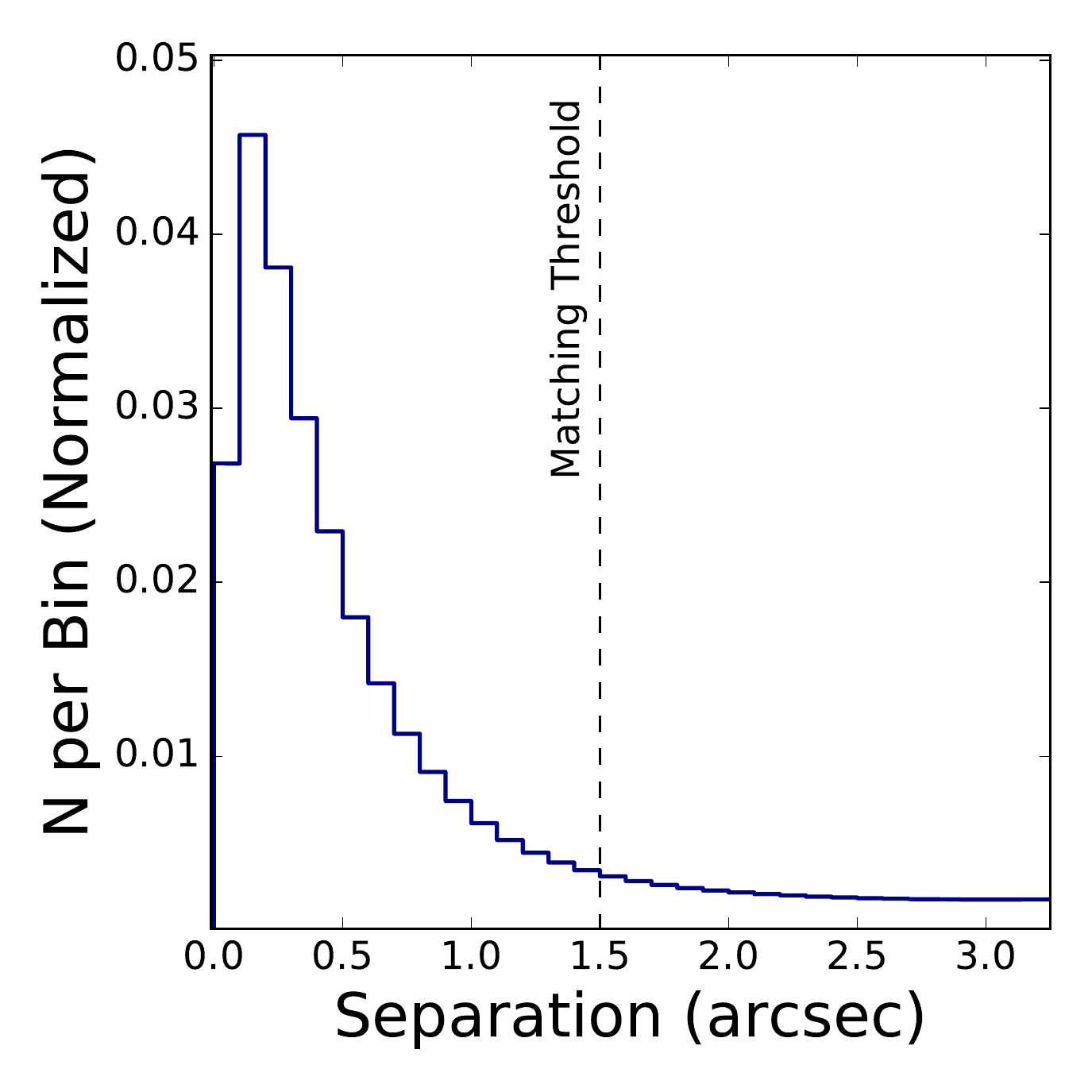}
\caption{The distribution of separations between \wise\ sources and the nearest \panstarrs\ source.
The vertical dashed line shows the 1.5$^{\prime\prime}$ search radius used to match \wise\ and \panstarrs\ sources in the \madcows{} cluster search.%
This matching radius is selected to be large enough to robustly associate true physical \wise---\panstarrs\ matches, while minimizing the rate of spurious associations.}
\label{fig:wise_sdss_separations}
\end{figure}

   To match the optical and \wise\ catalogs, we perform a nearest neighbor match for each \wise\ detection. 
We consider a match to be a true physical association if the separation is less than 1.5$^{\prime\prime}$ from each \wise\ detection. 
This matching radius was set empirically to be sufficiently large to robustly identify true matches, while minimizing the rate of spurious associations.  In Figure \ref{fig:wise_sdss_separations} 
we show the distribution of nearest neighbor matches for \wise\ sources. For associations within $3^{\prime\prime}$, 90\% of matches have separations less than our 1.5$^{\prime\prime}$ threshold. 
The AllWISE Explanatory Supplement (section II.5.b) quantifies the distribution of astrometric offsets between \wise\ and UCAC4 \citep{zacharias2013}, accounting for proper motions, finding a standard deviation $\sigma\simeq0.55^{\prime\prime}$ at \wone$=$16. Our matching radius is thus slightly less than the 3$\sigma$ astrometric uncertainty.

%% ---------------
%% WISE-SDSS Angular Separations
%% ---------------

   \subsection{Tiling the Sky} 
   Once the \wise\ and optical catalogs have been cross-matched, we apply the magnitude and color cuts described in \S \ref{sec:galcuts} and construct density maps from the remaining sources.
   For existing \spitzer\ searches for high-redshift galaxy clusters, which typically cover $<100$ deg$^2$, there is generally
   no need to subdivide the survey region.  In contrast, for \madcows\ it is necessary to develop a tiling strategy to subdivide
   the search region, enabling efficient handling of the catalogs and generation of density maps. The chosen approach is to
   conduct the search within $10^\circ\times10^\circ$ tiles, each of which overlaps with neighboring tiles by approximately 1$^\circ$. The overlap regions are used for validation in assessing the robustness of the search results.

   \subsection{Constructing the Density Maps}
   \label{subsec:maps}
   For each tile we generate a raw density map with a resolution of 15$^{\prime\prime}$ pix$^{-1}$. Each galaxy that passes the color, magnitude, and quality cuts described above is then inserted into the raw density map, using a smoothing kernel that assigns uniform weight over a width of two pixels. The result is a number-weighted projected galaxy density map. We note that one could instead attempt to use a flux-weighted map given the weak dependence of \wone\ upon redshift at $z\ga0.7$ (Figure \ref{fig:sdss_wise_phot_depth}). Such a flux-weighted approach has the advantage of giving greater weighting to blended galaxies in cluster cores that are undercounted in number-weighted maps; however, flux-weighted maps also amplify the impact of bright contamination from low-redshift interlopers. Moreover, increasing the importance of individual bright cluster galaxies for cluster detection is not necessarily desirable, as detection becomes more sensitive to omission of a single galaxy from the density map due to the photometric quality cuts.

An important element of generating the density maps is construction of corresponding masks to properly account for survey boundaries, regions around bright stars, and low coverage regions. 
For masking we use a two stage approach. 
First, we generate masks directly from the \wise\ catalog data in parallel with construction of the density maps. For every source that passes the quality cuts, the value for the coverage at that location is used as input to generate an initial coverage map at the same resolution as the density map. A smoothing kernel is applied to the map to interpolate the coverage map over pixels lacking sources.
These smoothed coverage maps are then converted into binary masks associated with each raw density map, effectively masking regions of low coverage. 
For coverage, we define a location as having low coverage if there are fewer than 20 single frame exposures in either \wone\ or \wtwo. For reference the standard two epoch coverage from AllWISE corresponds to twenty-two observations \citep{wise_allwise_exsup}.  
In practice, our low coverage restriction has little impact upon the \madcows\ survey because the AllWISE coverage in our survey region rarely falls below 20 exposures (see Figure 7 in section IV.2. of the AllWISE Explanatory Supplement \citep{wise_allwise_exsup}. At this stage we also mask regions that lie outside the footprint of the associated optical data set. 

Second, we use the \wise\ source catalog to mask regions near bright stars. Within the region of the scattered-light halo for bright stars, the photometry for fainter objects can be contaminated. It is therefore best to avoid these sources in the survey. Table 11 in section 4.4.g.ii.1.a of the All-Sky Explanatory Supplement provides coefficients relating the halo radius for scattered light halos to the magnitude of the source. Using this relation, we mask all sources with halo radii larger than 1$^{\prime}$ (\wone$<6.8$) out to the halo radius.

Once the raw density map and mask are generated, we next convolve both with the Difference-of-Gaussians kernel (Equation \ref{eq:dog}).
For the inner and outer Gaussians, we use a 6:1 ratio of outer to inner radii, setting $\sigma_1=38.2^{\prime\prime}$ and $\sigma_2=3.82^{\prime}$ (320 kpc and 1.9 Mpc at $z=1$, respectively).\footnote{These $\sigma$ values correspond to FWHM of 1.5$^{\prime}$ and 9$^{\prime}$.}
The value of $\sigma_1$ is similar to that used for the ISCS and IDCS surveys (400 kpc and 300 kpc, respectively), while for \madcows\ $\sigma_2$ is larger than for those surveys (1.6 Mpc and 1.2 Mpc, respectively). 
Physically, the larger $\sigma_2$ is designed to avoid oversubtraction for the most massive clusters, for which the signal can extend to larger radii.   Dividing the convolved density map by the convolved mask properly removes gradients in the smoothed images that arise from the masking.

   \subsection{Extracting Cluster Detections}
\label{sec:extract}

Within the smoothed density maps, we use Source Extractor \citep{bertin1996} to identify candidate clusters. Source Extractor is run on each tile with no background subtraction. Only a single pixel is required to exceed the detection threshold for a source to be selected. Specifically, we define the peak amplitude for a source to be the maximum pixel value associated with a detection in the smoothed density maps (which is equivalent to \texttt{FLUX\_MAX} in SExtractor), and only this peak amplitude must exceed the threshold for a source to be detected.
 Detections from all tiles are then combined to form a single catalog; within overlap regions detections are merged to eliminate duplicates.
For all cluster candidates we also calculate signal-to-noise based upon the peak amplitude and the RMS noise in the tile within which a cluster is detected.

\begin{figure*}
\begin{center}
\includegraphics[height=0.30\textwidth,keepaspectratio=true]{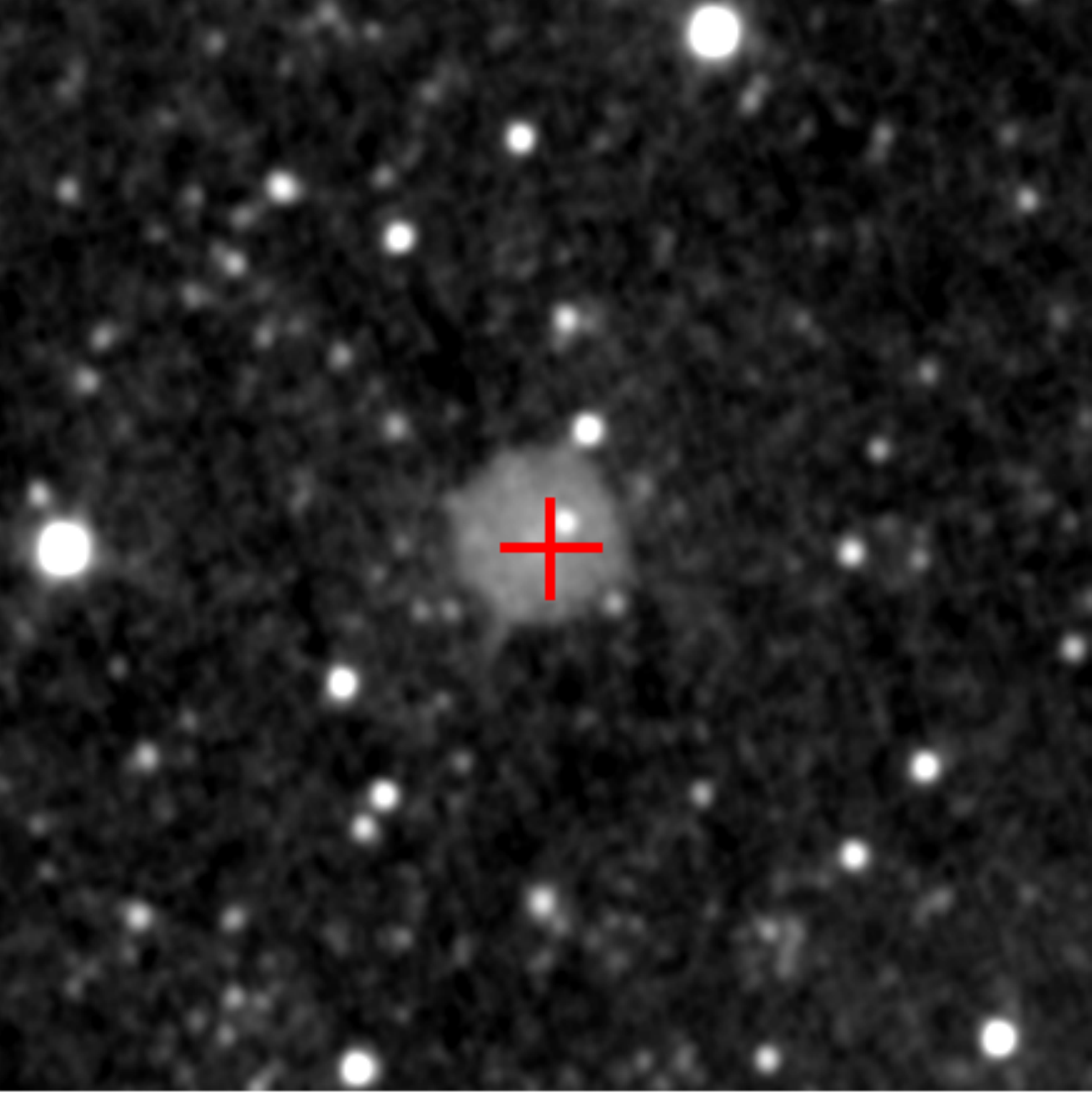}
\includegraphics[height=0.30\textwidth,keepaspectratio=true]{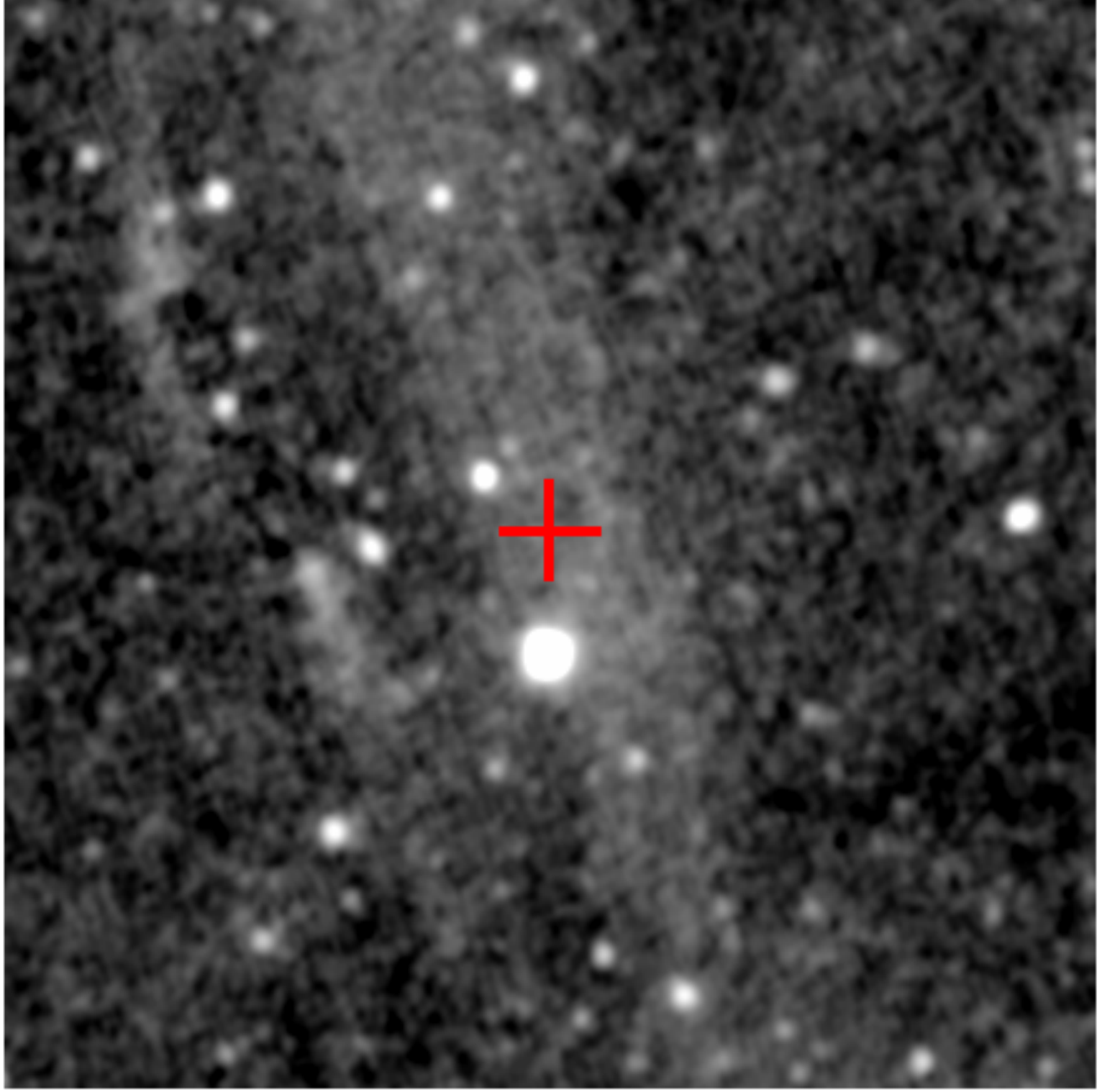}
\includegraphics[height=0.30\textwidth,keepaspectratio=true]{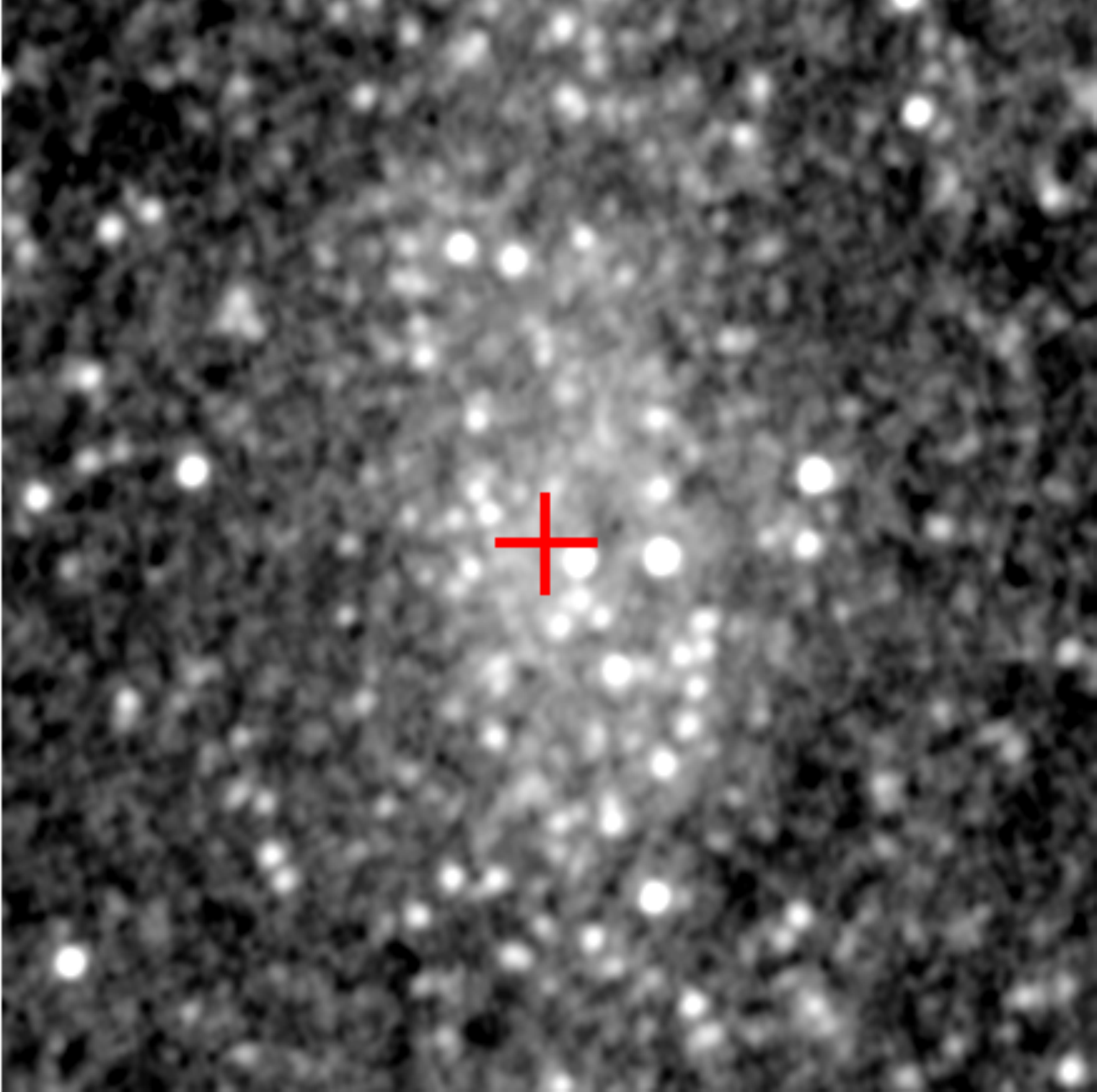}
\end{center}
\caption{\wtwo\ images ($10^\prime\times10^\prime$) showing examples of contamination that is removed by visual inspection. In each panel the cross denotes the location of the detection. The left, center, and right panels respectively correspond to spurious detections caused by an optical ghost, scattered light, and a nearby dwarf galaxy. In the latter case, the galaxy is Wolf-Lundmark-Melotte \citep[DDO 221;][]{wolf1909} at a distance of 933 kpc \citep{mcconnachie2012}.}
\label{fig:contamination}
\end{figure*}

From the remaining candidate list, we then search through the 2MASS Extended Source Catalog \citep{jarrett2000} and remove all candidates that lie within twice the total magnitude extrapolation radii ($\texttt{r\_ext}$) of the 2MASS extended sources. This cut, which is designed to remove peaks that may be associated with substructure in nearby galaxies, eliminates 8\% of candidates. 
We next impose the Galactic latitude restrictions mentioned in \S \ref{sec:data}. We restrict our search to $|b|>25^\circ$  for the \wise---\panstarrs\ and \wise---SDSS data sets, increasing the Galactic zone of avoidance to $|b|>30^\circ$ for cluster candidates at $300^\circ<l<360^\circ$ and $0^\circ<l<60^\circ$.
For the \wise---SuperCOSMOS search we opt to maintain a $|b|>30^\circ$ Galactic zone of avoidance at all $l$.
For the SuperCOSMOS search we also apply avoidance regions near the Magellanic Clouds. We impose the restriction that candidates cannot lie within $3^\circ$ of the SMC, or within an ellipse with semi-major axes of $13^\circ$ and $4.5^\circ$ for the LMC. In practice, this exclusion cut did not remove any candidates from the catalog presented below.

At this stage we also apply an automated rejection of all cluster candidates for which the peak flux lies in a pixel adjacent to a masked region (12\% of detections). While the majority of these sources are expected to be true clusters, these sources have an enhanced likelihood of being spurious due to contamination near diffraction spikes of bright stars or other subtle image artifacts. Moreover, the peak fluxes for clusters on mask edges will often be underestimated due to the masking. For these reasons we opt for a modest sacrifice in area for increased catalog fidelity and uniformity. 

Finally, our team visually inspects \wise\ cutouts of each candidate in \wone\ and \wtwo\ to identify any non-cluster sources of peaks in the wavelet maps. 
There are three main sources of such contamination, examples of which are shown in Figure \ref{fig:contamination}.
The first source is optical ghosts, which for \wise\ appear as ring-like structures at a fixed position from the parent star. While optical ghosts are flagged as artifacts during generation of the \wise\ catalog, we have found that there exist some instances where these sources are not flagged, resulting in clusters of sources that in catalog space mimic a cluster detection. Additional examples of \wise\ optical ghosts can be see in Figures 19--21 of section II.4.b.ii of the All Sky Explanatory Supplement. The second source of contamination arises from scattered light. Scattered light can yield anomalously red sources, and can induce spurious sources of a common color in the images.  
The third main source of contamination 
consists of local galaxies not present in the 2MASS Extended Source catalog.
All the above sources of contamination are easily identifiable visually. In addition to these three main contributors, we also remove a small number of detections associated with satellite trails and other rare anomalies. 
In total, visual inspection removes 6\% of the candidates which remain after automated rejection. 

\section{The Catalog}
\label{sec:catalog}
We describe in this section the properties of the ensemble of cluster candidates that remain after the detection and cleaning stages.  For both the \wise---\panstarrs\  and \wise---SuperCOSMOS searches we present catalogs of all sources
detected above thresholds in peak amplitude (see \ref{sec:peakheight}). The precise detection thresholds are set such that a cluster with a peak amplitude exceeding this threshold would have SNR$\ge 8$ in any survey tile. The motivation for this specific SNR criteria is simply that it yields a sample for which most of the \wise---\panstarrs\ clusters have \spitzer\ photometry.   For \wise---\panstarrs, the catalog includes \npanstarrs\ clusters, which are presented in Table \ref{tab:panstarrsclusters}. For the \wise---SuperCOSMOS search the noise levels are higher due to the shallower optical data, and the catalog is correspondingly smaller. We present the \nsupercosmos\ clusters in this region in Table \ref{tab:supercosmos}.
We also publish data for all clusters from our earlier \wise---SDSS search for which we have \spitzer\ imaging, 
but which are not detected above the threshold of the \wise---\panstarrs\ catalog. 
A key contributing factor in their omission from the \wise---\panstarrs\ catalog is that subsequent to the preliminary \wise---SDSS search increased masking was employed and the color cuts were tweaked, which together led to these clusters being either masked or detected below the peak amplitude threshold.
Spectroscopic redshifts, masses, and cross-identifications are provided in the comments when applicable.  The designation for \madcows\ candidates in all tables is MOO, which stands for Massive Overdense Object. In Table \ref{tab:panstarrsclusters} we include photometric redshifts and richnesses (as defined in \S \ref{sec:redshifts} and \ref{sec:richness}, respectively) for the \nspitzerpanstarrs\ clusters with \spitzer\ imaging. Similarly, in Table \ref{tab:supercosmos} we include photometric redshifts and richnesses for 64 clusters from the \wise---SuperCOSMOS search with \spitzer\ imaging that lie within the DES footprint.  In Table \ref{tab:catalogsummary} we summarize the total number of clusters and number of clusters with IRAC photometry for each of these catalogs.

\begin{deluxetable}{lcc}
\tablecolumns{3}
\tablewidth{\linewidth}
\tablecaption{Summary of Catalog Sample Sizes}
\tablehead{
\colhead{}   & \colhead{Clusters} &  \colhead{IRAC Subsample}   }  
\startdata
\wise---\panstarrs\ &  \npanstarrs\ & \nspitzerpanstarrs \\
\wise---SuperCOSMOS &  \nsupercosmos\ & \nspitzersupercosmos\tablenotemark{a} \\
\wise---SDSS &  \nsdss  & \nsdss 
\enddata
\tablenotetext{a}{Only 64 of these clusters have the requisite optical imaging from DES for photometric redshifts and richnesses.}
\label{tab:catalogsummary}
\end{deluxetable}

\subsection{Spatial Distribution}
\begin{figure*}
\scalebox{-1}[1]{\includegraphics[width=0.99\textwidth,keepaspectratio=true]{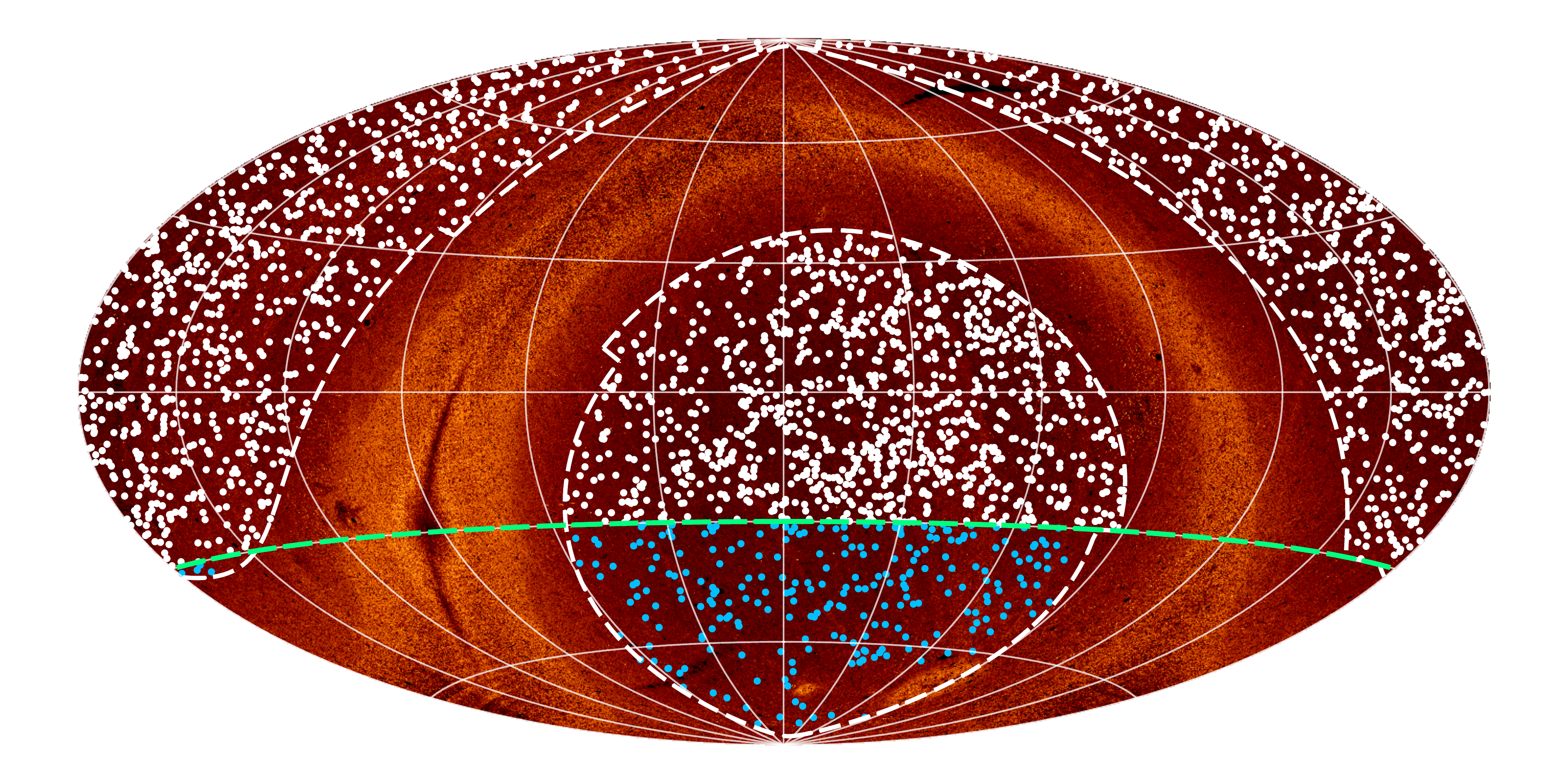}}
\caption{\footnotesize{Distributions of cluster candidates from the \wise---\panstarrs\ and \wise---SuperCOSMOS searches atop a \wise\ source density map. White circles denote the \npanstarrs\ candidates from the \wise---\panstarrs\ region that are presented in Table \ref{tab:panstarrsclusters}, while blue circles identify the \nsupercosmos\ highest amplitude candidates found outside this region using \wise\ and SuperCOSMOS data
(see Table \ref{tab:supercosmos}).
The white dashed curves delineate the Galactic zone of avoidance, which lies at $|b|<25^\circ$ for the \panstarrs\ region more than 30$^\circ$ in longitude from the Galactic center, and at $|b|<30^\circ$ near the Galactic center and within the SuperCOSMOS region.
The green curve at $\delta=-30^\circ$ corresponds to the southern limit of the Pan-STARRS survey.
 }} 
\label{fig:wise_sdss_density_maps}
\end{figure*}

Because of the Difference-of-Gaussians filtering, the \madcows\ cluster search is relatively insensitive to larger scale variations in the source counts, which can arise from a variety of observational (sensitivity gradients) and astrophysical (foreground extinction, large scale structure) effects.  In Figure \ref{fig:wise_sdss_density_maps}, we show the projected distribution of the \npanstarrs\ highest amplitude detections in the \wise---\panstarrs\ region and the \nsupercosmos\ highest amplitude detections over the rest of the extragalactic sky. The effective area of the \wise---\panstarrs\ region after accounting for masking (17,668 deg$^2$) constitutes 82\% of the combined area covered by the \wise---\panstarrs\ and \wise---SuperCOSMOS searches. 
As discussed in \S \ref{sec:extract}, we avoid $|b|\le 25^\circ$ over the full sky, and widen our Galactic zone of avoidance both for the SuperCOSMOS search and towards the Galactic center. 

%% ---------------
%% Density Maps
%% ---------------

\subsection{Peak Amplitudes} % and Richness}
\label{sec:peakheight}

\begin{figure}
\includegraphics[width=0.45\textwidth,keepaspectratio=true]{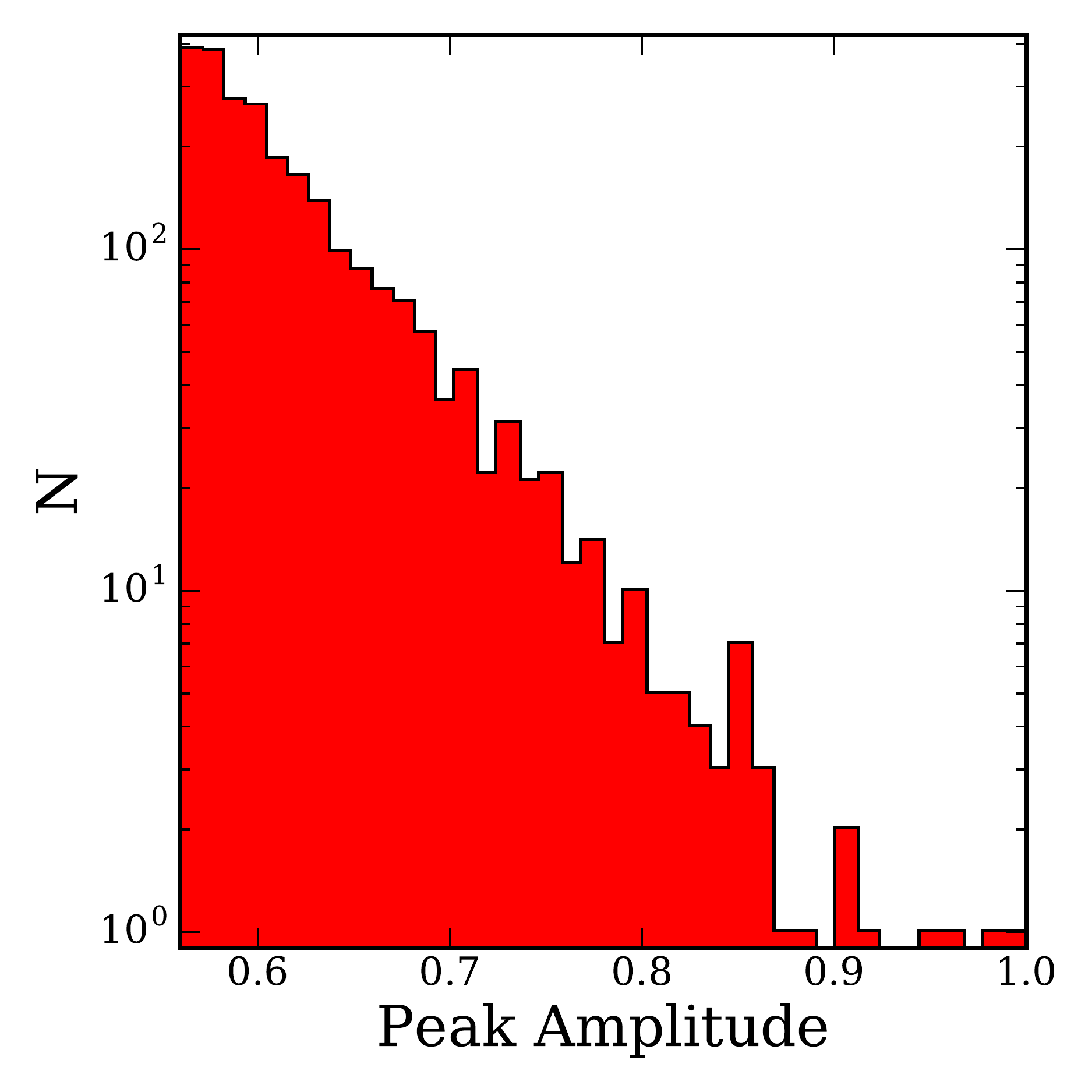}
\caption{ Histogram showing the distribution of the peak amplitudes for detections in the \panstarrs\ catalog, normalized so that the highest amplitude detection has an amplitude of 1.}
\label{fig:snr}
\end{figure}

The measured peak amplitude of an overdensity in the smoothed maps, as defined in section \ref{sec:extract}, is the observable quantity used to select clusters for the \madcows\ catalog. 
The distribution of peak amplitudes for the \wise---\panstarrs\ search, normalized such that the most significant peak has an amplitude of 1,
is shown in Figure \ref{fig:snr}. It is approximately a power law in number versus 
peak amplitude. For a given detection, the amplitude of a peak is determined by the number of galaxies associated with the cluster core and the physical
size of the smoothing kernel.

While this quantity provides the best direct observable for identifying clusters in the \madcows\ search, it is important to understand that peak amplitude is only a coarse tracer of the true cluster richness. We therefore expect broad dispersions in cluster richness and mass for a given observed peak amplitude.  There are several reasons for this scatter.
First, the number of galaxies contributing to a given overdensity in the smoothed maps will be dependent on the redshift of the cluster (due to both the optical magnitude and \wise\ color cuts, which have the greatest impact at lower redshifts, and the fixed limiting apparent magnitude). Second, the observed number of galaxies is affected by blending in the \wise\ data, which will be most pronounced for the richest and most centrally concentrated clusters. Third, the observed peak amplitude will also be affected by physically associated structures along the line of sight such as filaments. The net impact of this scatter is that for a catalog selected at a fixed peak amplitude threshold, the completeness at a fixed mass threshold is expected to be relatively low -- put succinctly, we detect massive clusters, but not in a statistically complete sense as would be needed for derivation of cosmological constraints.

\begin{figure}
\includegraphics[width=0.45\textwidth,keepaspectratio=true]{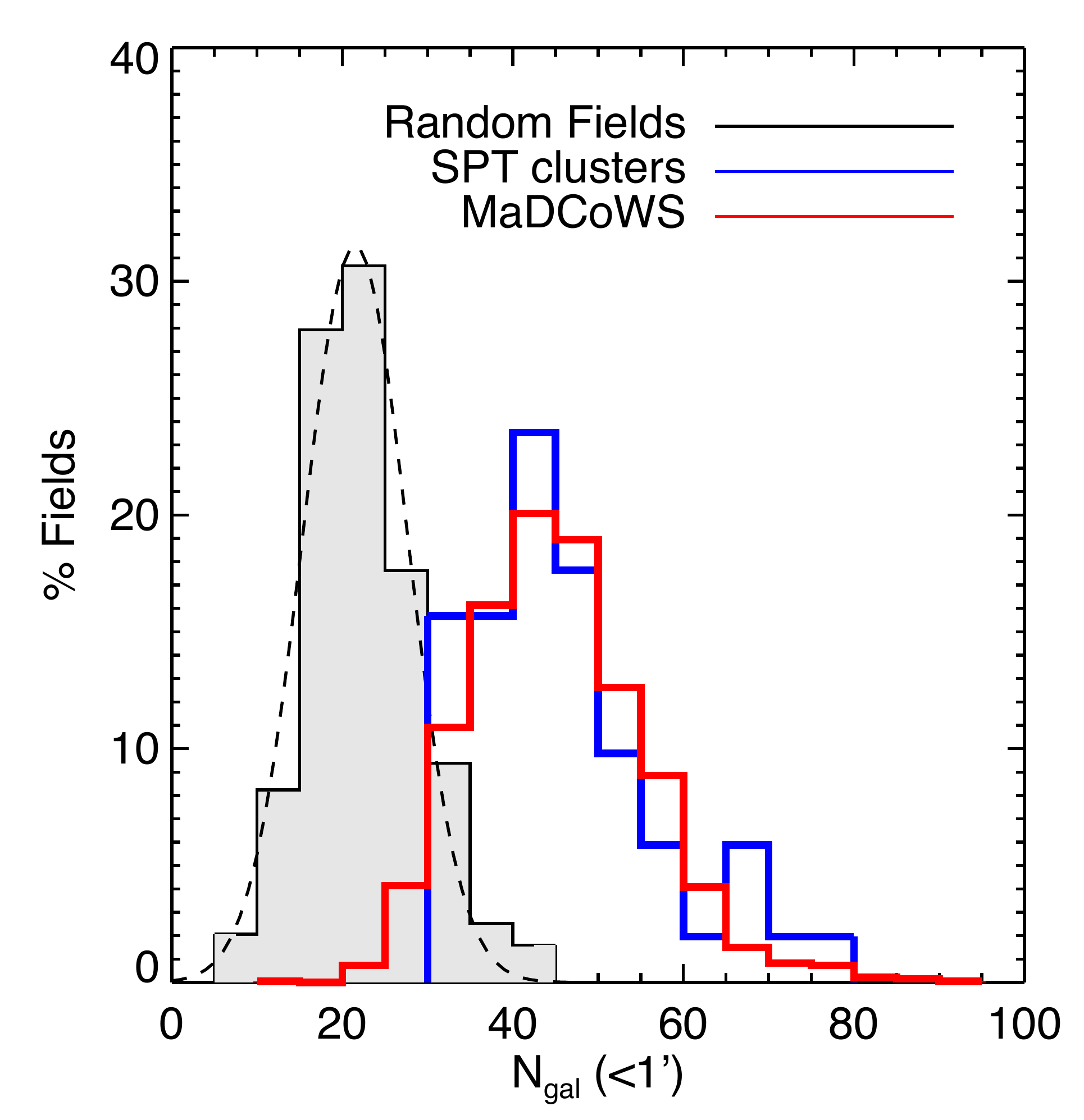}	
\caption{ 
Comparison of \irac\ richnesses, defined as the number of galaxies with red \irac\ colors that lie within a 1$^{\prime}$ circle of the cluster locations, for \madcows\ cluster candidates (red) with $z>0.9$ SPT-SZ clusters (blue), and with random locations in the SpUDS field survey (shaded grey, with the dashed black line denoting a best fit Gaussian).  The \madcows\ cluster candidates and SPT-SZ clusters on average have similar overdensities of red galaxies, with both samples significantly exceeding the random field distribution. 
}
\label{fig:richness}
\end{figure}

Keeping this limitation in mind, as an initial validation of our approach we use IRAC photometry to confirm that the \wise\ candidates selected via peak amplitude correspond to overdensities of red galaxies.  We directly counted the number of red galaxies (defined as \chone$-$\chtwo$>0.1$ Vega) within 1$^\prime$ of the cluster centroid defined by the IRAC data (see \S \ref{sec:astrometry})
for the \nspitzerpanstarrs\ clusters from the \panstarrs\ region with IRAC photometry.\footnote{The IRAC \chone$-$\chtwo\ color of a galaxy at $z\simeq0.8-1$ is $\sim0.06-0.12$ mag bluer than the \wone$-$\wtwo\ color. The definition of a red galaxy for this comparison is thus roughly similar to the \wise\ color cut used for cluster detection.}
For comparison, we applied the same criteria to derive the equivalent density of red galaxies for 50 massive clusters from the South Pole Telescope ($0.9<z_{phot}<1.3$) and for a distribution of random locations from the \spitzer\ Public Legacy Survey of the UKIDSS Ultra Deep Survey \citep[SpUDS;][]{kim2011}.  
We show the results of this comparison in Figure \ref{fig:richness}. By this \irac-based measure, both the SPT-SZ and \madcows\ clusters have distributions with significantly higher median values of $N_{gal,1^\prime}$ (43 and 44, respectively) 
 than the random field locations from SpUDS (6.6). 
This Figure indicates that \madcows\ is identifying true overdensities, but should be taken only as illustrative. In \S \ref{sec:richness} we derive a higher-fidelity richness estimator incorporating background subtraction, and we revisit the topic of the mass distribution of \madcows\ clusters in \S \ref{sec:mass}.

\subsection{Astrometric Precision}
\label{sec:astrometry}

There are two factors that limit the astrometric precision of the locations presented for the cluster candidates. The first is the resolution of the smoothed density maps. The coordinates presented correspond to the central value for the pixel with the peak flux associated with each detection, with no sub-pixel interpolation.  The precision of these coordinates is therefore limited by the 15$^{\prime\prime}$ pixel scale of the density maps. Second, the shot noise associated with each detection is significant, given that the detections are typically based upon only the $L\ga$\lstar\ galaxy population in the presence of both source confusion and foreground and background contamination. 
To quantitatively estimate the centering uncertainty associated with these positions, 
we use the \irac\ photometry to calculate  the centroid of the galaxy distribution as defined by the deeper \spitzer\ data
for the subset of galaxy clusters in the \wise\---\panstarrs\ catalog with existing \irac\ imaging. 

Details of the \spitzer\ centroiding will be described in an upcoming paper focused upon the \spitzer\ catalogs; %(Wylezalek et al., in prep); 
most pertinent for the current discussion is that the centroids are number-weighted and defined using galaxies detected at 3.6\micr\ down to the completeness limit of 10~$\mu$Jy, which corresponds to roughly a 0.3 \lstar\ galaxy at $z=1$ \citep{mancone2010}. 
Centroids correspond to the most significant density peaks of galaxies 
within 1$^{\prime}$ of the \madcows\ location. This matching radius corresponds to 500 kpc at $z=1$, 
and is set to be substantially larger than the expected centroiding error. For this centering comparison we apply no \chone$-$\chtwo\ color cut to the \irac\ photometry. This choice maximizes the signal for centroiding and avoids spurious centroids for any low-redshift clusters in the sample. We include in Table \ref{tab:panstarrsclusters} both the original detection coordinates and the \spitzer-derived centroids. 

\begin{figure}
\begin{center}
\includegraphics[width=0.45\textwidth,keepaspectratio=true]{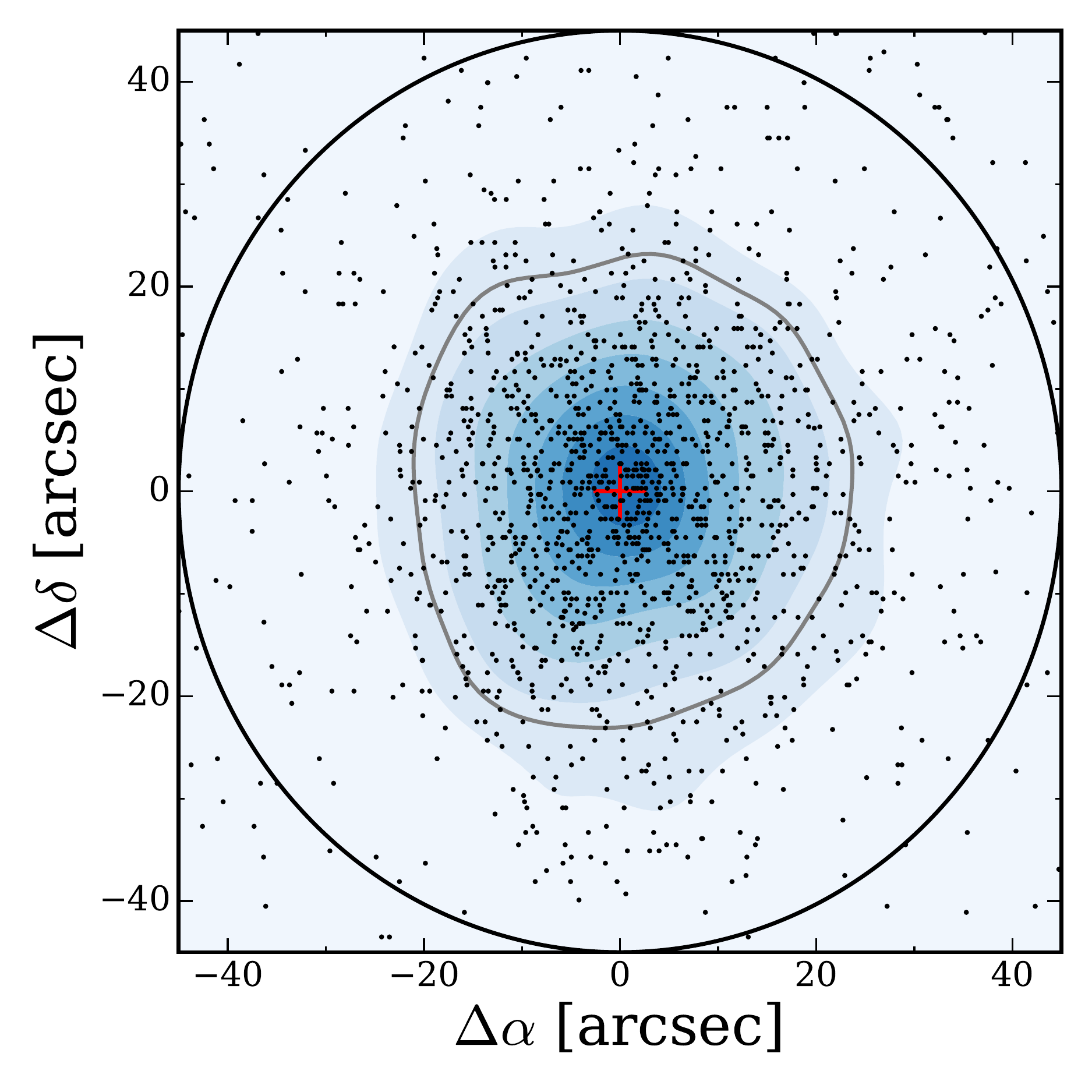}	
\end{center}
\caption{Distribution of offsets between the cluster positions from the \madcows\ search and the centroid of the galaxy distribution defined with \spitzer. For centroids within 45$^{\prime\prime}$ (91\%), which we take as the maximum separation for real matches (black circle), the rms scatter is $14.3^{\prime\prime}$ in Right Ascension and $15^{\prime\prime}$ in Declination, nearly identical to the 15$^{\prime\prime}$ pixel scale used in the cluster search. The red cross denotes zero offset. The shading corresponds to a smoothed density map generated using kernel density estimation, while the grey contour encloses 68.3\% of the clusters.
\label{fig:centering}}
\end{figure}

In Figure \ref{fig:centering} we show the distribution of offsets. The average catalog and centroid coordinates are co-centric to within 1$^{\prime\prime}$, 
with standard deviations $\sigma_\alpha=14.3^{\prime\prime}$ and $\sigma_\delta= 15^{\prime\prime}$ ($\sim 1$ pixel).
For clusters at $z=1$, the two-dimensional positional uncertainty of 21$^{\prime\prime}$ 
corresponds to a physical uncertainty of 175 kpc in the cluster position relative to the peak of the galaxy density distribution derived from \spitzer\ data. 

\section{Survey Characterization}
\label{sec:characterization}
In the previous section we presented the \madcows\ catalog and basic properties of the cluster candidates. We now proceed with a more extended discussion of derived properties of the candidates and sample based upon additional data obtained for subsets of the sample.

\subsection{Photometric Redshift Calibration}
\label{sec:redshifts}
We have previously reported spectroscopic redshifts for \madcows\ clusters in \citet{gettings2012}, \citet{stanford2014}, \citet{brodwin2015}, \citet{gonzalez2015}, and \citet{decker2018}. In this paper we provide spectroscopic confirmation for one additional cluster, MOO J1229+6521, which also appears in the \planck\ cluster catalog as PSZ2 G126.57+5161 \citep{planck2016_27}. Observational details and individual redshifts for newly confirmed members of this cluster are reported in Appendix \ref{appendix:redshifts}. Literature redshifts also exist for several known clusters \citep{hilton2007,hilton2018}. The full spectroscopic sample includes \nspecztot; the subset of  \nspecz\ clusters which have both spectroscopic data and IRAC photometry serves as the validation set for our photometric redshifts.

\begin{figure}
\includegraphics[width=0.45\textwidth,keepaspectratio=true]{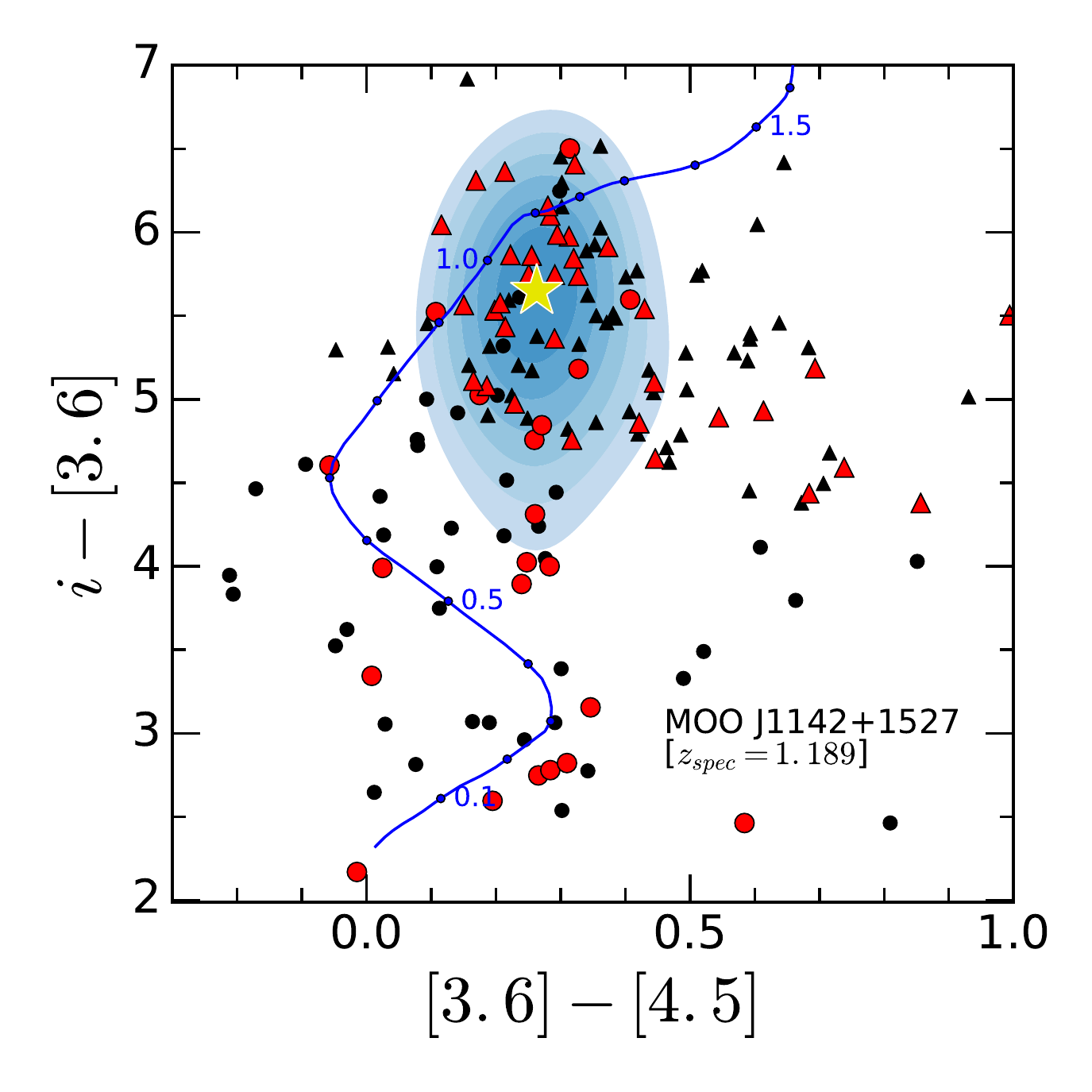}
\caption{Color-color diagram for MOO J1142+1527 ($z=1.189$; $M_{500}=5.7\pm0.5\times10^{14}$ M$_\odot$). All magnitudes are Vega. Only galaxies with $f_{4.5}>15~\mu$Jy that lie within 2$^\prime$ ($\sim 1$~Mpc) of the cluster center are shown. Circles denote galaxies detected in all bands; triangles indicate galaxies with lower limits in $i-$\chone. Red symbols indicate galaxies that lie within 1$^{\prime}$ of the cluster centroid. The light blue color map indicates the smoothed density distribution derived from the red points. The lowest density threshold corresponds to 40\% of the maximum height of the smoothed distribution, with color intervals spaces by 10\%. The dark blue curve shows the model track from $z=0-1.7$ for a passively evolving galaxy formed at $z=3$. The yellow star indicates the peak of the density distribution, which is used to determine the photometric redshift, $z_{phot}=1.10^{+0.05}_{-0.04}$. We note that the derived density distribution and yellow star are biased towards bluer $i-[3.6]$ color due to the inclusion of lower limits on $i-[3.6]$ when computing this distribution. 
This offset will be generally be true for \madcows\ clusters at $z\sim1$, for which the $i-$band data are only providing lower limits on the colors of cluster galaxies. It does not however yield a corresponding bias in the photometric redshifts because \chone$-$\chtwo\ increases monotonically at $z\ga0.7$. }

\label{fig:colorcolor}
\end{figure}

We derive photometric redshifts based upon the \chone$-$\chtwo\ colors of cluster galaxies, augmented by $i-$\chone\ color information.  This approach is similar to that of \citet{muzzin2013}, who used a combination of $z-$band and IRAC photometry to derive photometric redshifts. Figure \ref{fig:colorcolor} shows $i-$\chone\ versus \chone$-$\chtwo\ color of galaxies in the field of one of our spectroscopically-confirmed clusters, MOO J1142+1527 ($z=1.189$).  Also shown is a curve tracing the expected colors as a function of redshift for a passively evolving galaxy with solar metallicity formed via a single stellar burst at $z_f=3$, using {\tt EzGal} \citep[][\url{www.baryons.org/ezgal}]{mancone2012} and the Flexible Stellar Population Synthesis code \citep[FSPS,][]{conroy2009,conroy2010}. 

To compute the effective color of the ensemble of cluster galaxies,  we first select all galaxies with $f_{4.5}>15 \mu$Jy 
that lie within $1^{\prime}$ of the cluster centroid.\footnote{Note that this flux density threshold is higher than the 10$\mu$Jy completeness limit for the IRAC photometry. This higher threshold is chosen to both enhance the density contrast of cluster galaxies relative to the field and to decrease the impact of photometric uncertainties in the \chone$-$\chtwo\ colors upon the photometric redshift determinations.} 
 We then construct a smoothed density distribution using a kernel density estimation algorithm. The peak of this smoothed density distribution is taken as the representative color of cluster galaxies.  For the subset of candidates with multiple color peaks, we associate the brightest peak with the cluster but also calculate the colors of any secondary or tertiary peaks. We report the redshifts of these peaks only if the derived richnesses (see \S \ref{sec:richness}) exceed that of the primary peak.   In principle, the peak of the smoothed density distribution associated with the cluster should lie close to the model curve for passive cluster populations, and blueward of the curve in $i-$\chone\ for star-forming galaxies. In practice, the $i-$\chone\ peak color is not well-constrained because many cluster galaxies are non-detections in \panstarrs. Inclusion of galaxies with only magnitude limits in $i-$band results in the peak of the distribution being biased towards bluer $i-$\chone. 

To infer redshifts from the color distribution, we rely primarily on the more robust \chone$-$\chtwo\ color.  This color increases monotonically at $0.7<z<1.7$,  and within this redshift range we calculate the photometric redshift by determining the model redshift which yields the \chone$-$\chtwo\ color closest to that of the peak of the smoothed density distribution. 
While the \irac\ photometry alone is sufficient to derive low-scatter photometric redshifts for clusters at $z>0.7$,{\footnote{The code {\tt rsz}, which can be found at \url{https://github.com/gillenbrown/rsz} yields comparable scatter to our approach at $z>0.7$.}} the expected  \chone$-$\chtwo\ colors of cluster galaxies at $0.7<z<1.15$ are degenerate with those of galaxies at $z<0.7$.
We use the $i-$\chone\ color to break this degeneracy.  For low-redshift structures the galaxies are brighter and the $i-$\chone\ colors bluer, yielding detections rather than upper limits, and enabling robust determination of the low-redshift solution.  

The strongest peaks in the smoothed density maps correspond to $z<0.7$ for $\sim2$\% of the full ensemble of candidates with \spitzer/IRAC photometry. Using data from the Legacy Surveys \citep{dey2018} we visually inspected the subset of these 2\% that lie within the Dark Energy Camera Legacy Survey DR7 footprint.\footnote{\url{http://legacysurvey.org/dr7/}} In all cases, we find that the low-redshift peak in color space is a foreground cluster unassociated with the galaxies that contributed to the \madcows\ detection. For this reason, we impose a prior on the photometric redshift estimates, requiring that the solution lie at $z\ge0.6$ for the \wise---\panstarrs\ and \wise---SDSS catalogs.
In cases where there is a strong peak in the color distribution corresponding to a low redshift cluster, we note in the Tables the presence of a foreground structure. There are total of six clusters in the two catalogs (0.3\% of the \spitzer\ sample) for which it is not possible to recover a redshift and richness for the background cluster. In these cases we simply note the presence of the foreground structure.  For the \wise---SuperCOSMOS catalog, which lies outside the DR7 footprint and has less robust removal of foreground galaxies (see Fig. \ref{fig:wise_supercosmos_color_z}), we impose no prior.

\begin{figure}
\includegraphics[width=0.45\textwidth,keepaspectratio=true]{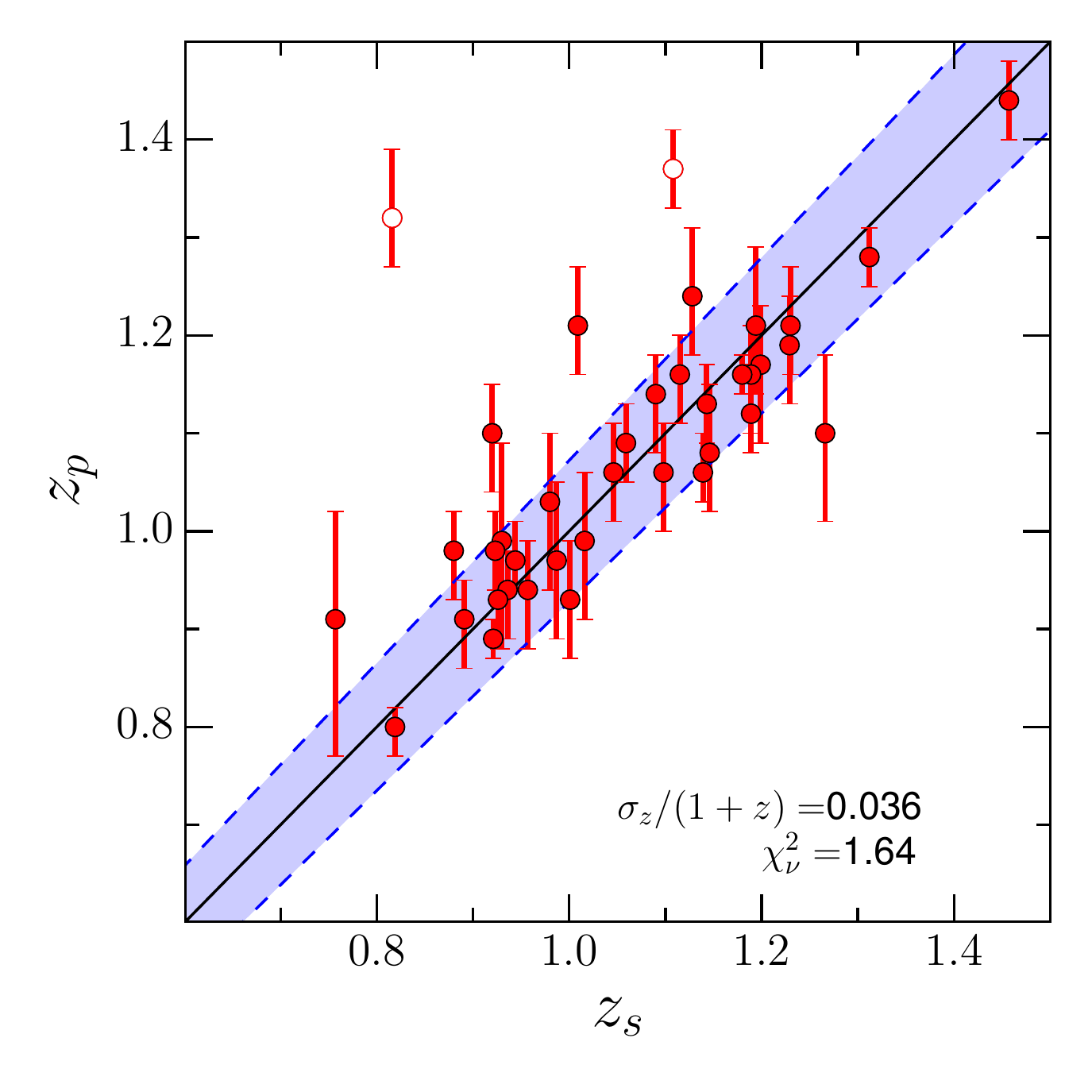}
\caption{Comparison of photometric and spectroscopic redshifts for confirmed \madcows\ clusters. The solid line is the one-to-one relation, while the shaded region corresponds to the interval $\sigma_z/(1+z)=\sigspecz$. 
The points denoted as open circles are the two clusters for which the photometric redshifts are $>5\sigma$ outliers.}
\label{fig:photzcal}
\end{figure}

Comparing with the \nspecz\ spectroscopic redshifts (Figure \ref{fig:photzcal}), we find two outliers for which the photometric redshifts are $>5\sigma$ from the spectroscopic redshift.{\footnote{The outliers are MOO J0224-0620 ($z_{spec}=0.816$, $z_{phot}=1.32^{+0.05}_{-0.07}$) and MOO J0113+1305 ($z_{spec}=1.108$, $z_{phot}=1.37\pm0.04$).}
For the rest of the sample
the scatter is 
$\sigma_z/(1+z)=$\sigspecz.} 
For all clusters with \spitzer/IRAC photometry, which is essential for achieving this fidelity in the redshift estimates, we include in Table \ref{tab:panstarrsclusters} the photometric redshifts and associated uncertainties. 

\subsection{Redshift Distribution}
\label{sec:zdist}
\begin{figure}
\includegraphics[width=0.45\textwidth,keepaspectratio=true]{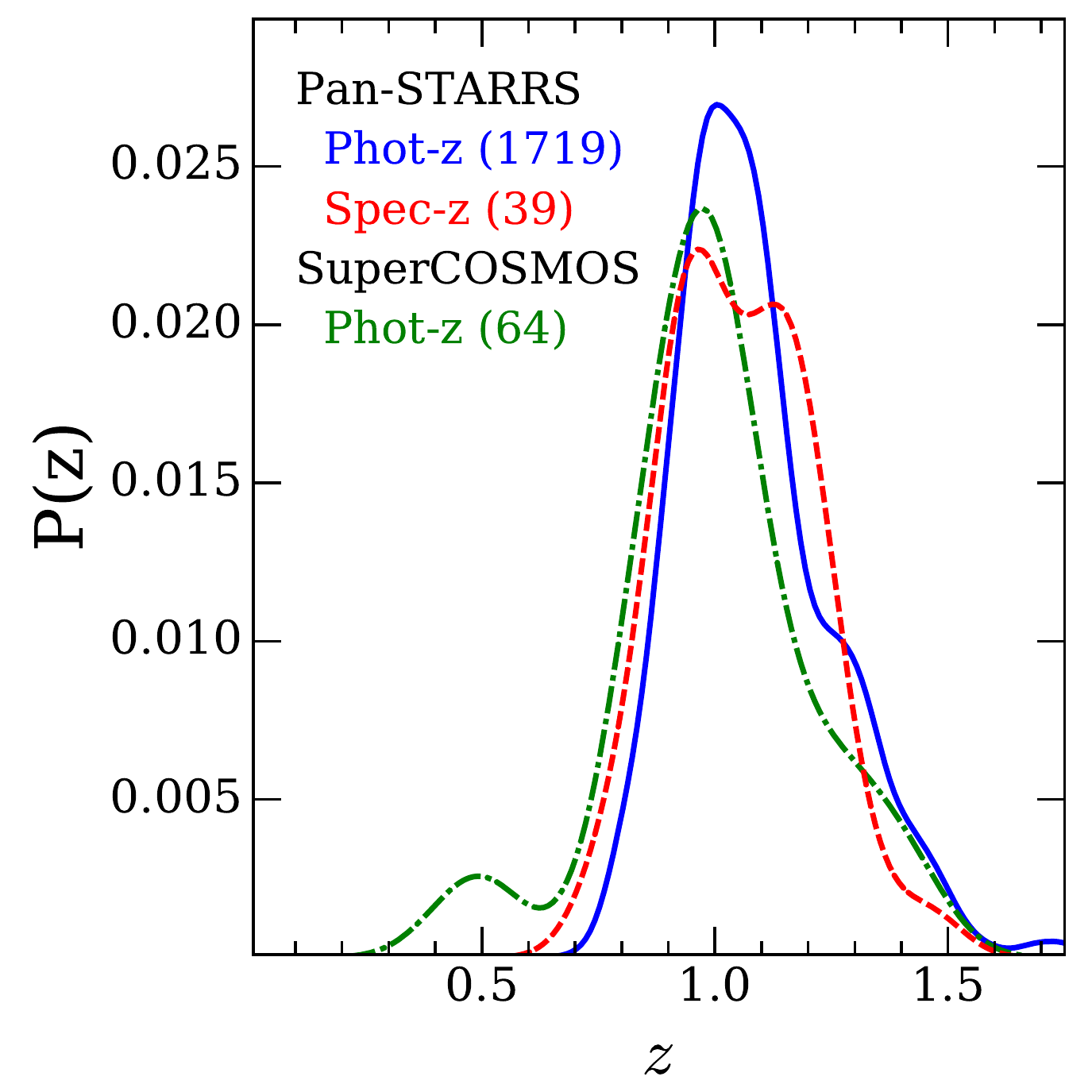}
\caption{Smoothed redshift probability distribution functions for clusters with spectroscopic redshifts (red dashed) and photometric redshifts from \spitzer\ (blue solid) in the \panstarrs\ region, and for those with photometric redshifts from \spitzer\ (green dot-dashed) in the SuperCOSMOS region. For the spectroscopic redshifts we include clusters with literature redshifts. The functional forms of the spectroscopic and photometric distributions are similar in width and mean redshift for the \wise---\panstarrs\ sample. The SuperCOSMOS   photometric redshift distribution is shifted to slightly lower redshift. The secondary peak in the SuperCOSMOS redshift distribution at $z\simeq0.5$ is due to the combination of lower fidelity rejection of low-redshift galaxies and omission for this catalog of the $z\ge0.6$ prior used for the \wise---\panstarrs\ photometric redshifts. These smoothed distributions are generated using Gaussian kernel density estimation.}
\label{fig:redshiftdistribution}
\end{figure}

In Figure \ref{fig:redshiftdistribution} we show the photometric redshift distribution 
for \madcows\ cluster candidates within the \panstarrs\ region with \spitzer\ photometry (blue solid curve). We also show the redshift distribution for all \madcows\ clusters within the \panstarrs\ region with spectroscopic redshifts (red dashed).
The curves shown are derived using Gaussian kernel density estimation, applying Scott's rule \citep{scott1992} to calculate the estimator bandwidth. The general similarity of the curves 
illustrates the robustness of the estimated redshift distribution.
The low-redshift cutoff seen in the full sample arises primarily from the magnitude and color cuts used in the initial galaxy selection for the cluster search,\footnote{The $z>0.6$ prior on the photometric redshifts impacts only 2\% of clusters.} while the high-redshift decline is due to a combination of a decrease in the number density of massive clusters and the \wone\ band not quite reaching constant stellar mass with increasing redshift.  

In Figure \ref{fig:redshiftdistribution} we also plot the photometric redshift distribution for \madcows\ clusters within the SuperCOSMOS footprint with \spitzer\ photometry (green dot-dashed). 
As expected, the redshift distribution is shifted to slightly lower redshift relative to the \panstarrs\ sample, with median redshifts of 0.98 and 1.06 for the two samples. For the \wise---SuperCOSMOS sample 6\% of the clusters have $z_{phot}<0.7$, compared to 0\% for the \wise\---\panstarrs\ sample. This difference is due to the combination of the weaker optical color cut, which retains more low-redshift galaxies during the cluster search, and the fact that we do not impose a $z\ge0.6$ prior on the photometric redshifts. The prior is omitted to reflect the fact that with the weaker color cut these low-redshift solutions may correspond to the cluster detections.

\subsection{Richness}
\label{sec:richness}

At a fundamental level, there are strong indications that robust cluster mass estimates are attainable directly from observations of the stellar content. Authors including \citet{girardi2000} and \citet{lin2003} provided early demonstrations that total baryon content scales with cluster mass. \citet{lin2003} for example found that the scatter in the relation between $K-$band luminosity ($L_K$) and \mfive\ from X-ray data was $\sim45$\%, with this scatter dominated by observational uncertainties. More recently, studies with much higher fidelity data and membership information have demonstrated convincingly that the intrinsic scatter is quite low. For example,
\citet{mulroy2014} determined that for the LoCuSS cluster sample the intrinsic scatter in the $L_K-M_{500}$ relation is $\sim10$\%. Consistent with these observations, multiple groups have also shown that at a fixed halo mass the ratio of gas mass in the ICM to stellar mass displays a remarkably small intrinsic scatter, indicative of the baryons being partitioned between these two phases with little variation between clusters at fixed \mfive\ \citep{lagana2008,zhang2011,gonzalez2013}. 

The challenge however lies in the reality that in contrast with the LoCuSS sample, membership information is not available for existing cluster surveys directly from the searches. As a result, interlopers can significantly degrade the fidelity of luminosity-based mass estimators. Cluster richnesses, defined based upon number counts rather than total luminosity, are more robust to such contamination. In recent years multiple groups have shown that it is possible to define richness measures that are robust mass proxies with low scatter \citep[e.g.,][and references therein]{rykoff2012,rozo2014,andreon2015,old2015,andreon2016}. Using mock galaxy catalogs to compare a suite of richness estimators, \citet{old2015} find a scatter of 0.18 dex in the \mtwo$-$richness relation for the best proxy. For samples of real, low-redshift clusters, \citet[][]{andreon2015}
and \citet[][]{rozo2014} 
define richness measures $n_{200}$ and $\lambda$, for which they find scatters of 0.16 dex and $\sim$0.11 dex, respectively.

Our practical goal for \madcows\ is to develop a similarly low-scatter mass proxy that can be applied to the full catalog. A limitation, as discussed in the previous section, is that the \wise\ data alone lack the spatial resolution and depth necessary for such a low-scatter estimator. We have therefore proceeded with the alternate approach of calibrating a \spitzer-based richness estimator that can be applied to the large fraction of the sample with IRAC data from either the archive or our programs in Cycles 9, 11, and 12.

\subsubsection{Richness Definition}

For \madcows\, we explored use of multiple richness measures to identify a suitable estimator for use with IRAC data. 
Similar to \citet{rettura2017}, we settled upon use of a fixed aperture for defining the richness. In contrast with that study, we employ a physical rather than angular aperture and incorporate optical data to minimize contamination and reduce scatter in the mass-richness relation.

Our first step in establishing a richness definition for \madcows\ is to set a uniform limiting \chtwo\ flux density for the IRAC input galaxy catalog of 15~$\mu$Jy ($m=17.7$ Vega). This 4.5$\mu$m-selection is designed to yield
an approximately constant stellar mass threshold at $0.7<z<1.5$ and hence minimize the redshift dependence of the richness measure. For this redshift range, 15~$\mu$Jy corresponds to a stellar mass of $\sim5\times 10^{10}$ M$_\odot$ assuming an FSPS model with a Chabrier IMF normalized to Coma, with only a modest dependence on star formation history. We also match all 4.5$\mu$m-selected sources to the \panstarrs\ PV2 catalog to obtain $i-$band magnitudes or upper limits for each galaxy. 

 A challenge that one encounters when using \spitzer\ imaging for this analysis is that 
the IRAC field-of-view extends to only $\sim1.3$ Mpc from the center of the cluster
for a galaxy cluster at $z\simeq1$. One consequence is that the total galaxy density does not necessarily reach the field level within the IRAC field-of-view \citep[for example, see][]{wylezalek2013}, precluding robust local background subtraction. For this reason, when calculating richnesses we use color cuts to minimize foreground contamination, and then use data from the \spitzer\ Deep, Wide-Field Survey \citep[SDWFS;][]{ashby2009} to estimate the background density. 
We isolate galaxies near the cluster redshift by combining an  
 $i-$\chone\ criteria with a second color cut in \chone$-$\chtwo. This additional cut helps compensate for the fact that the \panstarrs\ imaging is not deep enough to detect all IRAC-selected cluster galaxies at $z=1$. 

\begin{figure*}
%\plottwo{cc_0105+1324.png}{cc_1142+1527.png}
\plottwo{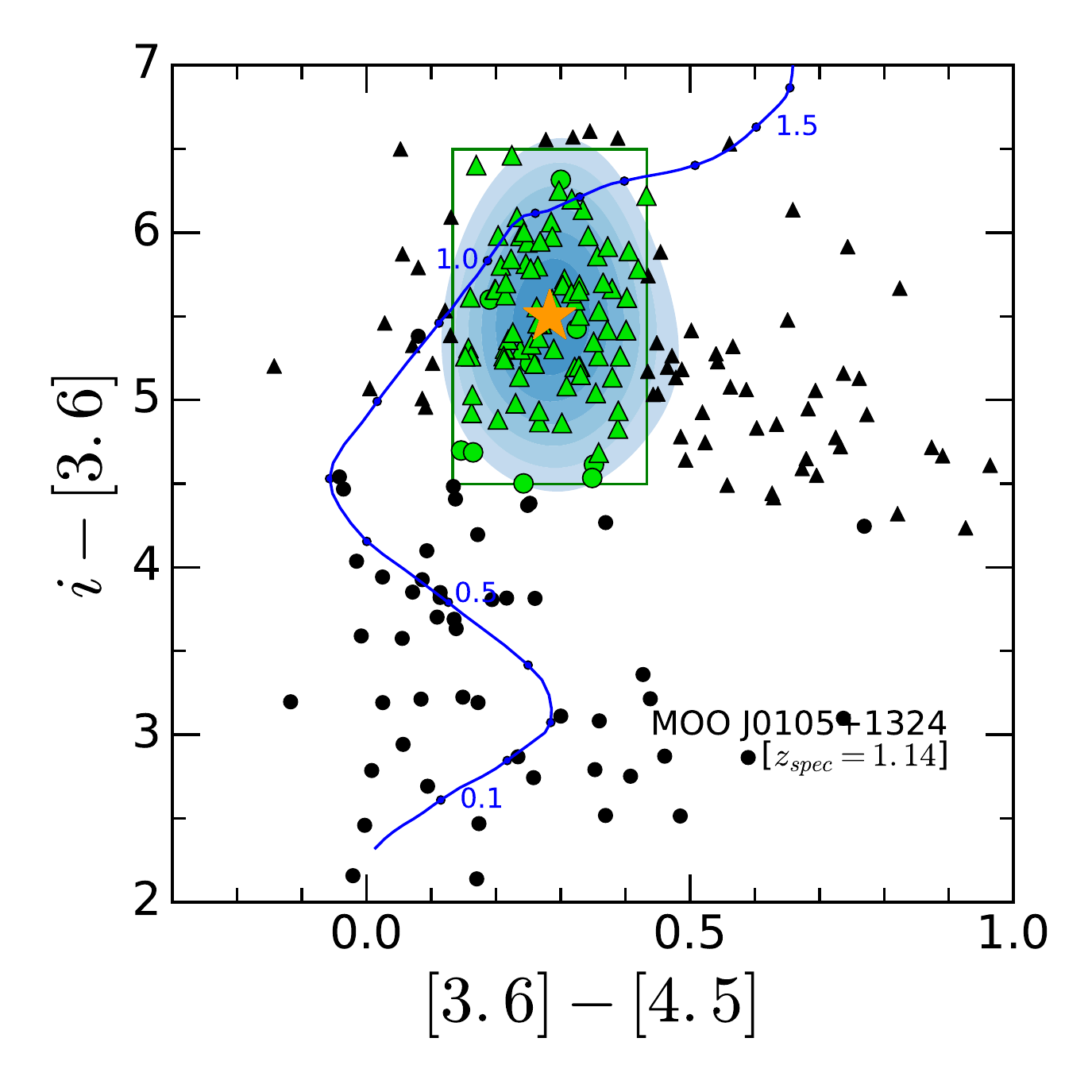}{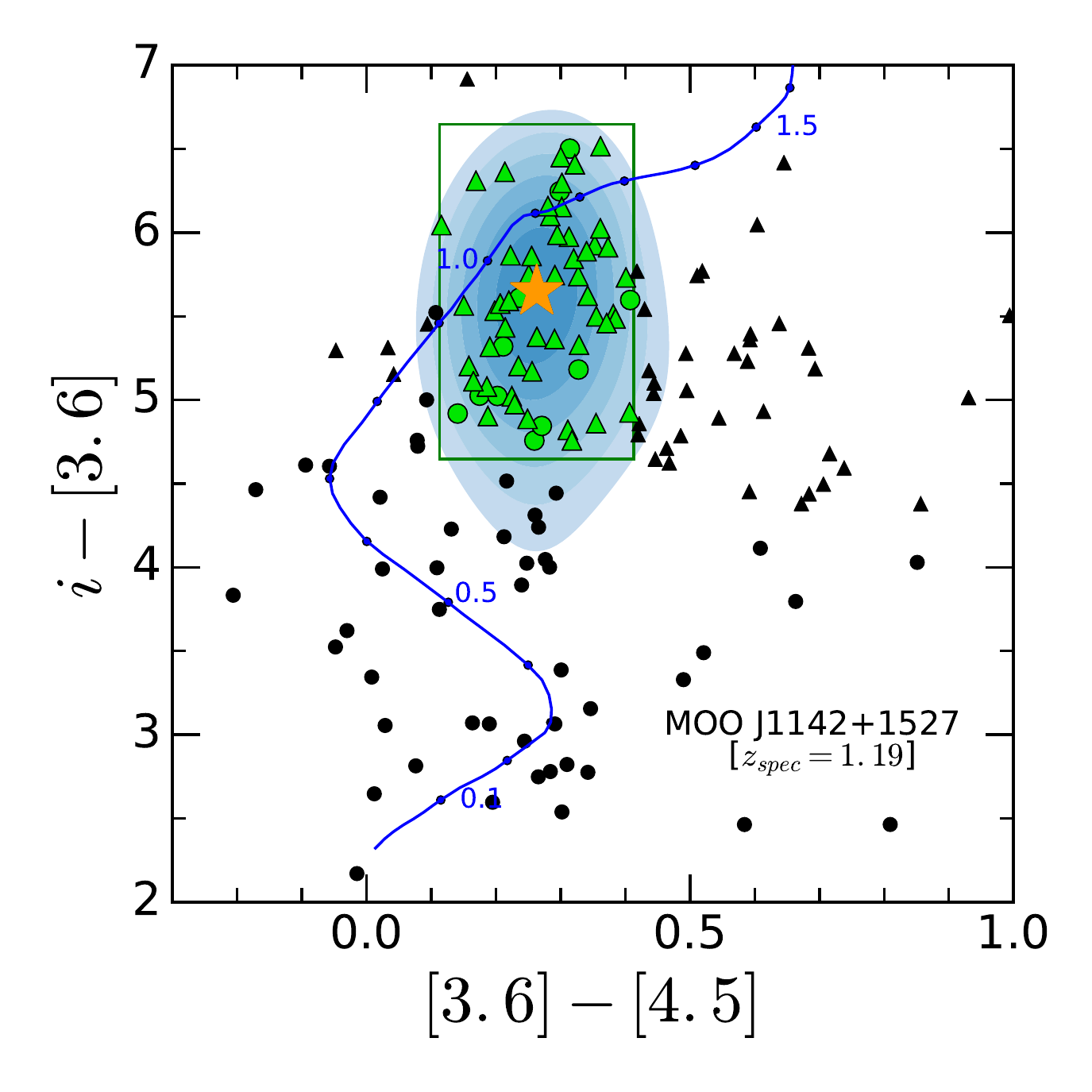}
\caption{Color-color diagrams for confirmed clusters MOO J0105+1324 \citep[$z=1.143$; $M_{500}=4.03^{+0.48}_{-0.45}\times10^{14}$ M$_\odot$,][]{gettings2012} and MOO J1142+1527 \citep[$z=1.189$; $M_{500}=5.45^{+0.58}_{-0.51}\times10^{14}$ M$_\odot$;][]{gonzalez2015}.  The symbols and shading are the same as described in Figure \ref{fig:colorcolor}, with the following exceptions. In this Figure green points correspond to galaxies that are included in calculating the richness based upon their color and physical distance from the cluster centroid ($<1$ Mpc).
The solid green lines further indicate the region in color space used to identify galaxies as possible cluster members. The color criteria are designed to retain cluster members while minimizing contamination. When calculating the cluster richness, counts within the SDWFS field are used to apply a statistical background correction. The smoothed density distribution used to determine the cluster redshift and color centroid is shown as the blue color map, as in Figure \ref{fig:colorcolor}.
}\label{fig:colorcuts}
\end{figure*}

Starting with the redshift for a given cluster, we use {\tt EzGal} to calculate the expected $i-$\chone\ and \chone$-$\chtwo\ color for a cluster galaxy. We calculate this color using the same passively evolving model as in \S \ref{sec:redshifts}.
We then consider galaxies to be possible cluster members if they are either detected in $i$ and less than one mag bluer in $i-$\chone\ than the fiducial color, or else are non-detections in $i$ and the lower limit on $i-$\chone\ is no more than one mag redder than the fiducial color. We additionally require that a galaxy
have a \chone$-$\chtwo\ color within $\pm0.15$ mag of the fiducial. The $i-$\chone\ color threshold is set such that 
this threshold will retain not only passive galaxies, but also star-forming galaxies with exponentially declining star formation histories ($\tau=1$ Gyr) and initial formation redshifts $z_f\ga 3$. The width of the color window in \chone$-$\chtwo\ minimizes exclusion of cluster members 
due to either photometric uncertainty or redder colors arising from moderate AGN contributions to the photometry, while still providing a meaningful reduction of the background contribution.
Examples of the implemented color cuts are shown in Figure \ref{fig:colorcuts} for two confirmed \madcows\ clusters at $z=0.99$ and $z=1.189$, respectively.  
The boxes in Figure \ref{fig:colorcuts} illustrate the color windows used for galaxies with $i-$band detections.

A second consequence of the field-of-view constraint is that the data do not uniformly reach to sufficiently large radii for us to use richness estimators extending to \rtwo -- motivating our use of a fixed, 1 Mpc radius metric aperture.  Green points in Figure \ref{fig:colorcolor} denote galaxies that lie within 1 Mpc of the \wise-based cluster centroid and satisfy the color criteria.  In defining the color cuts and apertures size, we use the photometric redshifts described in \S \ref{sec:redshifts}.

We define the richness $\lambda=N-N_{\mathrm{field}}$, where $N$ is the total number of color-selected galaxies within the metric aperture. In quoting values of $\lambda$ we also include as a subscript the threshold flux density, such that $\lambda_{15}$ denotes the richness calculated for sources $f_{4.5}>15 \mu$Jy.
For each cluster we calculate the expected field density, $N_{\mathrm{field}}$, by computing the average density of galaxies found in SDWFS for the same magnitude and color cuts and scaling to the appropriate aperture area.  In cases where the IRAC data are incomplete within the metric aperture, we apply a correction to account for the fractional area lost. We refrain however from quoting richnesses for clusters at $z<0.7$. For these clusters a 1 Mpc radius extends beyond the field of the IRAC imaging, and a fractional area correction would generally lead to a poor estimate of the true richness. For clusters with archival \spitzer\ data, we also avoid quoting richnesses for systems with low partical \irac\ coverage.  Richnesses are included in the catalog in Table \ref{tab:panstarrsclusters}.

The caveat with this approach is that photometric redshift scatter will increase the scatter between richness and mass, and  a catastrophic failure on the photometric redshift will result in a spurious richness estimate. 
As discussed in the Appendix, we find that the former effect is minor. Based upon our spectroscopic confirmation, catastrophic outliers are also rare (at the few percent level). When they do occur, the impact will be a mis-estimation of the richness due to shifting of the color-selection window away from the appropriate cluster color.

\subsubsection{The Relation Between Richness and Mass}
\label{sec:mr}

To provide an initial calibration of the mass-richness relation 
we consider a subset of \madcows\ clusters imaged with \spitzer\ with derived SZ mass estimates from CARMA. The \mfive\ measurements are for a total of 14 clusters, five of which have previously reported SZ detections in \citet{brodwin2015} and \citet{gonzalez2015}. For previously reported clusters we use updated mass estimates from \citet{decker2018}, which will provide a homogenous analysis for the full sample.  The list of clusters used for this analysis is presented in Table \ref{tab:massrichness}.

\begin{deluxetable}{llrr}
\tablecolumns{4}
\tablewidth{\linewidth}
\tablecaption{Clusters in Mass-Richness Calibration}
\tablehead{
\colhead{Name}   & \colhead{$z$} &  \colhead{$\lambda_{15}$}   & \colhead{\mfive}  \\
\colhead{}   & \colhead{} &  \colhead{}   & \colhead{($10^{14}$ M$_\odot$)}  
}
\startdata
MOO J0037+3306 & 1.139 & 54$\pm$8 & $2.34^{+0.65}_{-0.63}$ \\
MOO J0105+1324 & 1.143 & 87$\pm10$ & $4.03^{+0.48}_{-0.45}$ \\
MOO J0123+2545 & 1.229 & 41$\pm7$ & $3.90^{+0.89}_{-0.81}$ \\
MOO J0319-0025 & 1.194 & 34$\pm$6 & $3.11^{+0.53}_{-0.47}$ \\
MOO J1014+0038 & 1.230 & 44$\pm$7 & $3.26^{+0.32}_{-0.30}$ \\
MOO J1111+1503 & 1.36\tablenotemark{p} & 33$\pm$6 & 2.08$^{+0.30}_{-0.31}$ \\
MOO J1142+1527 & 1.189 & 58$\pm$8 & $5.45^{+0.58}_{-0.51}$ \\
MOO J1155+3901 & 1.009 & 33$\pm$6 & $2.61^{+0.56}_{-0.55}$ \\
MOO J1231+6533 & 0.99\tablenotemark{p} & 50$\pm$8 & $4.69^{+1.24}_{-1.00}$ \\
MOO J1335+3004 & 0.984 & 30$\pm$6 & $1.38^{+0.75}_{-0.74}$ \\
MOO J1514+1346 & 1.059 & 73$\pm$9 & $1.89^{+0.68}_{-0.79}$ \\
MOO J1521+0452 & 1.312 & 47$\pm$7 & $3.65^{+1.03}_{-0.94}$ \\
MOO J2206+0906 & 0.951 & 54$\pm$8 & $2.66^{+0.93}_{-0.74}$ \\		
MOO J2231+1130 & 0.80\tablenotemark{p} & 49$\pm$8 & $4.38^{+1.51}_{-1.37}$ 
\enddata
\tablecomments{We list in this table all clusters that are included in determination of the mass-richness calibration. All \mfive\ measurements are derived from CARMA SZ observations.}
\tablenotetext{p}{Photometric redshift.}
\label{tab:massrichness}
\end{deluxetable}

\begin{figure*}
%\plottwo{figs/richness/mr.png}{figs/richness/mr_nomergers.png}
\plottwo{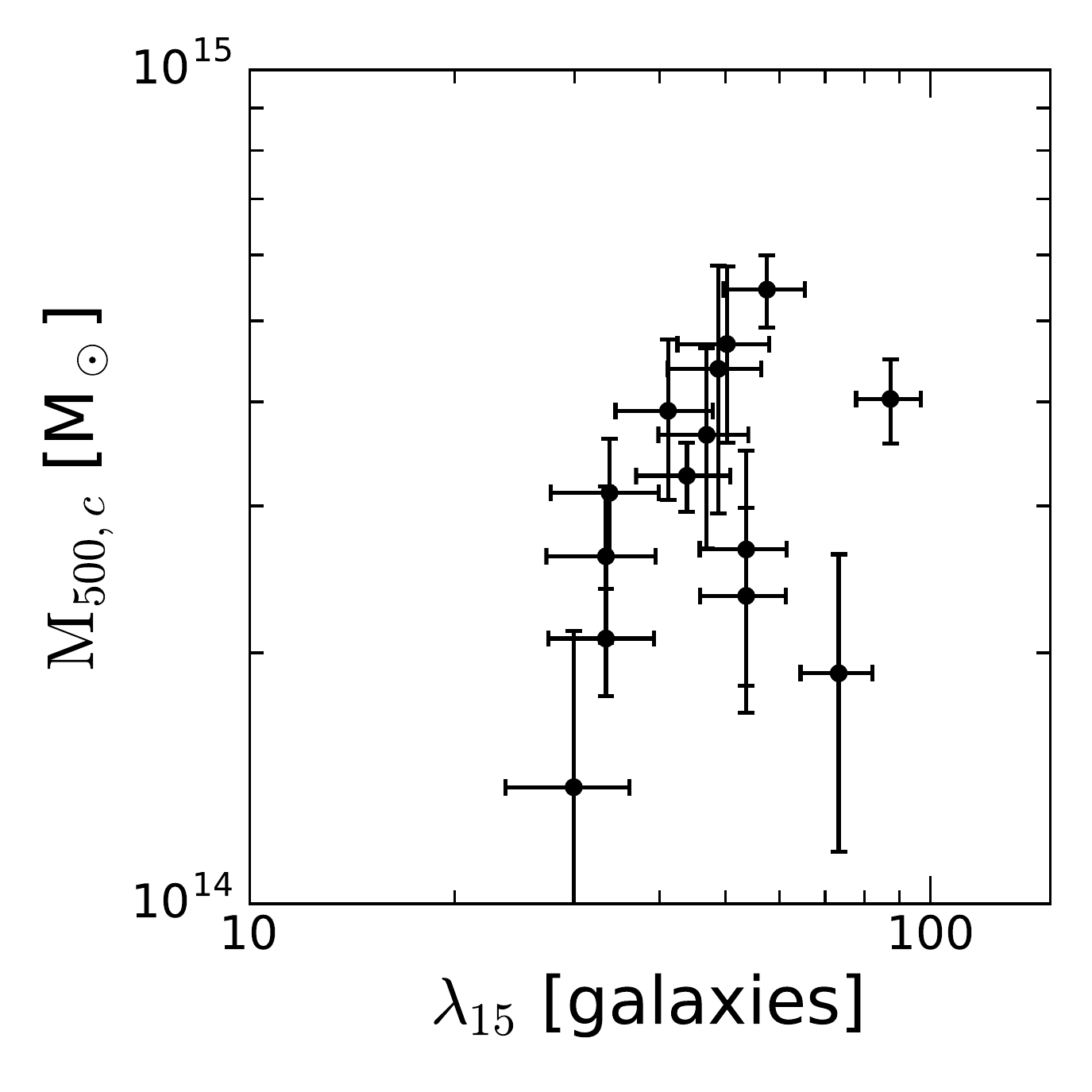}{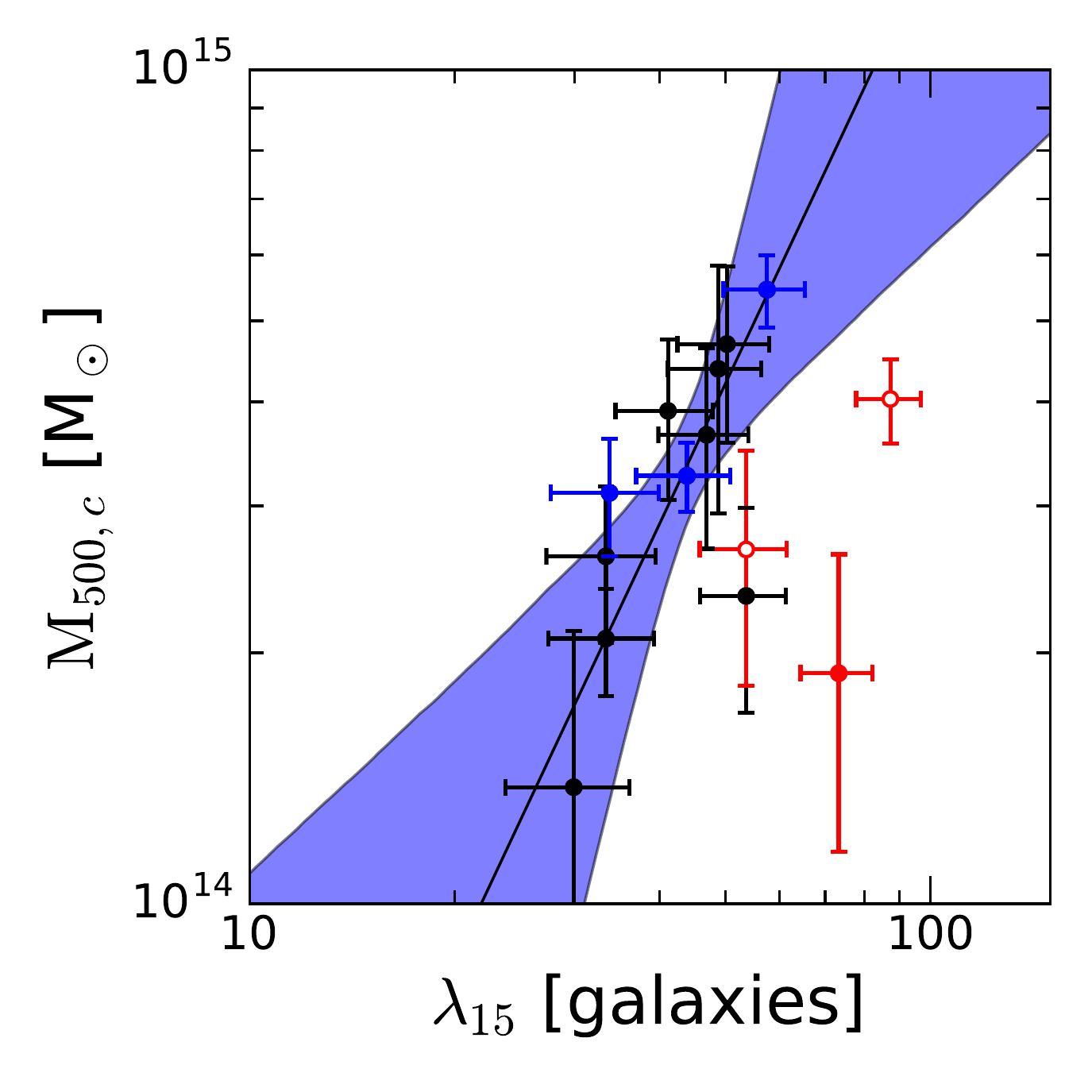}
\caption{ $Left:$ SZ-based \mfive\ versus
 $\lambda_{15}$, where $\lambda_{15}$ is defined as the number of galaxies within a 1 Mpc aperture centered on the cluster above a flux density threshold of 15$\mu$Jy. For all clusters we use the photometric redshifts to derive the richness; use of the spectroscopic redshifts has a negligible impact on the resulting richnesses. $Right:$ The same as in the left panel, except that systems that are known (likely) major mergers based upon \chandra\ observations are denoted as open (solid) red points and those with no evidence of major mergers from \chandra\ observations are plotted as blue points. This panel also includes  a best-fit relation is derived excluding the known and likely major mergers.  The best-fit relation is shown as a solid black line, while the shaded region denotes the 68\% confidence interval. The dispersion in the relation is $\sigma_{\log M|\lambda}=0.12$, or $\sigma_{\log M|\lambda}=0.07$ if one excludes MOO J0037+3306.}
\label{fig:massrichness}
\end{figure*}

We derive a best fit mass-richness relation, 
which we parameterize as
\begin{equation}
\log \frac {M_{500}}{10^{14} M_\odot}= \alpha \log \lambda_{15} + \beta,
\label{eqn:mr}
\end{equation}
using the python implementation{\footnote{See \url{linmix.readthedocs.io}}} 
of the Bayesian code {\tt linmix} \citep{kelly2007}. 
For the sake of uniformity, the richnesses used in this fit are calculated using the photometric redshifts to define the appropriate color window for selecting cluster members. 
We show the data, with richness calculated within a 1 Mpc diameter aperture, in the left panel of Figure \ref{fig:massrichness}. The scatter between mass and richness is large for the full ensemble; however, we note
that two of these clusters, MOO J0105$+$1323 and MOO J2206$+$0906, are clearly early-stage major mergers based upon \chandra\ observations that will be presented in a forthcoming paper. These two clusters are plotted as red open circles in the right panel of this figure. A third cluster, MOO J1514$+$1346 (red filled circle), which has the second highest \spitzer-derived richness of the clusters in the Figure, also shows tentative evidence of major merger activity. In the right panel we additionally plot in blue the clusters with existing \chandra\ data that exhibit no evidence for early-stage major merger activity. Overlaid, we  show a best fit mass-richness relation derived  excluding the red points.% three systems with evidence of major mergers.  
The best-fit relation is plotted as a solid line, with the shaded region indicating the 1$\sigma$ confidence interval. 

The best-fit values, which are not well-constrained given the limited dynamic range in mass and small sample size, are formally $\alpha=1.65^{+1.45}_{-0.96}$ and $\beta=-2.16^{+1.57}_{-2.38}$. The scatter in mass about the relation is $36\pm11$\% ($\sigma_{\log M|\lambda}=0.12$),  where the quoted uncertainty is derived via a bootstrap resampling of the data. It is clear from the right panel of Figure \ref{fig:massrichness} that a single cluster, MOO J0037+3306, is a significant contributor to this scatter. If we assume that this cluster, for which we currently lack \chandra\ data, is also a merging cluster, then we can re-fit the data and obtain a refined estimate of the scatter for the  other systems that lack similar evidence of ongoing major mergers. Doing so, the best-fit parameters change minimally ($\alpha=1.86^{+1.53}_{-0.88}$ and $\beta=-2.49^{+1.43}_{-2.50}$), while the scatter is reduced to  $16\pm6$\% ($\sigma_{\log M|\lambda}=0.07$).

To assess the sensitivity of this relation to photometric redshift uncertainties, positional offsets, and flux density thresholds, we repeat the above analysis varying these quantities. First, we use spectroscopic redshifts, which are available for all but three of these clusters.  The change in the richnesses is minimal and hence the fit and $\sigma_{\log M|\lambda}$ remain essentially unchanged. Second, we use the \spitzer-derived centers instead of the \wise\ cluster centers, again finding negligible change in $\sigma_{\log M|\lambda}$. Finally, we also test the use of a 10$\mu$Jy rather than 15$\mu$Jy threshold for the richness. This again does not appreciably alter the scatter, though it by definition does change the normalization of the relation.

It thus appears, perhaps not surprisingly, that there may exist a relatively tight underlying relation between mass and richness for non-merging clusters, while a subset of merging systems  are offset to lower SZ mass (or higher richness) 
than one would expect from this relation.  Multiple studies \citep[e.g.][]{poole2007,krause2012,yu2015} find in simulations that major mergers can systematically bias downward the masses inferred from $Y_{SZ}$. This bias is on average $\sim10-15$\% for \mtwo\ in \citet{krause2012}, but in some cases can be significantly larger.
Physically, this bias is due to the time required for the temperature to increase to the equilibrium level corresponding to the mass of the merged cluster. If the richness measure approaches the new level more quickly than the temperature, which is expected given the large 1 Mpc radius metric aperture used in this paper, then there will also be an offset of merging systems in the $\lambda_{15}-$\mfive\ plane.\footnote{As an aside, we note that \citet{saro2015} found that the merging cluster SPT-CL J0516-5430 is a similarly large outlier in the SPT $\lambda-$\mfive\ relation.}

The \madcows\ clusters with the highest \spitzer\ richnesses will therefore be comprised of a combination of the most massive clusters and those undergoing major mergers. ICM observations are necessary to discriminate between these two scenarios. It should also thus be expected that as major mergers become an increasing fraction of the total cluster population with increasing redshift, the observed scatter between SZ mass and richness will increase commensurately unless one identifies and exclude mergers.

We caution that the above is preliminary, being based upon a small number of clusters and not including CARMA non-detections. It therefore should be taken as indicative of the general trend rather than a definitive measure of the mass-richness relation. Ongoing SZ programs with ALMA (PI: Brodwin, programs \#2016.2.00014.S and \#2017.1.00961.S), MUSTANG-2 (PI: Brodwin, programs GBT 18A-272 and GBT 18B-215), and NIKA2 (PI: Brodwin, programs 095-17 and 095-18), plus a more thorough analysis of the CARMA observations including non-detections and stacking, are forthcoming. These efforts should yield a superior calibration and a better assessment of the total scatter.  

\begin{figure}
\includegraphics[width=0.45\textwidth,keepaspectratio=true]{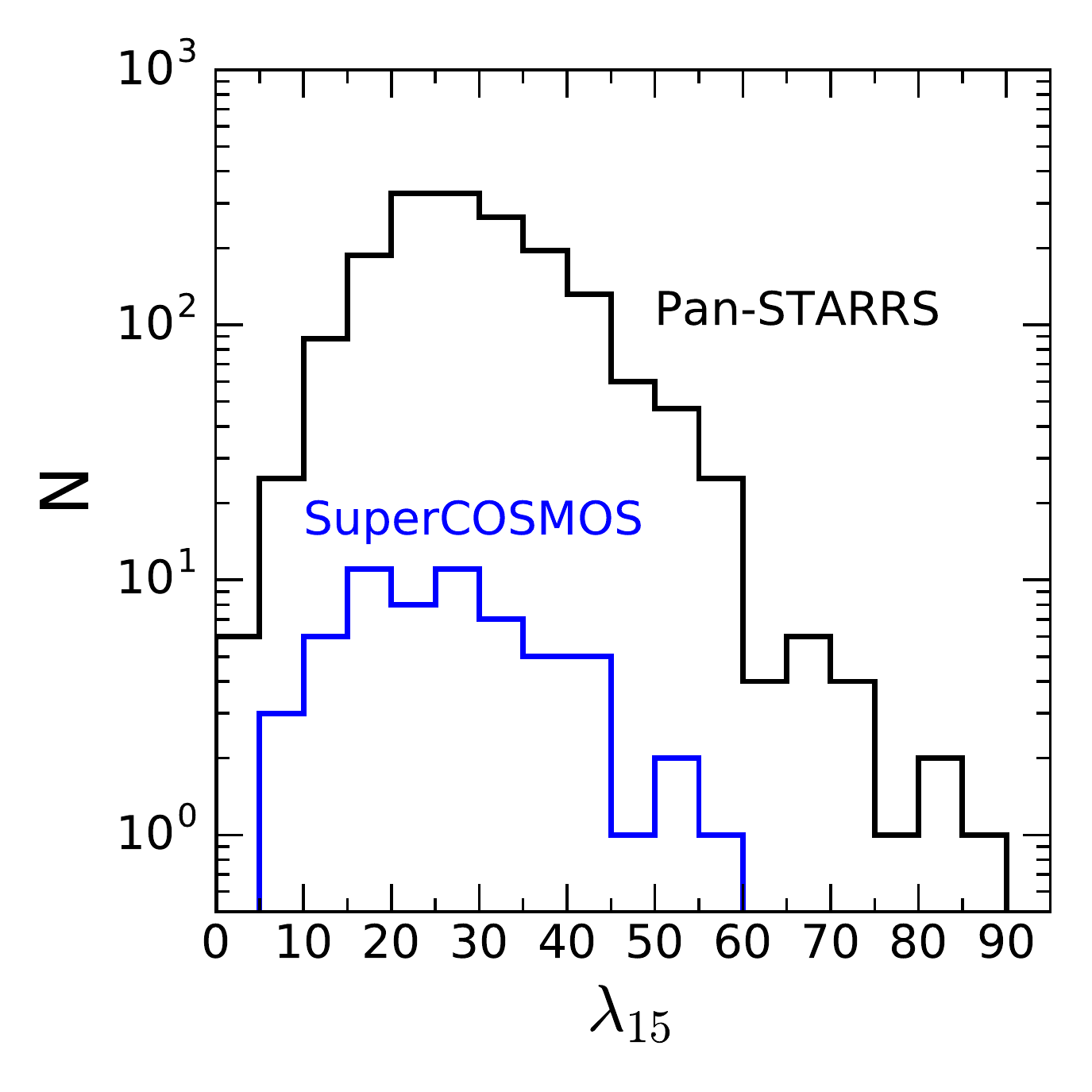}
\caption{Histograms showing the distribution in $\lambda_{15}$ for all \madcows\ clusters with IRAC photometry. The black histogram is for clusters from the \wise---\panstarrs\ region, while the blue histogram is for clusters from the southern \wise---SuperCOSMOS region. Both samples have similar median richnesses and richness distributions.}
\label{fig:rdist}
\end{figure}

\begin{figure}
\includegraphics[width=0.45\textwidth,keepaspectratio=true]{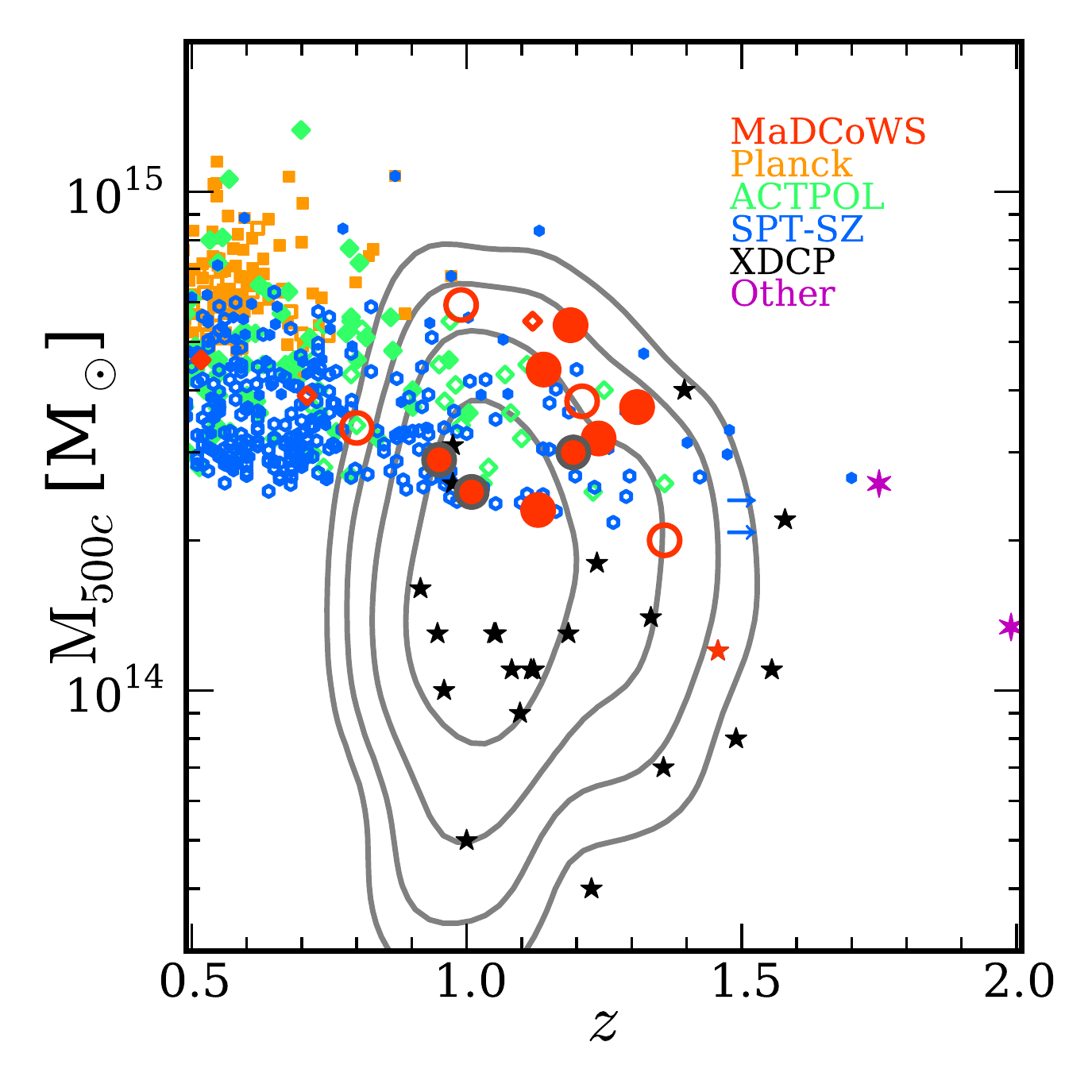}	
\caption{Comparison in the mass-redshift plane of \madcows~clusters with those
of other wide-area cluster surveys, including 
{\it Planck} (Planck Collaboration 2014), ACTPol \citep{hilton2018} 
and the South Pole Telescope Sunyaez-Zel'dovich effect survey  \citep[SPT-SZ][]{reichardt_2013,bleem2015}. For ACTPol we use the M$_{500c}^{Cal}$ masses, which are scaled by a weak lensing mass calibration factor. The \madcows~ clusters shown are those with existing SZ-based masses from CARMA \citep{brodwin2015,gonzalez2015,decker2018}, SPT-SZ \citep{bleem2015},  ACTPol \citep{hilton2018}, or X-ray-based masses for clusters from the XMM-Newton Distant Cluster Project \citep[XDCP][]{fassbender2011}. Filled circles denote clusters with spectroscopic redshifts (including those from Khullar et al. 2018 for SPT); open circles indicate photometric redshifts for all surveys. There are several \madcows\ clusters from early versions of the search that were confirmed, but did not make the final \panstarrs\ selection due to detection amplitude, masking, or Galactic plane restrictions. We denote these clusters with grey circles around the solid red points. For SPT, the clusters with lower limits on the redshifts are denoted by arrows.
We also include XLSSU J021744.1-034536 \citep[$z=1.99$]{mantz2014,mantz2017} and  IDCS J1426.5+3508 \citep[$z=1.75$]{stanford2012} as the highest redshift clusters with published SZ masses and redshifts. Finally, we note that \madcows\ clusters previously detected in the other samples are plotted as red symbols in the marker style corresponding to data points from the other survey. The Planck cluster detected by \madcows\ has no published mass and is therefore not shown. 
The contours show the estimated distribution of the full \madcows\ sample using photometric redshifts and richness-based mass estimates via the mass-richness relation presented in section \ref{sec:mr}. The contour spacing corresponds to factor of two changes in the number of clusters per unit redshift and $\log$ mass ($dN/dz/d\log M$). 
Considering the $1\sigma$ confidence interval on the mass-richness relation, the median mass of the \madcows\ sample is \mfive$\simeq0.9-2.2\times10^{14}$ M$_\odot$. 
These contours represent an extrapolation of the mass-richness relation for \mfive$<1.5\times10^{14}$ M$_\odot$. They should therefore be considered as only illustrative of the expected full distribution and interpreted with caution.
The \madcows\ sample  likely extends down to similar masses as are reached by deep X-ray studies, and probes lower masses than current or planned SZ surveys.
}

\label{fig:massredshift}
\end{figure}
\subsection{Mass and Richness Distributions of \madcows\ Clusters}
\label{sec:mass}
  In Figure \ref{fig:rdist} we plot the observed richness distribution for all clusters with IRAC photometry from both the \wise---\panstarrs\ and southern \wise---SuperCOSMOS searches. In both instances these histograms correspond to peak amplitude-limited subsamples, modulo the inclusion of a small number of clusters added from the \spitzer\ archive. As is evident from the figure, both samples have similar median richnesses and approximately power law distributions at higher richness, as might be expected if the distribution is probing the halo mass function at the high richness end with the survey selection function yielding a turnover in the number of clusters below $\lambda_{15}\sim 25$. 
Using the mass-richness calibration derived in section \ref{sec:richness}, the median richness for the \wise---\panstarrs\ sample corresponds to a mass \mfive$=1.6^{+0.7}_{-0.8}\times 10^{14}$ M$_\odot$. The equivalent number for the \wise--SuperCOSMOS sample is \mfive$=1.4\pm0.7\times10^{14}$ M$_\odot$.

We also present in Figure \ref{fig:massredshift} the current distribution in the mass-redshift plane of all \madcows\ clusters with masses from CARMA or the literature \citep{fassbender2011b,bleem2015,hilton2018}, comparing to existing wide-area SZ and X-ray surveys. We denote with open circles clusters for which we currently lack a spectroscopic redshift. These clusters are placed at their estimated photometric redshift. 
It is apparent from Figure \ref{fig:massredshift} that the \madcows\ sample includes clusters that span the mass range probed by the combination of existing SZ and X-ray surveys at this epoch, including several of the most massive clusters known at $z>1$. 
For comparison, we also plot contours showing the inferred distribution for all  \madcows\ clusters with IRAC photometry, where we use the photometric redshifts from \S \ref{sec:zdist} and richness-based mass estimates from \S \ref{sec:richness}. The density contours are spaced by powers of two, illustrating that the distribution is strongly peaked at $z\simeq1$ and $M\simeq1-2\times10^{14}$ M$_\odot$. We caution against over-interpretation of these contours, particularly  outside the range over which the mass-richness relation is calibrated (\mfive$\sim 1.5-5.4\times10^{14}$ M$_\odot$). These contours should be considered illustrative rather than definitive.

\subsection{Comparison with ACTPol}
As a test of our ability to recover known massive, high-redshift clusters, we compare our \madcows---\panstarrs\ results with the two-season ACTPol Sunyaev-Zel'dovich catalog \citep{hilton2018}.  ACTPol, which covers 987.5 deg$^2$, is the only published high-redshift SZ survey that overlaps with the \wise---\panstarrs\ region. The ACTPol catalog includes 19 clusters at $z>0.9$, four of which have spectroscopic redshifts, with weak-lensing calibrated masses $M^{Cal}_{500}\ga2.5\times10^{14}$ M$_\odot$. Of these 19 ACTPol clusters,
only one (ACT-CL J0125.2-0802) is in the \madcows\ catalog in Table \ref{tab:panstarrsclusters}. We investigate the cause of this minimal overlap, finding that it can be attributed to several factors. A minor factor is masking of bright stars in the \madcows\ search, which removes one of the 19 clusters (ACT-CL J0248.7-0019). The other two more significant factors are the high threshold for our catalog and the large scatter between peak amplitude and mass --  the latter also being the reason that IRAC imaging is required for determining richnesses. Two additional clusters are detected at SNR$>8$, but just below our peak amplitude threshold, and a total of  8 (12) out of the 18 unmasked clusters are detected at SNR$>5$ ($>3$). From a practical perspective, using the current approach it would not be possible to identify these clusters as the most massive among the larger ensemble of \madcows\ clusters in this region without deeper mid-infrared imaging such as we have obtained with \spitzer/IRAC for a subset of the \madcows\ clusters.

\subsection{Future Improvements}
\label{sec:future}
  The current \madcows\ search attempts to make optimal use of existing surveys, but there are several prospects for upcoming data sets that can yield an improved version of the \madcows\ search.  
One notable limitation of the current search is the limited depth of the SuperCOSMOS imaging outside the \panstarrs\ footprint. 
As described in \S \ref{sec:supercosmos}, the shallowness of this imaging yields higher foreground contamination (Fig. \ref{fig:wise_color_z}), resulting in a lower fidelity and lower median redshift catalog at $\delta<-30$. 
Several surveys are underway that will 
enable a uniform search comparable in quality to what is currently achieved in the \panstarrs\ region over the full extragalactic sky. Observations for the DES \citep{des2016} are expected to provide adequate data over $\sim5000$ deg$^2$, of which over half are at $\delta<-30^\circ$. 
Of particular note, this area includes the region of the SPT-SZ survey \citep{bleem2015}, enabling us to compare catalogs and better assess selection biases associated with the \madcows\ search. 
Two other surveys that also have the potential to enable a higher fidelity search in the south are 
the SkyMapper Southern Sky Survey Main Survey \citep[SMSS,][]{keller2007} and the Southern Photometric Local Universe Survey (S-PLUS).\footnote{See \url{https://confluence.astro.ufsc.br:8443}}
The SMSS is designed to cover the entire southern sky ($\delta<0^\circ$) to $u,g,v,r,i,z= 20.5,20.5,21.7,21.7,20.7,19.7$ (AB, $5\sigma$).\footnote{Depths from \url{skymapper.anu.edu.au/surveys}} 
These data will significantly improve rejection of low-redshift galaxies relative to the $R_F$ limit used for SuperCOSMOS, although the $i-$band depth is still shallow relative to \panstarrs.
S-PLUS meanwhile plans  to cover $\sim8000$ deg$^2$ in $ugriz$ and seven narrow-band filters. S-PLUS is designed to have shallower $i$-band photometry than \panstarrs\, but is expected to be sufficiently deep in $z-$band to enable an equivalent search.

A more fundamental limitation for the current \madcows\ search is the depth of the \wise\ photometric catalog that is used for the initial selection of galaxies. As was shown in Figure \ref{fig:sdss_wise_phot_depth}, the \wise\ photometry is only currently deep enough to identify at $>5\sigma$ individual $\sim$\lstar\ galaxies out to $z\simeq1$ in \wone, and only detects these galaxies at $\sim 2 \sigma$ in \wtwo. This depth threshold has multiple implications for the search. First, it necessitates that we use \wone\ for galaxy selection, reducing the sensitivity of the survey to the highest redshift clusters. Second, because we are only detecting the bright end of the luminosity function with \wise, cluster identification relies upon extracting a cluster signal generated by a small number of bright galaxies. The strength of the signal is therefore highly sensitive to statistical variations in the number counts of cluster galaxies, which can rise due to both blending of individual sources at the resolution of \wise\ and statistical variations in the luminosity function. Thus, while detections in the current survey catalog result from true overdensities, not all overdensities are detected as significant due to such statistical variance.

 There exists the potential for significant improvement on this front. For this paper we have used the \allwise\ catalog. This catalog, which was released in 2013, is the deepest currently available all-sky \wise\ catalog and incorporates all data prior to the end of the post-cryogenic mission in February 2011.  During this period \wise\ mapped the full sky twice in the short wavelength bands. In October 2013 the \wise\ satellite was reactivated for the NEOWISE mission \citep{mainzer2014}, resuming survey observations in \wone\ and \wtwo. The mission is currently scheduled to continue through December 2018, providing a factor of five or more increase in total exposure time in these bands over the full sky, and thus a factor of five increase in the total exposure time relative to \allwise\ images.  \citet{meisner2017} have demonstrated the potential gain in depth. Stacking three years worth of data from \wise\ and NEOWISE, they reach 0.56 (0.46) mag deeper in \wone\ (\wtwo) than \allwise\ data alone. 
The ``CatWISE'' effort, funded by NASA's Astrophysics Data Analysis Program, is adapting the \allwise\ data processing pipeline to generate a catalog from four years of \wise\ and \neowise\ data, with planned release in mid-2019.
Galaxy catalogs derived from full-depth stacks from the entire \wise\ and \neowise\ missions will have sufficent depth to detect \lstar\ galaxies in \wtwo\ out to $z\ga2$ and push much fainter than \lstar\ in \wone, enabling a higher completeness cluster search at $z\sim1$ and greater sensitivity to high-redshift ($z\simeq1.5-2$) clusters.  

Finally, from an algorithmic perspective, the increased sensitivity of CatWISE, coupled with the optical surveys, will enable a more sophisticated treatment of foreground rejection and should enable a detection observable that is a significantly lower scatter proxy for cluster mass.
The combination of a full-depth CatWISE catalog and the upcoming southern optical surveys together thus hold promise for a uniform, high-fidelity cluster search extending to $z>1.5$ and spanning the full extragalactic sky.

\section{Summary}
\label{sec:summary}

The Massive and Distant Clusters of \wise\ Survey (\madcows) is a program designed to identify massive galaxy clusters at $z\simeq 1$ over the full extragalactic sky using the combination of \wise\ imaging and ground-based optical photometry. \madcows\ uses the combination of optical rejection and infrared color-selection to isolate a 3.4$\mu$m flux-limited population of galaxies at $z\ga 0.8$, and then to search for overdensities on the expected physical scale of galaxy clusters at $z\sim1$. This approach provides
a large, wide-area sample of massive galaxy clusters at $z\simeq1$ for evolutionary studies, and allows the most massive galaxy clusters at this epoch to be identified over the full extragalactic sky.

The primary \madcows\ search covers the full extragalactic footprint of \panstarrs\ ($\delta>-30^\circ$). This search  uses the \allwise\ catalog coupled with  \panstarrs\ $i-$band photometry to effectively identify galaxy clusters at $0.7\la z\la 1.5$. The resultant catalog, which includes \npanstarrs\ cluster candidates, is  presented in Table \ref{tab:panstarrsclusters}. These clusters are selected based upon the peak amplitude in the cluster detection maps and all have SNR$>8$. 

We conduct a complementary \madcows\ search outside the \panstarrs\ footprint using the combination of \wise\ data and shallower SuperCOSMOS $r-$band photometry. This search yields \nsupercosmos\ cluster candidates, which are presented in Table \ref{tab:supercosmos}. 
The main limitation of this catalog, as discussed in the \S \ref{sec:galcuts}, is that it is more prone to contamination from lower-redshift clusters and chance projections due to the less efficient removal of low-redshift galaxies.

For the primary \wise---\panstarrs\ search, we have obtained follow-up \spitzer\ observations for \nspitzerpanstarrs\ clusters, enabling us to derive photometric redshifts and richness estimates. Using a subset of \nspecz\ clusters with spectroscopic redshifts and IRAC imaging, we find that these redshifts have an uncertainty of $\sigma_z/(1+z)=$\sigspecz. The median photometric redshift for the ensemble is $z=1.06$, and all photometric redshifts lie at $z>0.7$.
Similarly, photometric redshifts based upon \spitzer\ and DES observations for \nsupercosmosdes\ clusters imply a median redshift of $z=0.98$, with 94\% of candidates at $z>0.7$.

Sunyaev-Zel'dovich mass estimates for a subset of \ncarma\ clusters also enables us to derive an initial mass-richness relation for the \madcows\ sample. We find that the distribution of masses and redshifts is consistent with the majority of clusters obeying a tight relation ($\sigma_{\log M|\lambda}=0.06$), with a subset of merging systems offset to higher richness (or equivalently lower mass). Based upon this relation, we estimate that the median mass of the \wise---\panstarrs\  and \wise---SuperCOSMOS catalogs are \mfive$=1.6^{+0.6}_{-0.7}\times10^{14}$ M$_\odot$ and \mfive$=1.4\pm0.7\times10^{14}$ M$_\odot$, respectively.

Finally, we compare in the mass-redshift plane the distribution of \madcows\ clusters with other existing cluster samples (Figure \ref{fig:massredshift}).
The \madcows\ sample extends to comparably high redshifts as the published Sunyaev-Zel'dovich samples from ACT and SPT, while probing a range in cluster mass similar to the XDCP survey but over a much larger area.

Looking forward, the additional observations from the \neowise\ mission incorporated into the CatWISE catalog, coupled with upcoming data releases from southern optical surveys together promise to enable a second generation \madcows\ search extending towards $z\sim2$ and covering the full extragalactic sky. This second generation search will complement eROSITA and next generation SZ surveys.

\acknowledgements

The authors thank Alex Merson and Marc Postman for valuable discussions that facilitated completion of this project, and the anonymouse referee for suggestions that improved this paper.

Funding for this program is provided by NASA through the NASA Astrophysical Data Analysis Program, award NNX12AE15G. Parts of this work have also been supported through NASA grants associated with the Spitzer observations (PID 90177 and PID 11080) and with HST observations (HST-GO-14456), and by two NASA Keck PI Data Awards, administered by the NASA Exoplanet Science Institute. 

This publication makes use of data products from the Wide-field Infrared Survey Explorer, which is a joint project of the University of California, Los Angeles, and the Jet Propulsion Laboratory/California Institute of Technology, funded by the National Aeronautics and Space Administration.  This work is also based in part on observations made with the Spitzer Space Telescope, which is operated by the Jet Propulsion Laboratory, California Institute of Technology under a contract with NASA. The Millennium Simulation databases used in this paper and the web application providing online access to them were constructed as part of the activities of the German Astrophysical Virtual Observatory.

The Pan-STARRS1 Surveys (PS1) have been made possible through contributions of the Institute for Astronomy, the University of Hawaii, the Pan-STARRS Project Office, the Max-Planck Society and its participating institutes, the Max Planck Institute for Astronomy, Heidelberg and the Max Planck Institute for Extraterrestrial Physics, Garching, The Johns Hopkins University, Durham University, the University of Edinburgh, Queen's University Belfast, the Harvard-Smithsonian Center for Astrophysics, the Las Cumbres Observatory Global Telescope Network Incorporated, the National Central University of Taiwan, the Space Telescope Science Institute, the National Aeronautics and Space Administration under Grant No. NNX08AR22G issued through the Planetary Science Division of the NASA Science Mission Directorate, the National Science Foundation under Grant No. AST-1238877, the University of Maryland, and Eotvos Lorand University (ELTE).

The SZ results presented in this paper are based upon data from the Combined Array for Research in Millimeter-wave Astronomy. Support for CARMA construction was derived
from the Gordon and Betty Moore Foundation; the Kenneth T.
and Eileen L. Norris Foundation; the James S. McDonnell
Foundation; the Associates of the California Institute of
Technology; the University of Chicago; the states of California,
Illinois, and Maryland; and the National Science Foundation.
Ongoing CARMA development and operations are supported
by the National Science Foundation under a cooperative
agreement and by the CARMA partner universities.

The spectroscopic results presented in ths paper are based upon observations with the Gran Telescopio Canarias, Gemini Observatory, the W. M. Keck Observatory
This program included observations taken with the Gran Telescopio Canarias, Gemini Observatory, the W. M. Keck Observatory, and Combined Array for Research in Millimeter-wave Astronomy. 
Data from the GTC, located at the Observatorio del Roque de los Muchachos, were obtained through time allocated by the University of Florida.
Gemini Observatory is operated by the Association of Universities for Research in Astronomy, Inc. under a cooperative agreement with the NSF on behalf of the Gemini partnership: the National Science Foundation (United States), the National Research Council (Canada), CONICYT (Chile), Ministerio de Ciencia, Tecnolog\'{i}a e Innovaci\'{o}n Productiva (Argentina), and Minist\'{e}rio da Ci\^{e}ncia, Tecnologia e Inova\c{c}\~{a}o (Brazil).
Data from the W. M. Keck Observatory were obtained via time from telescope time allocated to the National Aeronautic and Space Administration through the agency's scientific partnership with the California Institute of Technology and the University of California. The Observatory was made possible by the generous financial support of the W.M. Keck Foundation. The authors wish to recognize and acknowledge the very significant cultural role and reverence that the summit of Mauna Kea has always had within the indigenous Hawaiian community.  We are most fortunate to have the opportunity to conduct observations from this mountain.

This research uses Dark Energy Survey (DES) data provided by the NOAO Data Lab. NOAO is operated by the Association of Universities for Research in Astronomy (AURA), Inc. under a cooperative agreement with the National Science Foundation.  Funding for the DES Projects has been provided by the U.S. Department of Energy, the U.S. National Science Foundation, the Ministry of Science and Education of Spain, the Science and Technology Facilities Council of the United Kingdom, the Higher Education Funding Council for England, the National Center for Supercomputing Applications at the University of Illinois at Urbana–Champaign, the Kavli Institute of Cosmological Physics at the University of Chicago, the Center for Cosmology and Astro-Particle Physics at the Ohio State University, the Mitchell Institute for Fundamental Physics and Astronomy at Texas A\&M University, Financiadora de Estudos e Projetos, Fundação Carlos Chagas Filho de Amparo à Pesquisa do Estado do Rio de Janeiro, Conselho Nacional de Desenvolvimento Científico e Tecnológico and the Ministério da Ciência, Tecnologia e Inovação, the Deutsche Forschungsgemeinschaft and the Collaborating Institutions in the Dark Energy Survey.

The Collaborating Institutions are Argonne National Laboratory, the University of California at Santa Cruz, the University of Cambridge, Centro de Investigaciones Enérgeticas, Medioambientales y Tecnológicas–Madrid, the University of Chicago, University College London, the DES-Brazil Consortium, the University of Edinburgh, the Eidgenössische Technische Hochschule (ETH) Zürich, Fermi National Accelerator Laboratory, the University of Illinois at Urbana-Champaign, the Institut de Ciències de l’Espai (IEEC/CSIC), the Institut de Física d’Altes Energies, Lawrence Berkeley National Laboratory, the Ludwig-Maximilians Universität München and the associated Excellence Cluster Universe, the University of Michigan, the National Optical Astronomy Observatory, the University of Nottingham, The Ohio State University, the OzDES Membership Consortium, the University of Pennsylvania, the University of Portsmouth, SLAC National Accelerator Laboratory, Stanford University, the University of Sussex, and Texas A\&M University.

Based in part on observations at Cerro Tololo Inter-American Observatory, National Optical Astronomy Observatory, which is operated by the Association of Universities for Research in Astronomy (AURA) under a cooperative agreement with the National Science Foundation.

%% Refs
%% -----------------------------------------------
\bibliographystyle{aasjournal}
\bibliography{all}

%% -------------------------------------------
\clearpage
\begin{turnpage}
%\LongTables
%\begin{landscape}
%\begin{deluxetable*}{lrrrrccccl}
\begin{deluxetable*}{lrrrrccrrrccl}
\centering
\tabletypesize{\scriptsize}
\tablewidth{0pt}
\tablecaption{Top \npanstarrs\ Candidate Clusters in the \wise---\panstarrs\ Region}
\tablehead{
%\colhead{Cluster}  & \colhead{Rank} & \colhead{$\alpha$} & \colhead{$\delta$} & \colhead{Peak  } & \colhead{SNR} &  \colhead{$z_{phot}$} &\colhead{$z$ (Members)}  & \colhead{$M_{500}$} & \colhead{Literature Names} & \colhead{References}\\
%\colhead{       }  & \colhead{    } & \colhead{[J2000] } & \colhead{[J2000] } & \colhead{Height} & \colhead{SNR} & \colhead{                  } & \colhead{           }  & \colhead{[$10^{14}\mathrm{M}_\odot$]} & \colhead{     } & \colhead{}
\colhead{Cluster}  & \colhead{$\alpha$} & \colhead{$\delta$} & \colhead{Peak  } & \colhead{SNR} & \colhead{$z_{phot}$} &\colhead{$z$ (Members)}  & \colhead{$\lambda_{15}$} & \colhead{$M_{500}$} & \colhead{\spitzer} &\colhead{Literature Names} & \colhead{Ref.}\\
\colhead{       }  & \colhead{[J2000] } & \colhead{[J2000] } & \colhead{Height} & \colhead{}    & \colhead{          } & \colhead{           }   &\colhead{}               & \colhead{[$10^{14}\mathrm{M}_\odot$]} & \colhead{} & \colhead{ \& Comments     } & \colhead{}
}
\startdata
MOO J0001+1428 &  00h01m09.1s & 14d28m57s & 0.59 & 9.0 & $1.14^{+0.13}_{-0.10}$ & \nodata & $15\pm4$  & \nodata & C11 & \nodata &  \\
MOO J0001+3644 &  00h01m09.8s & 36d44m38s & 0.56 & 9.9 & \nodata & \nodata & \nodata & \nodata & \nodata & \nodata &  \\
MOO J0001+3440 &  00h01m38.5s & 34d40m50s & 0.56 & 9.3 & \nodata & \nodata & \nodata & \nodata & \nodata & \nodata &  \\
MOO J0001$-$2447 &  00h01m49.4s & -24d47m32s & 0.58 & 8.4 & \nodata & \nodata & \nodata & \nodata & \nodata & \nodata &  \\
MOO J0001$-$2533 &  00h01m54.7s & -25d33m35s & 0.69 & 10.6 & $1.17^{+0.06}_{-0.05}$ & \nodata & $43\pm6$  & \nodata & C11 & \nodata &  \\
MOO J0002$-$0820 &  00h02m06.2s & -08d20m50s & 0.56 & 8.5 & \nodata & \nodata & \nodata & \nodata & \nodata & \nodata &  \\
MOO J0002+1751 &  00h02m29.5s & 17d51m30s & 0.60 & 9.5 & $1.33^{+0.04}_{-0.06}$ & \nodata & $26\pm5$  & \nodata & C11 & \nodata &  \\
MOO J0003$-$0903 &  00h03m01.2s & -09d03m24s & 0.66 & 10.0 & $0.94^{+0.07}_{-0.06}$ & \nodata & $18\pm5$  & \nodata & C11 & \nodata &  \\
MOO J0003$-$2925 &  00h03m28.3s & -29d25m58s & 0.84 & 12.9 & $1.05^{+0.10}_{-0.10}$ & \nodata & $22\pm5$  & \nodata & C11 & \nodata &  \\
MOO J0003$-$1341 &  00h03m37.3s & -13d41m16s & 0.69 & 10.5 & $0.85^{+0.06}_{-0.05}$ & \nodata & $32\pm6$  & \nodata & C11 & \nodata &  \\
MOO J0004$-$0232 &  00h04m26.7s & -02d32m44s & 0.56 & 8.5 & \nodata & \nodata & \nodata & \nodata & \nodata & \nodata &  \\
MOO J0004+0024 &  00h04m42.8s & 00d24m00s & 0.57 & 8.7 & \nodata & \nodata & \nodata & \nodata & \nodata & \nodata &  \\
MOO J0004+0108 &  00h04m52.1s & 01d08m36s & 0.61 & 9.3 & $0.94^{+0.09}_{-0.09}$ & \nodata & $21\pm5$  & \nodata & C11 & \nodata &  \\
MOO J0005+1329 &  00h05m28.6s & 13d29m32s & 0.58 & 8.8 & $0.94^{+0.08}_{-0.08}$ & \nodata & $34\pm6$  & \nodata & C11 & \nodata &  \\
MOO J0005+0024 &  00h05m29.2s & 00d24m00s & 0.56 & 8.6 & \nodata & \nodata & \nodata & \nodata & \nodata & \nodata &  \\
MOO J0005+1408 &  00h05m37.8s & 14d08m10s & 0.69 & 10.4 & $0.98^{+0.06}_{-0.06}$ & \nodata & $17\pm5$  & \nodata & C11 & \nodata &  \\
MOO J0005$-$0443 &  00h05m40.1s & -04d43m26s & 0.57 & 8.6 & \nodata & \nodata & \nodata & \nodata & \nodata & \nodata &  \\
MOO J0006+3050 &  00h06m29.3s & 30d50m55s & 0.65 & 10.9 & $1.02^{+0.06}_{-0.06}$ & \nodata & $42\pm7$  & \nodata & C11 & \nodata &  \\
MOO J0006$-$0244 &  00h06m34.3s & -02d44m42s & 0.57 & 8.7 & \nodata & \nodata & \nodata & \nodata & \nodata & \nodata &  \\
MOO J0006$-$0751 &  00h06m36.3s & -07d51m39s & 0.57 & 8.6 & \nodata & \nodata & \nodata & \nodata & \nodata & \nodata &  \\
MOO J0007$-$2108 &  00h07m13.6s & -21d08m36s & 0.56 & 8.4 & \nodata & \nodata & \nodata & \nodata & \nodata & \nodata &  \\
MOO J0008$-$1703 &  00h08m16.4s & -17d03m40s & 0.56 & 8.4 & \nodata & \nodata & \nodata & \nodata & \nodata & \nodata &  \\
MOO J0009$-$0750 &  00h09m34.3s & -07d50m12s & 0.62 & 9.4 & $1.14^{+0.17}_{-0.15}$ & \nodata & $22\pm5$  & \nodata & C11 & \nodata &  \\
MOO J0010+2027 &  00h10m18.9s & 20d27m28s & 0.63 & 10.0 & $0.88^{+0.12}_{-0.10}$ & \nodata & $14\pm5$  & \nodata & C11 & \nodata &  \\
  &    &   &     &     & $1.48^{+0.26}_{-0.10}$ & \nodata & $18\pm5$  & \nodata &   &   &   \\
MOO J0010+3142 &  00h10m47.7s & 31d42m14s & 0.66 & 10.9 & $1.41^{+0.05}_{-0.05}$ & \nodata & $33\pm5$  & \nodata & C11 & \nodata &  \\
MOO J0010+2751 &  00h10m54.1s & 27d51m10s & 0.58 & 9.7 & $0.98^{+0.10}_{-0.09}$ & \nodata & $20\pm5$  & \nodata & C11 & \nodata &  \\
MOO J0011$-$1414 &  00h11m14.4s & -14d14m46s & 0.60 & 9.0 & $1.03^{+0.08}_{-0.06}$ & \nodata & $35\pm6$  & \nodata & C11 & \nodata &  \\
MOO J0011$-$2530 &  00h11m31.3s & -25d30m55s & 0.56 & 8.7 & \nodata & \nodata & \nodata & \nodata & \nodata & \nodata &  \\
MOO J0012$-$0218 &  00h12m54.4s & -02d18m10s & 0.60 & 9.1 & $1.19^{+0.08}_{-0.07}$ & \nodata & $35\pm6$  & \nodata & C11 & \nodata &  \\
MOO J0012$-$1941 &  00h12m59.5s & -19d41m38s & 0.57 & 8.5 & \nodata & \nodata & \nodata & \nodata & \nodata & \nodata &  \\
MOO J0013+0700 &  00h13m48.1s & 07d00m52s & 0.59 & 8.9 & $1.09^{+0.07}_{-0.07}$ & \nodata & $33\pm6$  & \nodata & C11 & \nodata &  \\
MOO J0014$-$0909 &  00h14m33.3s & -09d09m23s & 0.61 & 9.2 & $0.89^{+0.05}_{-0.04}$ & \nodata & $16\pm6$  & \nodata & C11 & \nodata &  \\
MOO J0014$-$0459 &  00h14m40.5s & -04d59m23s & 0.62 & 9.5 & $1.02^{+0.04}_{-0.04}$ & \nodata & $32\pm6$  & \nodata & C9 & \nodata &  \\
MOO J0015+2059 &  00h15m06.2s & 20d59m58s & 0.57 & 9.1 & \nodata & \nodata & \nodata & \nodata & \nodata & \nodata &  \\
MOO J0015+2822 &  00h15m06.3s & 28d22m16s & 0.59 & 9.9 & $0.98^{+0.10}_{-0.09}$ & \nodata & $27\pm5$  & \nodata & C11 & \nodata &  \\
\label{tab:panstarrsclusters}
\tablecomments{The \npanstarrs\ highest significance galaxy cluster candidates in the \madcows\ catalog drawn from the \panstarrs\ region. The detection peak height is normalized such that the highest significance detection has a value of 1. %The relative ranking of the significance of all cluster detections is also given. 
The signal-to-noise ratio corresponding to each detection is also given in Column 5.
Column 6 lists the photometric redshifts from \spitzer\ when available, and column 7 provides spectroscopic redshifts and the number of confirmed members for clusters with spectroscopic data.  For clusters with \spitzer\ data, we also present the derived richness in Column 8 and indicate in Column 10 whether these data were from the Cycle 9 or Cycle 11 programs, or from the archive. We also include masses for clusters with SZ observations in Column 9.    This table is available in its entirety in machine-readable form.  A \textdagger\ in the reference column indicates that there is a foreground cluster along the line of sight. References: 
[1] \citet{postman1996}, [2] \citet{olsen1999}, [3] \citet{gonzalez2001}, [4] \citet{andreon2005}, [5] \citet{hilton2007}, [6] \citet{pacaud2007}, [7] \citet{andreon2008}, [8] \citet{muzzin2009}, [9] \citet{stern2010}, [10] \citet{durret2011}, [11] \citet{fassbender2011}, [12] \citet{gettings2012}, [13] \citet{mehrtens2012}, [14] \citet{brodwin2015}, [15] \citet{ford2014,ford2015}, [16] \citet{stanford2014}, [17] \citet{gonzalez2015}, [18] \citet{sifon2016}, [19] \citet{hilton2018}, [20] \citet{wh2018}, [21] \citet{decker2018}
}

\end{deluxetable*}
%\clearpage
%\end{landscape}
\end{turnpage}
%\clearpage
%\global\pdfpageattr\expandafter{\the\pdfpageattr/Rotate 90}

\begin{turnpage}
%\LongTables
%\begin{landscape}
%\begin{deluxetable*}{lrrrrccccl}
%\begin{deluxetable*}{lrrrcccccl}
%\centering
%\tabletypesize{\scriptsize}
%\tablewidth{0pt}
%\tablecaption{Candidate Clusters in the \wise---SuperCOSMOS region with \spitzer\ imaging}
%\tablehead{
%%\colhead{Cluster}  & \colhead{Rank} & \colhead{$\alpha$} & \colhead{$\delta$} & \colhead{Peak  } &  \colhead{$z$ (Members)}  & \colhead{$M_{500}$} & \colhead{Literature Names} & \colhead{References}\\
%%\colhead{       }  & \colhead{    } & \colhead{[J2000] } & \colhead{[J2000] } & \colhead{Height} &  \colhead{           }  & \colhead{[$10^{14}\mathrm{M}_\odot$]} & \colhead{     } & \colhead{}
%%\colhead{Cluster}  \colhead{Rank} & \colhead{$\alpha$} & \colhead{$\delta$} & \colhead{Peak  } &  \colhead{$z$ (Members)}  &  \colhead{$M_{500}$}                   &\colhead{Literature Names} & \colhead{Ref.}\\
%%\colhead{       }  & \colhead{    } & \colhead{[J2000] } & \colhead{[J2000] } & \colhead{Height} &  \colhead{           }    & \colhead{[$10^{14}\mathrm{M}_\odot$]} & \colhead{}           & \colhead{}
%}
\begin{deluxetable*}{lrrrrccrcl}
\centering
\tabletypesize{\scriptsize}
\tablewidth{0pt}
\tablecaption{Candidate Clusters in the \wise---SuperCOSMOS region with \spitzer\ imaging}
\tablehead{
%\colhead{Cluster}  & \colhead{Rank} & \colhead{$\alpha$} & \colhead{$\delta$} & \colhead{Peak  } & \colhead{SNR} &  \colhead{$z_{phot}$} &\colhead{$z$ (Members)}  & \colhead{$M_{500}$} & \colhead{Literature Names} & \colhead{References}\\
%\colhead{       }  & \colhead{    } & \colhead{[J2000] } & \colhead{[J2000] } & \colhead{Height} & \colhead{SNR} & \colhead{                  } & \colhead{           }  & \colhead{[$10^{14}\mathrm{M}_\odot$]} & \colhead{     } & \colhead{}
\colhead{Cluster}  & \colhead{$\alpha$} & \colhead{$\delta$} & \colhead{Peak  } & \colhead{SNR} & \colhead{$z_{phot}$}  & \colhead{$\lambda_{15}$} & \colhead{$M_{500}$} & \colhead{Literature Names} & \colhead{Ref.}\\
\colhead{       }  & \colhead{[J2000] } & \colhead{[J2000] } & \colhead{Height} & \colhead{}    & \colhead{          }  & \colhead{}                     & \colhead{[$10^{14}\mathrm{M}_\odot$]} & \colhead{\& Comments}  & \colhead{}
}
%MOO J0009$-$3411 &      270 &           00h09m42.21s     & -34d11m03.5s       & 0.602           & \nodata                    & \nodata                               & \nodata                                &  \nodata &  \\

\startdata
MOO J0002$-$3118 &  00h02m12.2s & -31d18m18s & 0.66 & 8.7 & \nodata & \nodata & \nodata  &  \nodata &  \\
MOO J0002$-$3419 &  00h02m24.6s & -34d19m32s & 0.71 & 9.3 & $1.42^{+0.05}_{-0.05}$ & $17\pm4$  & \nodata  &  \nodata &  \\
MOO J0003$-$7017 &  00h03m11.9s & -70d17m39s & 0.66 & 9.5 & \nodata & \nodata & \nodata  &  \nodata &  \\
MOO J0003$-$4725 &  00h03m21.9s & -47d25m48s & 0.80 & 11.0 & $1.45^{+0.05}_{-0.05}$ & $37\pm6$  & \nodata  &  \nodata &  \\
MOO J0007$-$3223 &  00h07m59.5s & -32d23m33s & 0.65 & 8.6 & \nodata & \nodata & \nodata  &  \nodata &  \\
MOO J0008$-$5332 &  00h08m07.9s & -53d32m49s & 0.65 & 8.9 & \nodata & \nodata & \nodata  &  \nodata &  \\
MOO J0008$-$4710 &  00h08m39.6s & -47d10m27s & 0.66 & 9.0 & \nodata & \nodata & \nodata  &  \nodata &  \\
MOO J0009$-$7206 &  00h09m56.2s & -72d06m23s & 0.61 & 8.8 & \nodata & \nodata & \nodata  &  \nodata &  \\
MOO J0010$-$4652 &  00h10m49.3s & -46d52m39s & 0.64 & 8.7 & \nodata & \nodata & \nodata  &  \nodata &  \\
MOO J0011$-$3126 &  00h11m20.7s & -31d26m37s & 0.63 & 8.4 & \nodata & \nodata & \nodata  &  \nodata &  \\
MOO J0012$-$4926 &  00h12m54.8s & -49d26m46s & 0.65 & 8.9 & \nodata & \nodata & \nodata  &  \nodata &  \\
MOO J0013$-$4452 &  00h13m57.7s & -44d52m31s & 0.65 & 8.8 & \nodata & \nodata & \nodata  &  \nodata &  \\
MOO J0016$-$6531 &  00h16m04.2s & -65d31m49s & 0.68 & 9.8 & \nodata & \nodata & \nodata  &  \nodata &  \\
MOO J0016$-$3302 &  00h16m48.4s & -33d02m25s & 0.73 & 9.7 & $1.05^{+0.09}_{-0.09}$ & $22\pm5$  & \nodata  &  \nodata &  \\
MOO J0019$-$6803 &  00h19m23.5s & -68d03m24s & 0.69 & 9.9 & \nodata & \nodata & \nodata  &  \nodata &  \\
MOO J0021$-$3240 &  00h21m32.8s & -32d40m22s & 0.61 & 8.1 & \nodata & \nodata & \nodata  &  \nodata &  \\
MOO J0028$-$4449 &  00h28m56.1s & -44d49m56s & 0.69 & 9.3 & $0.93^{+0.03}_{-0.03}$ & $29\pm6$  & \nodata  &  \nodata &  \\
MOO J0033$-$4629 &  00h33m15.5s & -46d29m41s & 0.61 & 8.4 & \nodata & \nodata & \nodata  &  \nodata &  \\
MOO J0033$-$5913 &  00h33m57.0s & -59d13m50s & 0.63 & 8.6 & \nodata & \nodata & \nodata  &  \nodata &  \\
MOO J0036$-$7637 &  00h36m45.8s & -76d37m37s & 0.65 & 9.5 & \nodata & \nodata & \nodata  &  \nodata &  \\
MOO J0037$-$5318 &  00h37m38.3s & -53d18m20s & 0.65 & 8.9 & \nodata & \nodata & \nodata  &  \nodata &  \\
MOO J0038$-$3134 &  00h38m46.4s & -31d34m48s & 0.63 & 8.3 & \nodata & \nodata & \nodata  &  \nodata &  \\
MOO J0039$-$3743 &  00h39m42.2s & -37d43m55s & 0.64 & 8.8 & \nodata & \nodata & \nodata  &  \nodata &  \\
MOO J0041$-$6154 &  00h41m56.9s & -61d54m04s & 0.62 & 8.6 & \nodata & \nodata & \nodata  &  \nodata &  \\
MOO J0042$-$5328 &  00h42m20.0s & -53d28m50s & 0.65 & 8.9 & \nodata & \nodata & \nodata  &  \nodata &  \\
MOO J0045$-$5226 &  00h45m21.4s & -52d26m17s & 0.62 & 8.5 & \nodata & \nodata & \nodata  &  \nodata &  \\
MOO J0046$-$4510 &  00h46m21.6s & -45d10m12s & 0.64 & 8.7 & \nodata & \nodata & \nodata  &  \nodata &  \\
MOO J0048$-$4110 &  00h48m44.1s & -41d10m16s & 0.61 & 8.3 & \nodata & \nodata & \nodata  &  \nodata &  \\
MOO J0054$-$3141 &  00h54m23.9s & -31d41m45s & 0.70 & 9.2 & $0.82^{+0.07}_{-0.05}$ & $18\pm6$  & \nodata  &  \nodata &  \\
MOO J0057$-$5107 &  00h57m34.8s & -51d07m41s & 0.62 & 8.5 & \nodata & \nodata & \nodata  &  \nodata &  \\
MOO J0100$-$5005 &  01h00m10.4s & -50d05m46s & 0.61 & 8.4 & \nodata & \nodata & \nodata  &  \nodata &  \\
MOO J0102$-$3116 &  01h02m41.5s & -31d16m39s & 0.68 & 8.9 & $1.02^{+0.10}_{-0.10}$ & $43\pm7$  & \nodata  &  \nodata &  \\
MOO J0105$-$7752 &  01h05m01.5s & -77d52m48s & 0.66 & 9.6 & \nodata & \nodata & \nodata  &  \nodata &  \\
MOO J0105$-$3355 &  01h05m05.3s & -33d55m25s & 0.74 & 9.8 & $0.98^{+0.10}_{-0.10}$ & $18\pm5$  & \nodata  &  \nodata &  \\
MOO J0107$-$3738 &  01h07m18.6s & -37d38m58s & 0.61 & 8.4 & \nodata & \nodata & \nodata  &  \nodata &  \\
\enddata
\label{tab:supercosmos}
\tablecomments{This table includes basic information for clusters from the SuperCOSMOS search for which there is existing \spitzer\ imaging. The columns are similar to Table \ref{tab:panstarrsclusters} with a few notable exceptions. Foremost, the peak height is normalized relative to the most significant peak in the SuperCOSMOS region. It is {\it not} directly comparable to the \panstarrs\ peak height column. %Similarly, the rankings are relative only to other clusters in this region.
%In this table we also do not quote a richness or photometric redshift. 
All quoted richnesses and photometric redshifts in this Table use DES for the optical photometry.
%It is not possible to these quantities outside the \panstarrs\ footprint due to the lack of sufficiently deep optical imaging. 
This table is available in its entirety in machine-readable form. Finally, the \spitzer\ column is omitted because all listed clusters have imaging from our Cycle 11-12 program, and the spectroscopic redshift column is omitted due to a lack of spectroscopic confirmation for this sample. References:
[1] \citet{bleem2015}.
 }
\end{deluxetable*}
%\clearpage
%\end{landscape}
\end{turnpage}
%\clearpage
%\global\pdfpageattr\expandafter{\the\pdfpageattr/Rotate 90}

\begin{turnpage}
%\LongTables
%\begin{landscape}
%\begin{deluxetable*}{lrrrrccccl}
\begin{deluxetable*}{lrrclrrcl}
\centering
\tabletypesize{\scriptsize}
\tablewidth{0pt}
\tablecaption{Additional Clusters from the Preliminary \wise---SDSS Search}
\tablehead{
\colhead{Cluster}   & \colhead{$\alpha$} & \colhead{$\delta$} &  \colhead{$z_{phot}$} &\colhead{$z$ (Members)}  & \colhead{$\lambda_{15}$} & \colhead{$M_{500}$} & \colhead{Literature Names} & \colhead{Ref.}\\
\colhead{       }   & \colhead{[J2000] } & \colhead{[J2000] } & \colhead{                  } & \colhead{           } &\colhead{} & \colhead{[$10^{14}\mathrm{M}_\odot$]} & \colhead{\& Comments     } & \colhead{}
}
\startdata
MOO J0005$-$0944 & 00h05m49.7s & -09d44m00s &  $1.49^{+0.26}_{-0.09}$ & \nodata & $27\pm5$  & \nodata &  &  \\
MOO J0012+1602 & 00h12m13.5s & 16d02m48s &  $0.97^{+0.03}_{-0.04}$ & 0.944 (23) & $67\pm8$  & \nodata &  & [1,3] \\
MOO J0022+0452 & 00h22m37.5s & 04d52m40s &  $0.96^{+0.05}_{-0.07}$ & \nodata & $25\pm6$  & \nodata &  &  \\
MOO J0023+0557 & 00h23m07.6s & 05d57m00s &  $1.02^{+0.10}_{-0.09}$ & \nodata & $43\pm7$  & \nodata &  &  \\
MOO J0026+1856 & 00h26m42.4s & 18d56m03s &  \nodata & \nodata & \nodata & \nodata &  &  \\
MOO J0036$-$2111 & 00h36m31.7s & -21d11m34s &  \nodata & \nodata & \nodata & \nodata & \textdagger &  \\
MOO J0045+0919 & 00h45m34.2s & 09d19m14s &  $0.81^{+0.07}_{-0.06}$ & \nodata & $34\pm6$  & \nodata &  &  \\
MOO J0045$-$0534 & 00h45m47.5s & -05d34m12s &  $0.97^{+0.09}_{-0.09}$ & \nodata & $20\pm5$  & \nodata &  &  \\
MOO J0121+0353 & 01h21m41.4s & 03d53m08s &  $1.29^{+0.05}_{-0.05}$ & \nodata & $36\pm6$  & \nodata &  &  \\
MOO J0123+0752 & 01h23m02.2s & 07d52m19s &  $0.93^{+0.07}_{-0.08}$ & \nodata & $27\pm6$  & \nodata &  &  \\
MOO J0134+0122 & 01h34m03.3s & 01d22m45s &  $1.28^{+0.06}_{-0.09}$ & \nodata & $37\pm6$  & \nodata &  &  \\
MOO J0140+2913 & 01h40m07.9s & 29d13m35s &  \nodata & \nodata & \nodata & \nodata & \textdagger &  \\
MOO J0153$-$0616 & 01h53m35.7s & -06d16m00s &  $0.93^{+0.08}_{-0.08}$ & \nodata & $36\pm6$  & \nodata &  &  \\
MOO J0204$-$1918 & 02h04m21.2s & -19d18m14s &  $1.10^{+0.05}_{-0.04}$ & \nodata & $47\pm7$  & \nodata &  &  \\
MOO J0207+0636 & 02h07m11.5s & 06d36m20s &  $0.86^{+0.09}_{-0.06}$ & \nodata & $30\pm6$  & \nodata &  &  \\
MOO J0211+2024 & 02h11m23.1s & 20d24m17s &  $1.08^{+0.04}_{-0.05}$ & \nodata & $45\pm7$  & \nodata &  &  \\
MOO J0222+1402 & 02h22m41.1s & 14d02m20s &  $0.99^{+0.07}_{-0.06}$ & \nodata & $20\pm5$  & \nodata &  &  \\
MOO J0222$-$0329 & 02h22m54.4s & -03d29m28s &  $1.14^{+0.10}_{-0.08}$ & \nodata & $30\pm5$  & \nodata &  &  \\
MOO J0224$-$0620 & 02h24m51.8s & -06d20m40s &  $1.32^{+0.05}_{-0.07}$ & 0.816 (7) & $38\pm6$  & \nodata & CFHTLENS 362136-6.33379 & [2,3] \\
MOO J0237+2809 & 02h37m10.8s & 28d09m54s &  $0.94^{+0.07}_{-0.07}$ & \nodata & $45\pm7$  & \nodata &  &  \\
MOO J0237$-$0806 & 02h37m46.4s & -08d06m45s &  $1.14^{+0.08}_{-0.06}$ & \nodata & $35\pm6$  & \nodata &  &  \\
MOO J0245+2018 & 02h45m08.4s & 20d18m22s &  $0.91^{+0.14}_{-0.11}$ & 0.757 (14) & $19\pm6$  & \nodata &  & [3] \\
MOO J0313+0014 & 03h13m35.3s & 00d14m40s &  $0.97^{+0.12}_{-0.12}$ & \nodata & $22\pm5$  & \nodata &  &  \\
MOO J0319$-$0025 & 03h19m24.7s & -00d25m24s &  $1.21^{+0.07}_{-0.08}$ & 1.194 (20) & $33\pm6$  & $ 3.0 ^{+0.4}_{-0.5}$ &  & [1,3] \\
MOO J0328$-$0532 & 03h28m52.5s & -05d32m01s &  $0.90^{+0.11}_{-0.09}$ & \nodata & $17\pm5$  & \nodata &  &  \\
MOO J0352+1123 & 03h52m05.6s & 11d23m01s &  $1.12^{+0.06}_{-0.07}$ & \nodata & $37\pm6$  & \nodata &  &  \\
MOO J0527+0057 & 05h27m56.3s & 00d57m27s &  $0.86^{+0.06}_{-0.04}$ & \nodata & $53\pm7$  & \nodata &  &  \\
MOO J0738+2757 & 07h38m39.9s & 27d57m03s &  $1.06^{+0.08}_{-0.07}$ & \nodata & $29\pm5$  & \nodata &  &  \\
MOO J0824+2251 & 08h24m56.1s & 22d51m27s &  $0.80^{+0.06}_{-0.05}$ & \nodata & $21\pm6$  & \nodata &  &  \\
  &   &   &  $1.21^{+0.07}_{-0.07}$ &   & $25\pm6$  &   &   &   \\
MOO J0826+4522 & 08h26m39.9s & 45d22m01s &  $1.26^{+0.05}_{-0.08}$ & \nodata & $37\pm6$  & \nodata &  &  \\
MOO J0828+1839 & 08h28m33.5s & 18d39m40s &  $0.88^{+0.09}_{-0.08}$ & \nodata & $4\pm4$  & \nodata &  &  \\
MOO J0830+2635 & 08h30m11.0s & 26d35m04s &  $1.10^{+0.10}_{-0.09}$ & \nodata & $22\pm5$  & \nodata &  &  \\
MOO J0838+2146 & 08h38m13.0s & 21d46m09s &  $0.81^{+0.08}_{-0.07}$ & \nodata & $12\pm5$  & \nodata &  &  \\
  &   &   &  $1.33^{+0.07}_{-0.07}$ &   & $25\pm5$  &   &   &   \\
MOO J0842+0033 & 08h42m24.6s & 00d33m13s &  $0.98^{+0.07}_{-0.06}$ & \nodata & $47\pm7$  & \nodata &  &  \\
MOO J0846+2747 & 08h46m46.1s & 27d47m10s &  $1.32^{+0.04}_{-0.06}$ & \nodata & $36\pm6$  & \nodata &  &  \\
\enddata
\label{tab:sdssclusters}
\tablecomments{This table lists clusters from the preliminary \wise---SDSS search that are not contained in Table \ref{tab:panstarrsclusters} but for which we have obtained spectroscopic redshifts, SZ masses, or Cycle 9 \spitzer\ imaging. Some of these clusters were lost due to more conservative masking in the \panstarrs\ search, while others lie below the peak height threshold for Table \ref{tab:panstarrsclusters}. For consistency within the table, we omit the peak height and SNR columns from the previous table. We also omit the column indicating the source of the \spitzer\ observations, since all clusters were observed in Cycle 9. 
This table is available in its entirety in machine-readable form. A \textdagger\ in the reference column indicates that there is a foreground cluster along the line of sight. References:
[1] \citet{brodwin2015}, [2] \citet{ford2014,ford2015}, [3] \citet{stanford2014}, [4] \citet{decker2018}, [5] \citet{wh2018}
}
\end{deluxetable*}
%\clearpage
%\end{landscape}
\end{turnpage}
%\clearpage
%\global\pdfpageattr\expandafter{\the\pdfpageattr/Rotate 90}

\clearpage
\appendix
\section{\spitzer-only Coordinates, Photometric Redshifts, and Richnesses}
\label{appendix:richness}
For the primary \wise---\panstarrs\ catalog in Table \ref{tab:panstarrsclusters} the quoted photometric redshifts and richnesses are derived using the cluster coordinates derived directly from the cluster search. As discussed in \S \ref{sec:astrometry}, for clusters with \spitzer\ IRAC imaging we also derived centroids using the \spitzer\ data. For completeness, we list in  Table \ref{tab:spitzer_appendix} these coordinates. We also present the photometric redshifts and richnesses that one would obtain using these centroids rather than the \wise\ coordinates. This is intended to enable consistency checks, but we emphasize that Table \ref{tab:panstarrsclusters} should be considered the fiducial catalog for the \panstarrs\ region. In this Table, we only include clusters for which the \spitzer\ centroiding algorithm was able to successfully recover a peak within 60$^{\prime\prime}$ of the \wise\ position, and only include photometric redshifts and richnesses when this association was within 45$^{\prime\prime}$.
%\begin{turnpage}
%\LongTables
%\begin{landscape}
%\begin{deluxetable*}{lrrrrccccl}
\begin{deluxetable*}{lrrrcr}
\centering
\tabletypesize{\scriptsize}
\tablewidth{0pt}
\tablecaption{Quantities Derived Using \spitzer\ Coordinates for \panstarrs\ Clusters}
\tablehead{
\colhead{Cluster} & \colhead{$\alpha_{S}$} & \colhead{$\delta_{S}$} & \colhead{\wise-\spitzer} & \colhead{$z_{phot,S}$}        & \colhead{$\lambda_{15,S}$} \\
\colhead{} & \colhead{[J2000]} & \colhead{[J2000]}  & \colhead{Separation[$^{\prime\prime}$]} & \colhead{                  } & \colhead{           }
}
\startdata
MOO J0001+1428 & 00h01m09.75s & 14d29m25.4s & 29 & $1.14^{+0.13}_{-0.10}$ & $15\pm4$ \\
MOO J0001$-$2533 & 00h01m54.22s & -25d33m21.9s & 15 & $1.17^{+0.06}_{-0.05}$ & $43\pm6$ \\
MOO J0002+1751 & 00h02m27.93s & 17d51m51.6s & 31 & $1.33^{+0.04}_{-0.06}$ & $26\pm5$ \\
MOO J0003$-$0903 & 00h03m01.49s & -09d03m21.5s & 5 & $0.94^{+0.07}_{-0.06}$ & $18\pm5$ \\
MOO J0003$-$2925 & 00h03m28.27s & -29d26m07.2s & 9 & $1.05^{+0.10}_{-0.10}$ & $22\pm5$ \\
MOO J0003$-$1341 & 00h03m38.64s & -13d41m54.4s & 43 & $0.85^{+0.06}_{-0.05}$ & $32\pm6$ \\
MOO J0003$-$1341 & 00h03m38.64s & -13d41m54.4s & 43 & $0.85^{+0.06}_{-0.05}$ & $32\pm6$ \\
MOO J0005+1329 & 00h05m28.65s & 13d29m28.0s & 4 & $0.94^{+0.08}_{-0.08}$ & $34\pm6$ \\
MOO J0005+1408 & 00h05m37.17s & 14d08m01.4s & 12 & $0.98^{+0.06}_{-0.06}$ & $17\pm5$ \\
MOO J0006+3050 & 00h06m26.82s & 30d51m08.9s & 34 & $1.02^{+0.06}_{-0.06}$ & $42\pm7$ 
\enddata
\label{tab:spitzer_appendix}
\tablecomments{As discussed in \S \ref{sec:astrometry}, for clusters with \spitzer\ imaging we derived \spitzer-based centroids. We list in this table these centroids and also the alternate  photometric redshifts and richness values that result if these centroids are used rather than those from the \wise\ search.  By construction, the \spitzer\ centroid must lie within 45$^{\prime\prime}$ of the \wise\ location. We only quote photometric redshifts ($z_{phot,S}$) and richnesses ($\lambda_{15,S}$) in cases where a \spitzer\ peak is identified within this distance. This table is available in its entirety in machine-readable form. }
\end{deluxetable*}
%\clearpage
%\end{landscape}
%\end{turnpage}
%\clearpage
%\global\pdfpageattr\expandafter{\the\pdfpageattr/Rotate 90}

\section{Spectroscopic Confirmation of PSZ2 G126.57+5161}
\label{appendix:redshifts}

This paper includes a new spectroscopic redshift for one cluster, PSZ2 G126.57+5161, which appears in our catalog as MOO J1229+6521. PSZ2 G126.57+5161 has a published redshift of $z=0.815$ based upon a single galaxy \citep{burenin2018}. Here we provide an improved redshift based upon multi-object spectroscopy. The cluster was observed on UT 2017 Mar 28 with OSIRIS on the GTC during $0\farcs6$ seeing. We obtained $3\times920$~s exposures on a single slit mask using the R2500R grism. Reductions were performed using standard {\tt IRAF} routines. 

In Table \ref{tab:planckcluster} we present the resulting redshift determinations for individual cluster galaxies. The quality flag can be interpreted as follows. Quality 3 indicates that a redshift is robust, with multiple well-determined features. Quality 2 redshifts are based upon at least one well-detected, unique feature. Quality 1 indicates that the redshift is based upon a single, weak emission line detection, and hence the redshift is uncertain.   From the six galaxies with quality 2 and 3 redshifts, we calculate a cluster redshift of 0.819 using a biweight average.

\begin{deluxetable}{lllcc}
\centering
\tablewidth{0pt}
\tablecaption{Spectroscopically Confirmed Members for MOO J1229+6521 (PSZ2 G126.57+5161)}
\tablehead{
\colhead{$\alpha$}  & \colhead{$\delta$} &\colhead{$z$ } & \colhead{Features} &\colhead{Quality} 
}
\startdata
12:29:47.52 & 65:21:13.8& 0.8163 & Ca HK & 3 \\	% 21
12:29:50.89 & 65:20:56.7& 0.8181 & Ca HK & 3 \\	% 17
12:29:52.94 & 65:22:19.9& 0.828  & D4000 & 2 \\	% 19
12:29:58.88 & 65:21:15.5& 0.829  & D4000 & 3 \\	% 16
12:30:01.60 & 65:21:03.6& 0.836  & [OII]$\lambda$3727 & 1 \\ %15
12:30:06.12 & 65:24:39.8& 0.8150 & Ca HK & 3 \\	% 13
12:30:11.41 & 65:20:08.6& 0.8127 & [OII]$\lambda$3727,D4000 & 3 	%slit id 11
\enddata
\label{tab:planckcluster}
\end{deluxetable}

\end{document}